\documentclass[twocolumn]{aastex631}

\usepackage{natbib}
\usepackage[style=american]{csquotes}
\newcommand{\tess}{\emph{TESS }}

\usepackage{scrextend}
\usepackage{changepage} 
\usepackage{rotating}

\revised{\today}
\accepted{----}

\usepackage[bulgarian,english]{babel}
\usepackage[T2A]{fontenc}
\shorttitle{Warm Jupiters around Solar analogues}
\shortauthors{Eberhardt et al.}
\graphicspath{{./}{figures/}}

\usepackage{array}
\usepackage{booktabs}
\usepackage{threeparttable}
\usepackage{graphicx}
\usepackage{siunitx}

\begin{document}


\title{Three Warm Jupiters around Solar-analog stars detected with \tess \footnote{Based on observations collected at the European Organization for Astronomical Research in the Southern Hemisphere under MPG programmes  0104.A-9007(A) and 0106.A-9014(A)}}

\author[0000-0003-3130-2768]{Jan Eberhardt}
\affil{Max-Planck-Institut für Astronomie,
              Königstuhl 17,
              D-69117 Heidelberg, Germany}

\author[0000-0002-5945-7975]{Melissa J.\ Hobson}
\affiliation{Max-Planck-Institut für Astronomie,
              Königstuhl 17,
              D-69117 Heidelberg, Germany}
\affil{Millennium Institute for Astrophysics, Chile}

\author[0000-0002-1493-300X]{Thomas Henning}
\affiliation{Max-Planck-Institut für Astronomie,
              Königstuhl 17,
              D-69117 Heidelberg, Germany}

\author[0000-0002-0236-775X]{Trifon Trifonov}
\affiliation{Max-Planck-Institut für Astronomie,
              Königstuhl 17,
              D-69117 Heidelberg, Germany}
\affiliation{Department
 of Astronomy, Sofia University ``St Kliment Ohridski'', 5 James Bourchier Blvd, BG-1164 Sofia, Bulgaria}
\affiliation{Landessternwarte, Zentrum f\"ur Astronomie der Universit\"at Heidelberg, K\"onigstuhl 12, D-69117 Heidelberg, Germany}


\author[0000-0002-9158-7315]{Rafael Brahm}
\affil{Facultad de Ingeniera y Ciencias, Universidad Adolfo Ib\'{a}\~{n}ez, Av. Diagonal las Torres 2640, Pe\~{n}alol\'{e}n, Santiago, Chile}
\affil{Millennium Institute for Astrophysics, Chile}

\author[0000-0001-9513-1449]{Nestor Espinoza}
\affil{Space Telescope Science Institute, 3700 San Martin Drive, Baltimore, MD 21218, USA}

\author[0000-0002-5389-3944]{Andr\'es Jord\'an}
\affil{Facultad de Ingeniera y Ciencias, Universidad Adolfo Ib\'{a}\~{n}ez, Av. Diagonal las Torres 2640, Pe\~{n}alol\'{e}n, Santiago, Chile}
\affil{Millennium Institute for Astrophysics, Chile}

\author[0000-0002-5113-8558]{Daniel Thorngren}
\affil{Department of Physics \& Astronomy, Johns Hopkins University, Baltimore, MD, USA}

\author[0000-0002-9020-7309]{Remo Burn}
\affiliation{Max-Planck-Institut für Astronomie,
              Königstuhl 17,
              D-69117 Heidelberg, Germany}


\author[0000-0003-3047-6272]{Felipe I.\ Rojas}
\affil{Instituto de Astrof\'isica, Facultad de F\'isica, Pontificia Universidad Cat\'olica de Chile, Chile}
\affil{Millennium Institute for Astrophysics, Chile}

\author[0000-0001-8128-3126]{Paula Sarkis}
\affiliation{Max-Planck-Institut für Astronomie,
              Königstuhl 17,
              D-69117 Heidelberg, Germany}

\author[0000-0001-8355-2107]{Martin Schlecker}
\affil{Department of Astronomy/Steward Observatory, The University of Arizona, 933 North Cherry Avenue, Tucson, AZ 85721, USA}
\affiliation{Max-Planck-Institut für Astronomie,
              Königstuhl 17,
              D-69117 Heidelberg, Germany}

\author{Marcelo Tala Pinto}
\affil{Facultad de Ingeniera y Ciencias, Universidad Adolfo Ib\'{a}\~{n}ez, Av. Diagonal las Torres 2640, Pe\~{n}alol\'{e}n, Santiago, Chile}
\affil{Millenium Institute of Astrophysics, Santiago, Chile}


\author[0000-0003-1464-9276]{ Khalid Barkaoui} 
\affil{Astrobiology Research Unit, Universit\'e de Li\`ege, All\'ee du 6 Ao\^ut 19C, B-4000 Li\`ege, Belgium}
\affiliation{Department of Earth, Atmospheric and Planetary Sciences,
Massachusetts Institute of Technology, Cambridge, MA 02139, USA}
\affiliation{Instituto de Astrof\'isica de Canarias (IAC), E-38200 La Laguna, Tenerife, Spain}

\author[0000-0001-8227-1020]{Richard P. Schwarz} 
\affiliation{Center for Astrophysics \textbar \ Harvard \& Smithsonian, 60 Garden St, Cambridge, MA 02138, USA}

\author{Olga Suarez} 
\affiliation{Universit\'e C\^ote d'Azur, Observatoire de la C\^ote d'Azur, CNRS, Laboratoire Lagrange, Bd de l'Observatoire, CS 34229, F-06304 Nice cedex 4, France}

\author[0000-0002-7188-8428]{Tristan Guillot} 
\affiliation{Universit\'e C\^ote d'Azur, Observatoire de la C\^ote d'Azur, CNRS, Laboratoire Lagrange, Bd de l'Observatoire, CS 34229, F-06304 Nice cedex 4, France}

\author[0000-0002-5510-8751]{Amaury H.M.J. Triaud}
\affiliation{School of Physics \& Astronomy, University of Birmingham, Edgbaston, Birmingham B15 2TT, United Kingdom}

\author[0000-0002-3164-9086]{Maximilian N. G\"unther} 
\affiliation{European Space Agency (ESA), European Space Research and Technology Centre (ESTEC), Keplerlaan 1, 2201 AZ Noordwijk, The Netherlands}

\author[0000-0002-0856-4527]{Lyu Abe} 
\affiliation{Universit\'e C\^ote d'Azur, Observatoire de la C\^ote d'Azur, CNRS, Laboratoire Lagrange, Bd de l'Observatoire, CS 34229, F-06304 Nice cedex 4, France}

\author[0009-0009-2966-7507]{Gavin Boyle} 
\affiliation{El Sauce Observatory, Coquimbo Province, Chile}
\affil{Cavendish Laboratory, J J Thomson Avenue, Cambridge, CB3 0HE, UK}

\author[0000-0002-6477-1360]{Rodrigo Leiva} 
\affil{Millennium Institute for Astrophysics, Chile}
\affiliation{Instituto de astrofísica de Andalucía, CSIC, Glorieta de la Astronomía s/n, E-18008 Granada, Spain}

\author[0000-0001-7070-3842]{Vincent Suc} 
\affiliation{Facultad de Ingeniera y Ciencias, Universidad Adolfo Ib\'{a}\~{n}ez, Av. Diagonal las Torres 2640, Pe\~{n}alol\'{e}n, Santiago, Chile}
\affiliation{El Sauce Observatory, Coquimbo Province, Chile}

\author[0000-0002-5674-2404]{Phil Evans} 
\affiliation{El Sauce Observatory, Coquimbo Province, Chile}

\author{Nick Dunckel} 
\affiliation{El Sauce Observatory, Coquimbo Province, Chile}

\author{Carl Ziegler} 
\affiliation{Department of Physics, Engineering and Astronomy, Stephen F. Austin State University, 1936 North Street, Nacogdoches, TX 75962}

\author{Ben Falk} 
\affil{Space Telescope Science Institute, 3700 San Martin Drive, Baltimore, MD 21218, USA}

\author{William Fong} 
\affiliation{Department of Physics and Kavli Institute for Astrophysics
and Space Research, Massachusetts Institute of Technology, Cambridge, MA
02139, USA}

\author{Alexander~Rudat} 
\affiliation{Department of Physics and Kavli Institute for Astrophysics
and Space Research, Massachusetts Institute of Technology, Cambridge, MA
02139, USA}

\author[0000-0002-1836-3120]{Avi Shporer} 
\affiliation{Department of Physics and Kavli Institute for Astrophysics
and Space Research, Massachusetts Institute of Technology, Cambridge, MA
02139, USA}

\author[0009-0008-5145-0446]{Stephanie Striegel}
\affiliation{SETI Institute, Mountain View CA 94043 USA/NASA Ames Research Center, Moffett Field CA 94035 USA}

\author[0000-0002-3555-8464]{David Watanabe} 
\affiliation{Planetary Discoveries, Valencia CA, USA}

 

\author[0000-0002-4715-9460]{Jon M. Jenkins} 
\affiliation{NASA Ames Research Center, Moffett Field, CA 94035, USA}

%
 
\author[0000-0002-6892-6948]{Sara Seager} 
\affiliation{Department of Physics and Kavli Institute for Astrophysics
and Space Research, Massachusetts Institute of Technology, Cambridge, MA
02139, USA}
\affiliation{Department of Earth, Atmospheric and Planetary Sciences,
Massachusetts Institute of Technology, Cambridge, MA 02139, USA}
\affiliation{Department of Aeronautics and Astronautics, MIT, 77
Massachusetts Avenue, Cambridge, MA 02139, USA}

\author[0000-0002-4265-047X]{Joshua N. Winn} 
\affiliation{Department of Astrophysical Sciences, Princeton University,
NJ 08544, USA}



\begin{abstract}
 
We report the discovery and characterization of three giant exoplanets orbiting solar-analog stars, detected by the \tess space mission and confirmed through ground-based photometry and radial velocity (RV) measurements taken at La Silla observatory with \textit{FEROS}. TOI-2373\,b is a warm Jupiter orbiting its host star every $\sim$ 13.3 days, and is one of the two most massive known exoplanet with a precisely determined mass and radius around a star similar to the Sun, with an estimated mass of m$_p$ = $9.3^{+0.2}_{-0.2}\,M_{\mathrm{jup}}$, and a radius of $r_p$ = $0.93^{+0.2}_{-0.2}\,R_{\mathrm{jup}}$. With a mean density of $\rho = 14.4^{+0.9}_{-1.0}\,\mathrm{g\,cm}^{-3}$, TOI-2373\,b is among the densest planets discovered so far. 
TOI-2416\,b orbits its host star on a moderately eccentric orbit with a period of $\sim$ 8.3 days and an eccentricity of $e$ = $0.32^{+0.02}_{-0.02}$.
TOI-2416\,b is more massive than Jupiter with $m_p$ = 3.0$^{+0.10}_{-0.09}\,M_{\mathrm{jup}}$, however is significantly smaller with a radius of $r_p$ = $0.88^{+0.02}_{-0.02},R_{\mathrm{jup}}$, leading to a high mean density of $\rho = 5.4^{+0.3}_{-0.3}\,\mathrm{g\,cm}^{-3}$.
TOI-2524\,b is a warm Jupiter near the hot Jupiter transition region, orbiting its star every $\sim$ 7.2 days on a circular orbit. It is less massive than Jupiter with a mass of $m_p$ = $0.64^{+0.04}_{-0.04}\,M_{\mathrm{jup}}$, and is consistent with an inflated radius of $r_p$ = $1.00^{+0.02}_{-0.03}\,R_{\mathrm{jup}}$, leading to a low mean density of $\rho = 0.79^{+0.08}_{-0.08}\,\mathrm{g\,cm}^{-3}$. The newly discovered exoplanets TOI-2373\,b, TOI-2416\,b, and TOI-2524\,b have estimated equilibrium temperatures of $860^{+10}_{-10}$ K, $1080^{+10}_{-10}$ K, and $1100^{+20}_{-20}$ K, respectively, placing them in the sparsely populated transition zone between hot and warm Jupiters.

\end{abstract}



\section{Introduction}
\label{sec:Introduction}
The discovery of an exoplanet around a solar-type star \citep{Mayor1995}, was one of the great achievements of modern astronomy. To date, the field of exoplanetary science has rapidly progressed, leading to the detection of over 5400 exoplanets\footnote{\url{http://exoplanet.eu}}, discovered with various astronomical techniques, such as radial velocity (RV) measurements, transit photometry, and direct imaging. The continued study of exoplanets has provided valuable insights into the diversity and frequency of planetary systems in the Galaxy.
With these systems showing a great diversity in physical and orbital characteristics, their study is fundamental for understanding planet formation and evolution. Especially interesting for testing theories of planet formation is the detection and  characterization of hot and warm Jupiter-mass planets \citep[see e.g.,][]{Dawson2018,Emsenhuber2021a}. 
Such massive planets in short-period orbits are easier to detect with the transit and RV methods than their long-period counterparts, which allow us to 
study their properties in depth. The orbits of hot Jupiters however, are influenced by tidal interactions with their host stars, leading to circularization on short timescales, erasing important information about their formation history. On the other hand, warm Jupiter planets are located at greater distances from their host stars, and are much more likely to maintain their postformation and migration orbital eccentricity, since the tidal interactions are weaker. Therefore, warm Jupiters provide an opportunity to  study the past interaction with the protoplanetary disk during planet formation and migration.

\begin{figure*}[tp]
    \centering
    \includegraphics[width=0.32\textwidth]{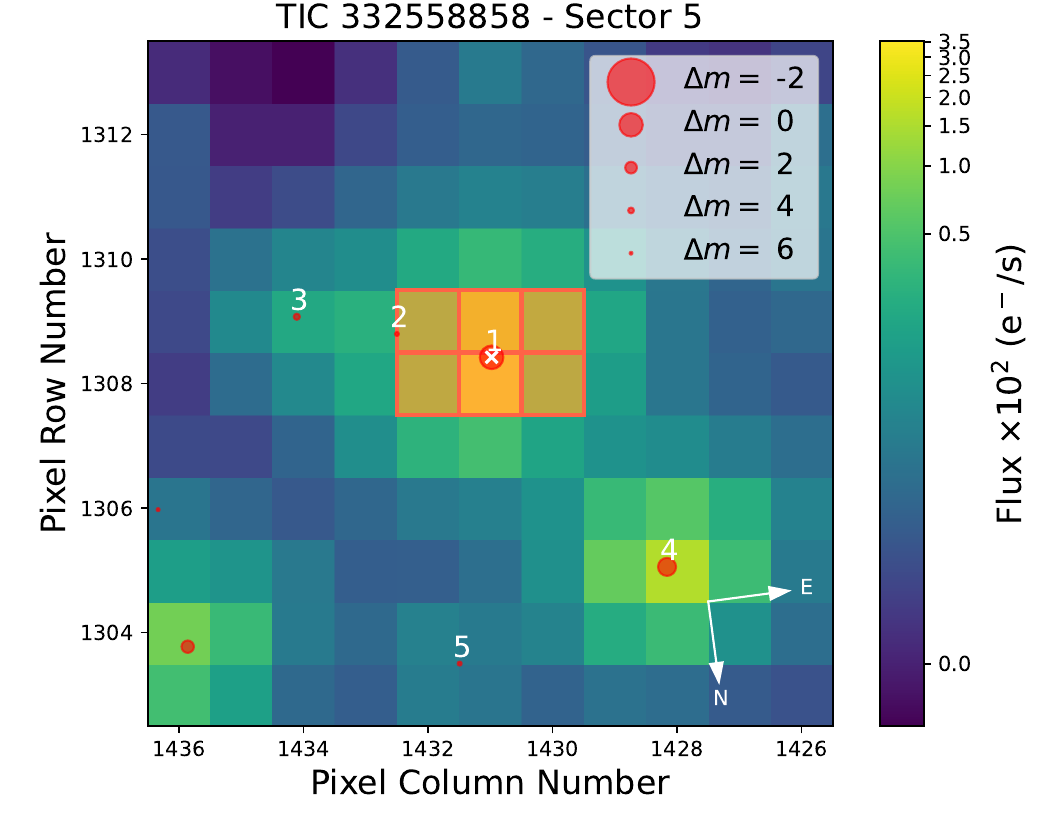}
    \includegraphics[width=0.32\textwidth]{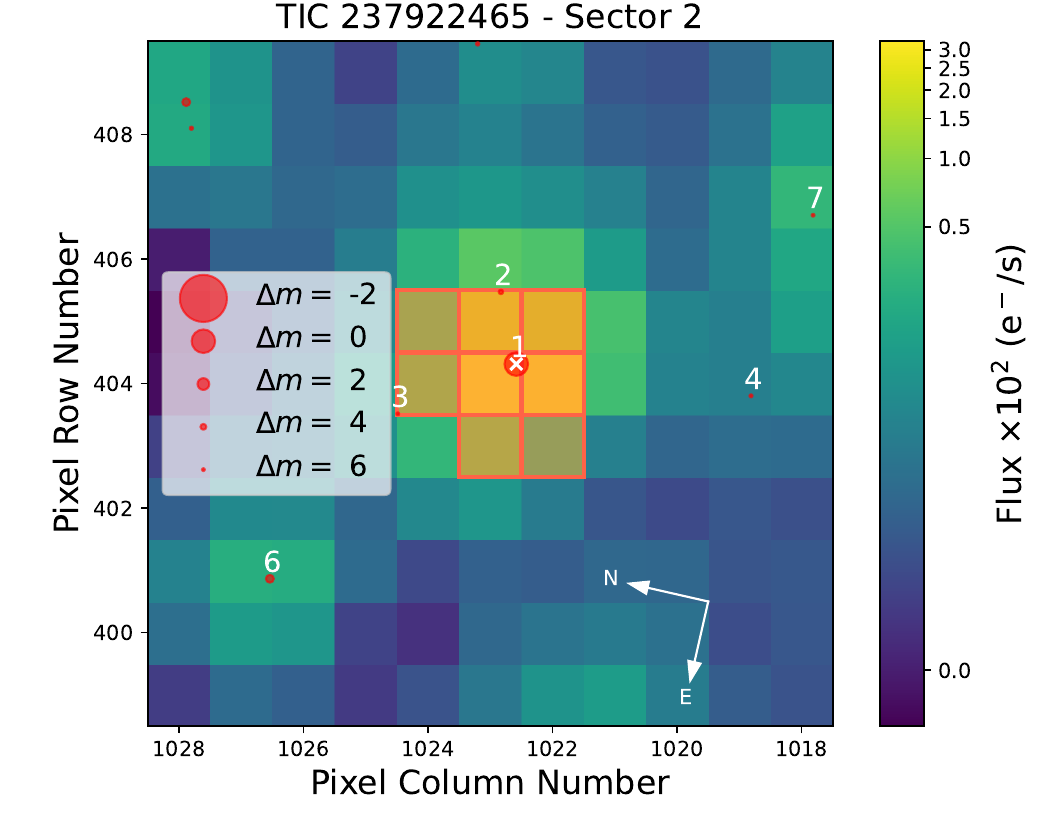}
    \includegraphics[width=0.32\textwidth]{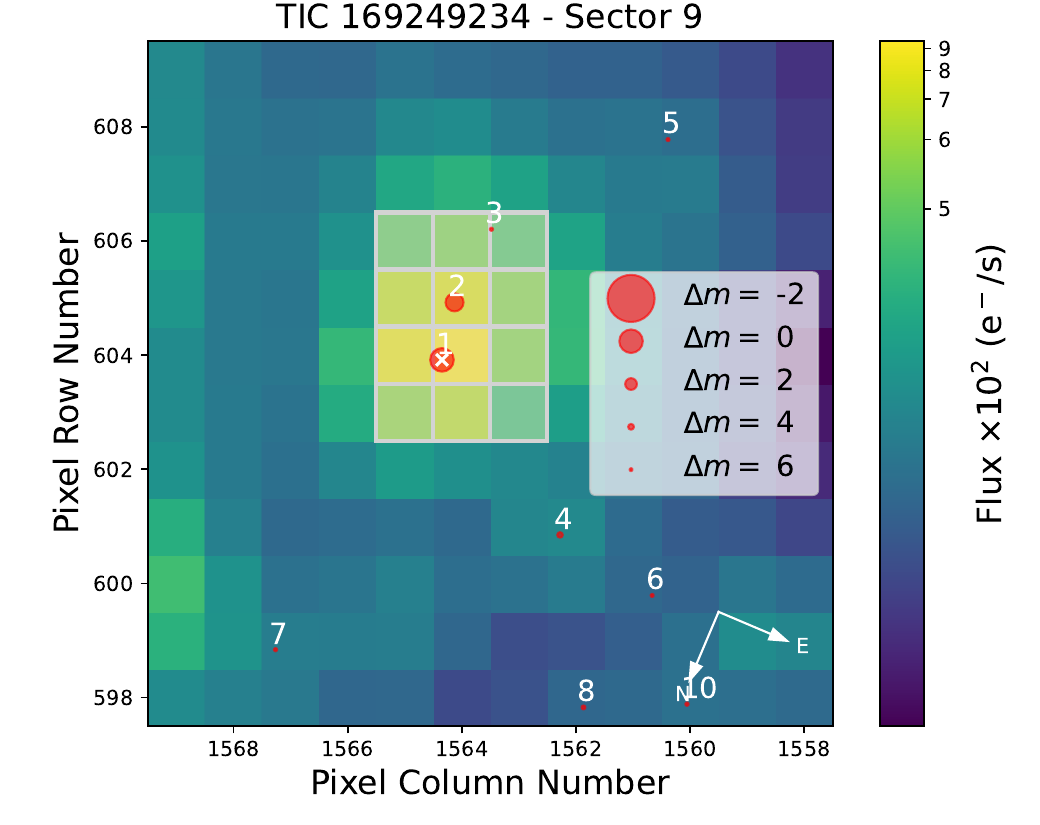}
    \caption{\tess TPFs of TOI-2373 (TIC\,332558858; \textit{left}), TOI-2416 (TIC\,237922465; \textit{middle}) and TOI-2524 (TIC\,169249234; \textit{right}), created using \texttt{tpfplotter}. Orange overlays show the apertures used to determine the flux (a white overlay indicates an automatically detected threshold aperture). Gaia DR2 catalog objects are shown as filled red circles sized according to their brightness relative to the target (marked with a white cross).}
    \label{fig:TPF}
\end{figure*}

\begin{figure*}[tp]
    \centering
    \includegraphics[width=0.32\textwidth]{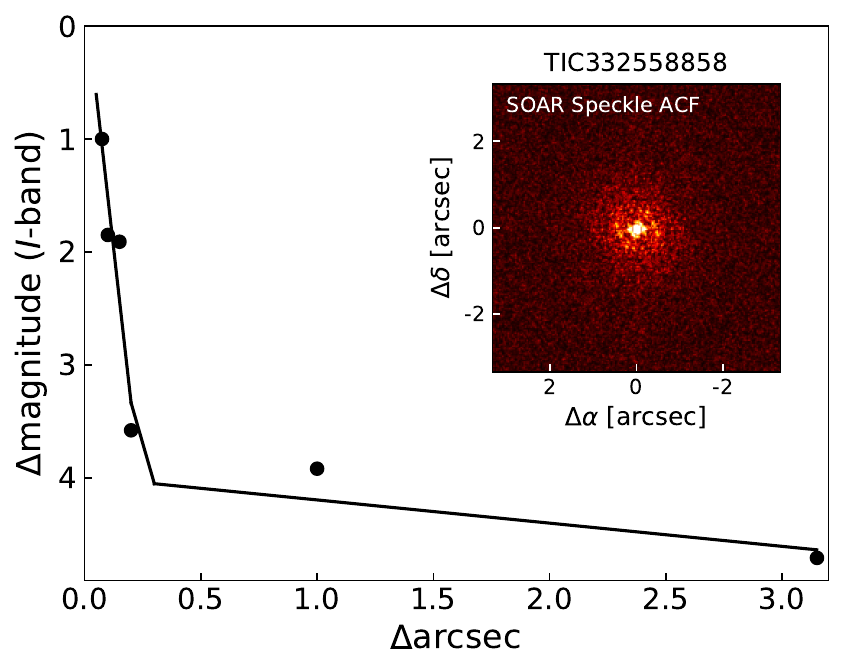}
    \includegraphics[width=0.32\textwidth]{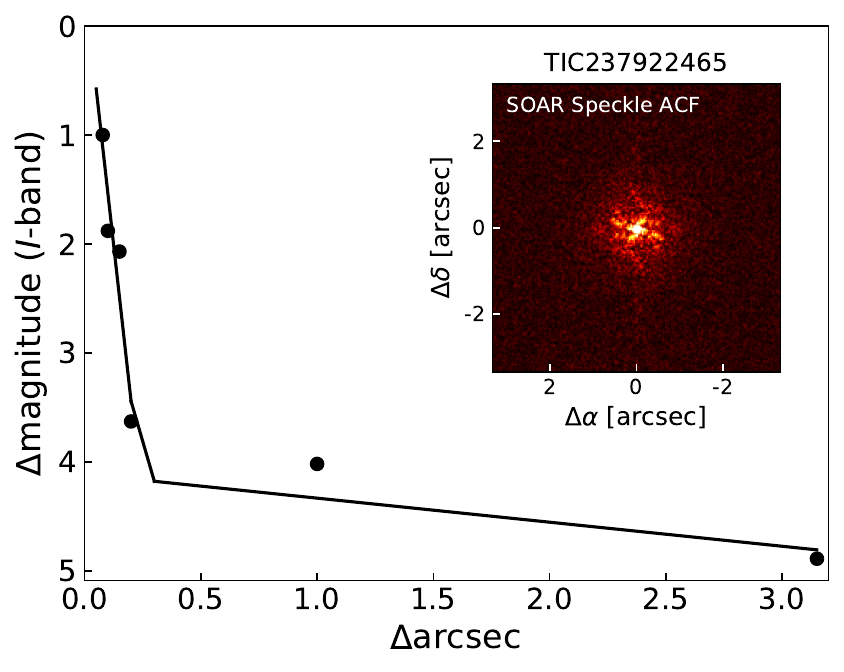}
    \includegraphics[width=0.32\textwidth]{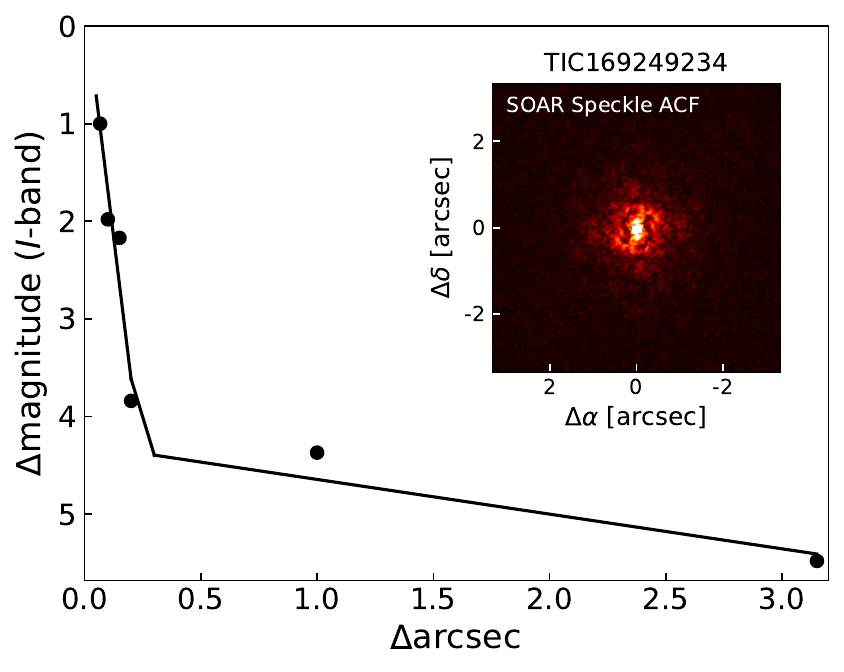}
    \caption{Speckle observations from HRCam at SOAR. The black points and solid curves represent the $5\sigma$ contrast curves. The insets show the speckle ACFs.}
    \label{fig:high_contrast}
\end{figure*}

With the {\bf W}arm g{\bf I}a{\bf N}ts with t{\bf E}ss (WINE) survey\footnote{\url{https://sites.google.com/view/wine-exoplanets/home}}, we aim for the confirmation and characterization of warm gas giants first identified in photometric data obtained from the Transiting Exoplanet Survey Satellite \citep[\tess,][]{Ricker2015}. This survey uses various spectroscopic facilities to provide Doppler validation for \tess planet candidates. The WINE follow-up survey has been highly successful to date, having detected and characterized many giant planets \citep[see e.g.,][]{Brahm2019,Kossakowski2019,Espinoza2020,Jordan2020,Brahm2020,Schlecker2020,Trifonov2021a,Hobson2021}, including 
a highly eccentric warm Jupiter \citep{Schlecker2020} and multiple planet systems consistent with a pair of two warm Jupiters near mean motion resonance 
\citep{Trifonov2021a,Bozhilov2023,Trifonov2023a}.\\

This work reports the discovery of three single, warm giant exoplanets as part of the WINE survey. The exoplanet companions are designated as TOI-2373\,b, TOI-2416\,b, and TOI-2524\,b, and are found in the transition phase between hot and warm Jupiters and orbit around stars similar to the Sun.
Adopting the definitions of \cite{Lehmann2022} for solar analogs \footnote{4.04\,$\mathrm{cm\,s}^{-2}<\log g < 4.84\,\mathrm{cm\,s}^{-2}$, $-0.3<$ [Fe/H] $<0.3$, 5472 K $<T_{\mathrm{eff}}<$ 6072 K} 
and solar twins \footnote{4.24\,$\mathrm{cm\,s}^{-2}<\log g < 4.64\,\mathrm{cm\,s}^{-2}$, $-0.1<$ [Fe/H] $<0.1$, 5672 K $<T_{\mathrm{eff}}<$ 5872 K}
a search in the TEPCAT catalog \citep{Southworth2011} reveals 276 exoplanets with known masses and radii orbiting solar analogs, of which only 25 orbit solar twins.
Therefore, the newly discovered TESS planets 
provide valuable insights into the formation and evolution of planets around G dwarf stars similar to the Sun.

In \ref{sec:Data}, we present the photometric and spectroscopic observational data used for our analysis. \ref{sec:AnalysisResults} describes the derivation of the stellar parameters from Fiber-fed Extended Range Optical Spectrograph (FEROS) spectra, as well our global analyses based on \tess photometry and FEROS precise Doppler measurements. Our results are discussed in \ref{sec:Discussion}, and finally, we provide a summary and conclusions of our work in \ref{sec:SummaryConclusion}.

\section{Observations}
\label{sec:Data}

\subsection{TESS photometry}
\label{sec:TESS}

During the first year of the \tess mission, the targets TOI-2373, TOI-2416, and TOI-2524 were observed in the 30 minute cadence mode in the following Sectors: 2, 3, 4, 8 (TOI-2416); 5 (TOI-2373); and 9 (TOI-2524).
These three systems were identified as candidates based on the analysis of the \texttt{tesseract}\footnote{\url{https://github.com/astrofelipe/tesseract}} (F. I. Rojas et al. 2023 in preparation.) generated light curves, where we automatically search for individual transit-like features produced by giant planets \citep[e.g.][]{Schlecker2020}.
In the extended \tess mission these targets were observed with a 2 minute cadence in the following Sectors: 28, 29, 30, and 38 (TOI-2416); 31 (TOI-2373, TOI-2416); and 35, 45, and 46 (TOI-2524).
The Science Processing Operations Center \citep[SPOC;][]{SPOC} at NASA Ames Research Center processed the 2 min data for each of these targets to calibrate pixels, provide photometry, and conducted single and multisector transiting planet searches using an adaptive, noise-compensating matched filter \citep{Jenkins2002,Jenkins2010,Jenkins2020}. The SPOC detected the transit signatures of TOI-2373, TOI-2416, and TOI-2524 on 2020 December 11, 2020 September 6, 2021 and December 9, respectively. An initial limb-darkened transit model was fitted \citep{Li2019} for the transit signatures, and a suite of diagnostic tests were conducted to help make or break the planetary nature of the signal \citep{Twicken2018}. The transit signatures all passed the diagnostic tests presented in the Data Validation reports. According to the difference image centroiding tests, the host star is located within \SI{0.74 \pm 2.64}{\arcsecond} of the transit signal source for TOI-2373, \SI{0.51 \pm 2.58}{\arcsecond} of the transit signal source for TOI-2416, and \SI{0.898 \pm 2.57}{\arcsecond} of the transit signal source for TOI-2524. The SPOC failed to find additional transiting planet signatures in each case. The transit signatures for all three targets were also detected in searches of Full Frame Image (FFI) data by the Quick Look Pipeline (QLP) at MIT \citep{Huang2020a,Huang2020b} as community-provided \tess Objects of Interest (TOIs). The TESS Science Office (TSO) reviewed the vetting information and issued an alert on 2020 November 11 for TOI-2416 based on the SPOC detection. TOI-2373 and TOI-2524 were alerted as CTOIs by the WINE collaboration on 2020 October 23 and 2021 March 2, respectively.

We obtained the light curves for the 2 minute cadence data by querying the Mikulski Archive for Space Telescopes (MAST)\footnote{\url{https://mast.stsci.edu/portal/Mashup/Clients/Mast/Portal.html}}. We use the data calculated by the SPOC \citep[][]{SPOC} at NASA Ames Research Center, which provides Simple Aperture Photometry (SAP) and systematics-corrected Presearch Data Conditioning \citep[PDC,][]{Smith2012, Stumpe2012, Stumpe2014} photometry. 

To account for possible contamination from other sources, we studied the Target Pixel Files (TPFs), generated with \texttt{tpfplotter} \citep{Aller2020}, adopting a standard magnitude limit of $\Delta m=$6. \ref{fig:TPF} shows the TPF plots for the sectors in which the first transits of TOI-2373, TOI-2416, and TOI-2524 were detected. The TPF plots show the field around the target observed by \tess, overplotted with an aperture grid, showing the pixels used to determine the flux, as well as nearby Gaia DR2 sources. The remaining TPF plots for the three targets can be inspected in \ref{fig:TPF_2373}, \ref{fig:TPF_2416}, and \ref{fig:TPF_2524}. Within the aperture of TOI-2373 we found one other source (TIC\,686510210), while the apertures of TOI-2416 and TOI-2524 each contained two additional sources (TIC 237922464 and TIC 650468769, and TIC 169249237 and TIC 169249240, respectively). We determined dilution factors for each 30 minute cadence \tess sector according to \cite{Espinoza2019}. The 2 minute PDCSAP cadence data from MAST  had already been corrected  for contamination from nearby stars and instrumental systematics, thus no dilution correction was applied by our team.

\subsection{Ground-based photometry}
Due to the \tess cameras' relatively large pixel scale of \SI{21}{\arcsecond} pixel$^{-1}$, nearby companions can contaminate the photometry. To confirm that the observed TESS signals were associated with the correct host stars, we used various ground-based facilities to observe TOI-2373, TOI-2416, and TOI-2524. The facilities are described in the following subsections.

\subsubsection{ASTEP}
Antarctica Search for Transiting ExoPlanets \citep[ASTEP;][]{Guillot2015, Mekarnia2016,Dransfield2022} is a 40 cm telescope located on the East Antarctic Plateau. It features a FLI Proline 16800E 4k $\times$ 4k CCD camera with a $\SI{1}{^{\circ}} \times \SI{1}{^{\circ}}$ field of view and a pixel scale of \SI{0.93}{\arcsecond} pixel$^{-1}$. ASTEP observed a full transit of TOI-2416\,b on 2021 June 28, which was not observed by \textit{TESS}.

\subsubsection{LCOGT}
Las Cumbres Observatory Global Telescope Network \citep[LCOGT,][]{Brown2013} is a worldwide network of \SI{1}{m} telescopes, equipped with 4096 $\times$ 4096 SINISTRO cameras. The cameras have a pixel scale of $\SI{0.389}{\arcsecond}$ pixel$^{-1}$, resulting in a $\SI{26}{\arcmin} \times \SI{26}{\arcmin}$ field of view. A full transit of TOI-2373\,b on 2021 November 10 was observed by the telescopes located at South Africa Astronomical Observatory (SAAO), Teide and Cerro Tololo Inter-American Observatory (CTIO), with SAAO having observed the ingress, CTIO the egress and Teide the full transit. CTIO already observed a partial transit on 2021 September 3.
All LCO science images were calibrated by the standard LCOGT {\tt BANZAI} pipeline \citep{McCully_2018SPIE10707E}, and photometric measurements were extracted using {\tt AstroImageJ} \citep{Collins2017}.

\subsubsection{Evans 0.36 m telescope at El Sauce}
El Sauce is a private observatory located in the Rio Hurtado province in Chile. TOI-2524 was observed with the Evans \SI{0.36}{m} telescope on 2021 March 31, obtaining a full transit light curve with the $R_c$ filter.
The telescope is equipped with an STT 1603-3 CCD camera with 1536 $\times$ 1024 pixels with an image scale of \SI{1.47}{\arcsecond} pixel$^{-1}$ when binned 2 $\times$ 2. The data were processed with the {\tt AstroImageJ} package \citep{Collins2017}.

TOI-2416 was observed with the CDK24ND telescope on 2020 October 23, obtaining an egress. The CDK24ND system consists of a PlaneWave \SI{0.61}{m} CDK (Corrected Dall-Kirkham) on a PlaneWave L-600 Direct-Drive mount using a Finger Lakes Instrumentation ProLine PL16803 CCD camera with no filter.

\subsection{Observatoire Moana}
Observatoire Moana (OM) is a global network of robotic telescopes. The station located at El Sauce Observatory (OM-ES), which  consists of a \SI{0.6}{m} CDK telescope coupled to an Andor iKon-L 936 deep depletion 2k $\times$ 2k CCD with a pixel scale of \SI{0.67}{\arcsecond} pixel$^{-1}$, was used on 2022 March 10 to obtain an egress for TOI-2524 using the Sloan $r'$ filter. The adopted exposure time was of 33 seconds. This same station was used on 2021 November 9 to obtain an egress of TOI-2373 using the  $r'$ filter and exposure times of 50 s.
The OM station located in SSO (OM-SSO), which consists of a \SI{0.5}{m} RCOS Ritchey Chretien telescope coupled to an FLI ML16803 4k $\times$ 4k CCD with a pixel scale of \SI{0.47}{\arcsecond} pixel$^{-1}$ operating with 2$\times$2 binning, was used to obtain an ingress for TOI-2416 on 2021 October 30. The adopted exposure time was of 49 s and an Astrodon Exoplanet (Clear Blue Blocking) filter was used. Data for both OM stations were processed with a dedicated pipeline that performs the CCD reduction steps along with the computation of the aperture photometry for all stars in the field of view and the generation of the final light curve by selecting the optimal comparison stars.

\subsection{High-resolution imaging}
\label{sec:HRimaging}

TOI-2373, TOI-2416, and TOI-2524 have all been observed as part of the SOAR \tess survey \citep{Ziegler2020}, in which high-resolution images of \tess planet candidate host stars have been acquired using speckle imaging with the high-resolution camera (HRCamera), mounted to the \SI{4.1}{m} Southern Astrophysical Research (SOAR) telescope, located at Cerro Pachón, Chile \citep{Tokovinin2018}. These high-resolution images aided in the identification and rejection of false-positive \tess candidates.

TOI-2373 and TOI-2416 were observed on 2020 December 3, while TOI-2524 was observed on 2022 April 15. For all three targets, no nearby sources were identified within \SI{3}{\arcsecond}. The contrast curves and auto-correlation functions (ACFs) are shown in Fig. \ref{fig:high_contrast}.

\begin{table*}[ht]

    \caption{Stellar parameters for the three targets discussed in this work along with their 1$\sigma$ uncertainties and floor uncertainties aaccording to \cite{Tayar2022} in parantheses.}
    \label{table:stellar_param}
    
    \centering
    \begin{tabular}{ p{4.0cm} p{5.0cm}  p{5.0cm} l}
    \hline\hline  \noalign{\vskip 0.5mm}
      Parameter & TOI-2373  & TOI-2416   &  TOI-2524\\
    \hline    \noalign{\vskip 0.5mm}
      $T_{\mathrm{eff}}$ (K) & 5651 $\pm$ 80 (113)& 5808 $\pm$ 80 (116)& 5831 $\pm$ 80 (116)\\ \noalign{\vskip 0.9mm}
      [Fe/H] (dex) & 0.3 $\pm$ 0.05 & 0.32 $\pm$ 0.05 & 0.06 $\pm$ 0.05 \\ \noalign{\vskip 0.9mm}
      Distance  (pc) & 496 $\pm$ 10 & 542 $\pm$ 8 & 429 $\pm$ 11 \\ \noalign{\vskip 0.9mm}
      Age (Gyr) & $5.9^{+1.7}_{-1.7}$ (1.2)& $4.9^{+1.2}_{-1.1}$ (1.0)& $6.7^{+1.7}_{-1.6}$ (1.3)\\ \noalign{\vskip 0.9mm}
      $v$ $\sin{i}$ (km\,s$^{-1}$) & 2.7 $\pm$ 0.5 & 2.4 $\pm$ 0.5 & 2.2 $\pm$ 0.5 \\ \noalign{\vskip 0.9mm}
      Mass ($M_{\odot}$) & $1.041^{+0.032}_{-0.028}$ (0.052)& $1.118^{+0.029}_{-0.027}$ (0.056)& $1.007^{+0.032}_{-0.029}$ (0.050)\\ \noalign{\vskip 0.9mm}
      $\log g$ (cm\,s$^{-2}$) & $4.371^{+0.024}_{-0.023}$ & $4.303^{+0.019}_{-0.020}$ & $4.344^{+0.026}_{-0.025}$ \\ \noalign{\vskip 0.9mm}
      Radius   ($R_{\odot}$) & $1.102^{+0.019}_{-0.019}$ (0.044) & $1.236^{+0.018}_{-0.018}$ (0.049)& $1.119^{+0.024}_{-0.023}$ (0.045)\\ \noalign{\vskip 0.9mm}
      Luminosity ($L{_\odot}$) & $1.125^{+0.058}_{-0.046}$ (0.023)& $1.574^{+0.077}_{-0.061}$ (0.031)& $1.307^{+0.079}_{-0.065}$ (0.026)\\ \noalign{\vskip 0.9mm}
      $\rho$ ($\mathrm{g\,cm}^{-3}$) & $1.096^{+0.083}_{-0.073}$ & $0.835^{+0.049}_{-0.048}$ & $1.013^{+0.083}_{-0.075}$ \\ \noalign{\vskip 0.9mm}
      $A_{\mathrm{V}}$ (mag) & $0.097^{+0.075}_{-0.062}$ & $0.097^{+0.070}_{-0.059}$ & $0.130^{+0.089}_{-0.077}$ \\ \noalign{\vskip 0.9mm}

    \hline\hline \noalign{\vskip 0.5mm}
    \end{tabular}
    
    \end{table*}

\subsection{FEROS Spectroscopy}
\label{sec:FEROS}

FEROS \citep[][]{Kaufer1999} is an echelle spectrograph mounted on the \SI{2.2}{m} MPG/ESO\footnote{MPG = {\em Max-Planck-Gesellschaft} = Max Planck Society; ESO = European Southern Observatory).} telescope at ESO's La Silla Observatory, Chile. FEROS is a highly efficient spectrograph with a resolving power of $R = 48,000$, covering the visible wavelength range from $\sim350$ to $\sim920$\,nm. FEROS's achievable radial velocity (RV) precision is approximately 10\,m\,s$^{-1}$, which makes it an ideal instrument for confirming the planetary nature of massive exoplanets discovered by \tess in the Southern Hemisphere.
During the time from 2020 February to 2021 March , we obtained 16 FEROS spectra for TOI-2373, 17 for TOI-2416, and eight for TOI-2524. From these spectra we extracted the RV measurements and the stellar activity indicators bisector span (BIS), H$\alpha$, He\,I, $\log{\mathrm{RHK}}$ and Na\,II using the Collection of Elemental Routines for Echelle Spectra (CERES) pipeline \citep[][]{Brahm2017}.
The FEROS RVs and stellar activity measurements for our targets are listed in Table \ref{tab:FEROS}.


\section{Analysis and Results}
\label{sec:AnalysisResults}

\subsection{Stellar Parameters}
\label{sec:stellar_parameters}

The stellar parameters for the three studied targets have been estimated using a similar method to previous discoveries from the WINE survey \citep[e.g.,][]{Brahm2019,Schlecker2020,Trifonov2021a,Hobson2021}.
 We computed the atmospheric parameters of the three targets from the coadded FEROS spectra by using the \texttt{zaspe} package \citep{zaspe}, which provides the effective temperature $T_{\mathrm{eff}}$, surface gravity $\log{g}$, metallicity [Fe/H], and the projected rotational velocity $v\sin{i}$. \texttt{zaspe} determines these by comparing the coadded spectra with a grid of synthetic spectra that were generated from the ATLAS9 model atmospheres \citep{atlas9}. The parameters of the best-fit model atmosphere are adopted as the parameters of the observed star.

To determine the physical parameters of the stars, we followed the procedure described in \cite{Brahm2019}. We used the Gaia DR2 parallaxes \citep{Gaia_Collaboration2016, Gaia_Collaboration2018b} to convert broadband photometric measurements from publicly available catalogs into absolute magnitudes and compared them to synthetic magnitudes from the PARSEC stellar evolutionary models \citep{Bressan2012}. For the employed stellar models, we fixed the stellar metallicities to those we found with \texttt{zaspe}, while using the effective temperature from \texttt{zaspe} as a prior. We determined the age, mass, luminosity, density, and extinction using the \texttt{emcee} package \citep{ForemanMackey2013} to sample the posterior distribution. We further obtained more precise values for $T_{\mathrm{eff}}$ and $\log{g}$ compared to the values obtained from \texttt{zaspe}. 
Table \ref{table:stellar_param} lists the atmospheric and physical properties of TOI-2373, TOI-2416, and TOI-2524, alongside their $1\sigma$ uncertainties.

\subsection{Period search analysis}
\label{sec:periodogram}

\begin{figure*}[tp]
    \centering
    \includegraphics[width=\textwidth]{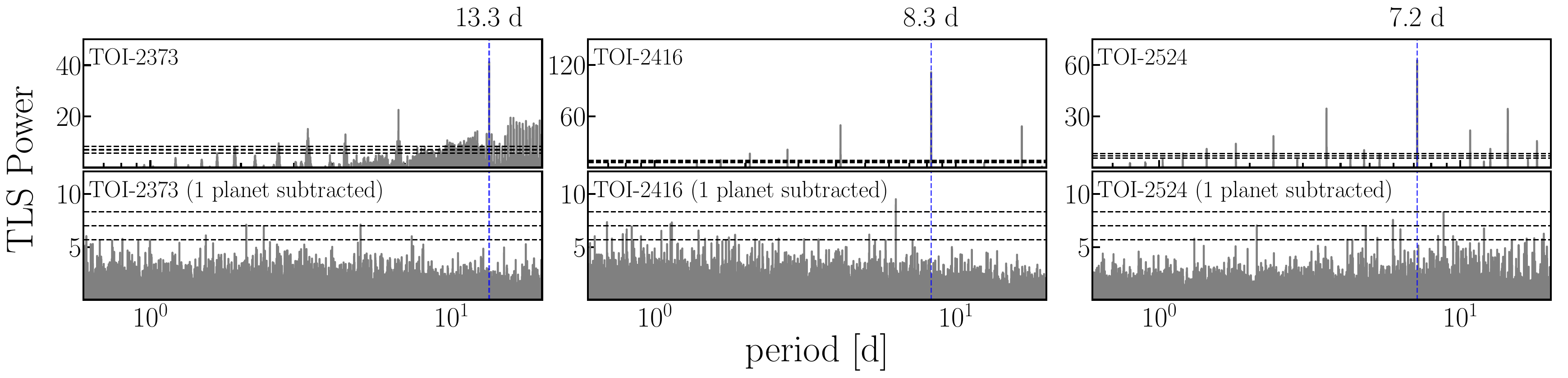}
    \caption{TLS analysis of TOI-2373, TOI-2416, and TOI-2524. The top panels show the TLS results of the raw light curves, the bottom panels show the TLS results after subtracting a photometric transit model. The dashed blue vertical lines show the transit signal periods. Dashed horizontal lines correspond to false-alarm-probability (FAP) levels of 10\%, 1\%, and 0.1\%.}
    \label{fig:TLS_periodograms}
\end{figure*}

We detrended the \tess transit light curves using the \texttt{wotan} package \citep{Hippke2019}. We chose a Gaussian Processes model with a Matern 3/2 kernel of size 5 days, while also taking dilution factors for each individual 30 minute \tess sector into account.
After detrending the \tess light curves (see Sections \ref{sec:TOI-2373} - \ref{sec:TOI-2524}), we performed a period search using the \texttt{transitleastsquares} package \cite[TLS;][]{Hippke2019b}.  The top panel of \ref{fig:TLS_periodograms} shows the results from the TLS analysis of the raw light curves after detrending, whereas the bottom panel shows the TLS analysis of the residuals after fitting of a best-fit transit model applied to the light curves.
\ref{fig:TLS_periodograms} indicates we detected very significant TLS signals for the three targets at periods of $\sim13.3$\,days, $\sim8.3$\,days, and $\sim7.2$\,days indicated by dashed blue lines. No further significant signals were detected in the residuals from our best-fit transit model, which indicates only one transiting planet in each target could be detected with \tess. \looseness=-4

\begin{figure*}
    \centering
    \includegraphics[width=\textwidth]{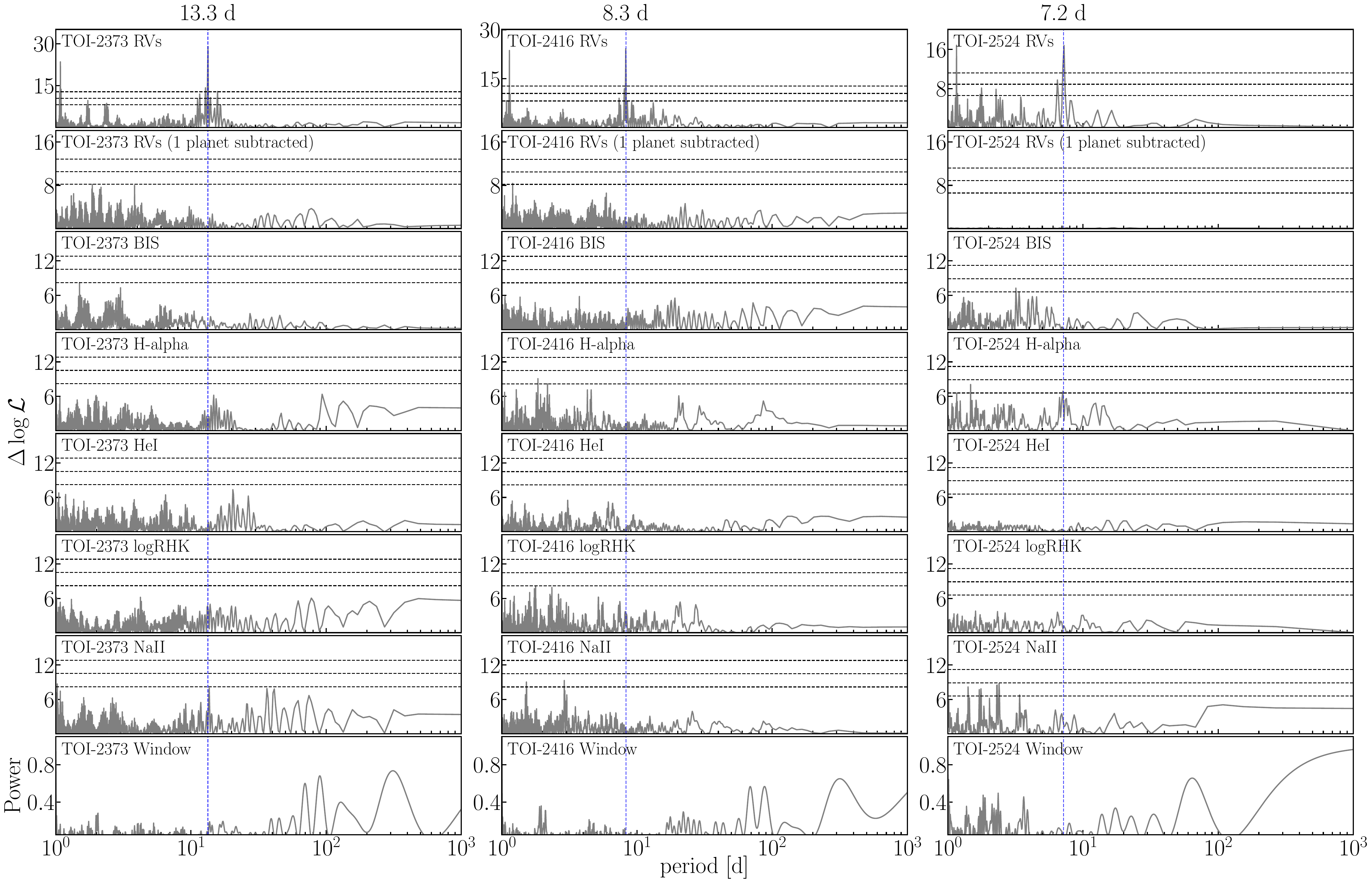}
    \caption{Maximum likelihood periodogram analysis of TOI-2373, TOI-2416, and TOI-2524. The top two panels show the periodograms of the RV data, before and after subtracting a Keplerian model, and the following panels show the periodograms of stellar activity indicators and the RV window function. Shown as dashed blue vertical lines are the transit signal periods. Dashed horizontal lines correspond to FAP levels of 10\%, 1\%, and 0.1\%.}
    \label{fig:RV_periodograms}
\end{figure*}

We employed the maximum likelihood periodograms \cite[MLPs;][]{Baluev2009,Zechmeister2019} to inspect the FEROS RV and stellar activity measurements for significant periodic signals.  \ref{fig:RV_periodograms} shows the results for the available RV and activity data. The top two panels of \ref{fig:RV_periodograms} show the power spectra for the  RVs and the RV residuals from a one-planet Keplerian fit, respectively. The following panels show the periodograms of the stellar activity data, and the window function of the FEROS RVs in the bottom panel. The orbital periods of the planet candidates are indicated as dashed blue lines. We determined the maximum possible rotational periods of the host stars $P_{\mathrm{rot}}/\sin{i}=2\pi R_{\star}/v\sin{i}$ and their 1$\sigma$ uncertainties by using the stellar radii and projected rotational velocities from Table \ref{table:stellar_param}. Using our determined orbital inclinations of the systems, and assuming orbits and stellar spins are aligned, we removed the dependency on the inclination and found rotational periods of $P_{\mathrm{rot}}=20.63_{-0.01}^{+0.01}$\,days, $26.04_{-0.06}^{+0.01}$\,days, and $25.71_{-0.02}^{+0.01}$\,d for TOI-2373, TOI-2416, and TOI-2524, respectively. The orbital periods of all three planet candidates reside outside the uncertainties of their host stars' rotational periods.

The MLPs of the RVs reveal significant signals at periods fully consistent with the results from the TLS, i.e., at $\sim13.3$\,days, $\sim8.3$\,days, and $\sim7.2$\,days for TOI-2373, TOI2416, and TOI-2524, respectively. Our analysis showed no counterparts to the RV signals in the stellar activity periodograms. While this is not necessarily excluding a stellar origin of the signals, the lack of stellar activity and the strong RV signals in phase with the detected transits convinces us that the signals are of planetary nature.

\subsection{Global Modelling}

For the combined analyses of the RV and photometry data for TOI-2373, TOI-2416, and TOI-2524, we used the \texttt{Exo-Striker}\footnote{\url{https://github.com/3fon3fonov/exostriker}} \citep{Trifonov2019_es} exoplanet toolbox. \texttt{Exo-Striker} employs the \texttt{batman} package \citep{Kreidberg2015} for light-curve transit models, and the formalism in \citet{Lee2003} for RV models. For posterior analysis, the \texttt{Exo-Striker} uses the nested sampling (NS) algorithm  \citep{Skilling2004} in conjunction  with the \texttt{dynesty} package \citep{Speagle2020}. The NS setup in this work is similar to that adopted in \citet{Trifonov2021a}. We run 100 "live-points" per parameter using a "dynamic" NS scheme. The priors for the parameters were estimated from consecutive NS runs, starting from parameters derived from TLS and MLP runs.

In the NS scheme, the orbital elements, transit, and RV data parameters, and transit light-curve parameters were modeled simultaneously. The fitted parameters are namely the RV semiamplitude $K$, the orbital period $P$, the eccentricity $e$, the argument of periastron $\omega$, 
the orbital inclination $i$, the time of midtransit $t_0$, and the relative semimajor axis and planetary radius $a/R_{\star}$ and $r/R_{\star}$. As well, we adopt quadratic limb-darkening coefficients $u_1$ and $u_2$ for the 
light-curve data. Additional parameters in our NS modeling are the transit and RV data offsets, and jitter parameters. The latter are added in quadrature to the instrumental data uncertainties to account for the unknown variance of the data, while we optimize the maximum $-\ln\mathcal{L}$ function of our combined model. From the resulting posterior probability distribution, we derive posteriors of the mass $m_p$, semimajor axis $a_p$, radius $r_p$, mean density $\rho_p$, and equilibrium temperature $T_{\mathrm{eq}}$ for each planet. 


For each target, we tested two competing models, a circular fit (i.e., $e=0$, while $\omega$ is undefined and forced to $0^{\circ}$) and a more complex, full Keplerian fit allowing eccentric orbits. We compared the competing models based on their Bayesian log-evidence $\ln{\mathcal{Z}}$ from the results of the NS fitting, computing $\Delta\ln{\mathcal{Z}}$ as $\ln{\mathcal{Z}_{\mathrm{ecc}}} - \ln{\mathcal{Z}_{\mathrm{circ}}}$. This model comparison was based on \citet{Trotta2008}. Two models are not considered distinguishable if their Bayesian log-evidence difference satisfies $\Delta \ln{\mathcal{Z}}\lesssim 2$. For $\Delta\ln{\mathcal{Z}}>2$, a model is moderately favored over another, while a difference of $\Delta\ln{\mathcal{Z}}>5$ indicates a strongly favored model. Below, we introduce each target analysis results individually, whereas \ref{tab:results} summarizes the relevant physical and orbital parameter estimates for TOI-2373, TOI-2416, and TOI-2524. The complete list of posterior estimates and $\ln\mathcal{Z}$ parameters are given in Tables \ref{tab:results_TOI-2373} - \ref{tab:results_TOI-2524}.


\subsubsection{TOI-2373}
\label{sec:TOI-2373}


 Our model comparison resulted in a difference of $\Delta\ln{\mathcal{Z}}=16.71$, thus favoring the eccentric model strongly against the circular model. Our final estimates show that TOI-2373\,b is a warm Jupiter with a mass of $m_p=9.3_{-0.2}^{+0.2}\,M_{\mathrm{jup}}$, a radius of $r_p=0.93_{-0.2}^{+0.2}\,R_{\mathrm{jup}}$, an orbital eccentricity of $e=0.112_{-0.009}^{+0.007}$, and a bulk density of $\rho_p=14.4_{-1.0}^{0.92}\,\mathrm{g\,cm}^{-3}$. It orbits its host star with an orbital period of $P=13.34$\,days corresponding to a distance of $a_p=0.11$\,au. The planetary equilibrium temperature is $T_{\mathrm{eq}}=860_{-10}^{+10}$ K.

\ref{fig:TIC332_TS} shows a time series of the photometric and RV data, along with the residuals underneath. The top panel shows the \tess light curves for Sectors 5 and 31, together with the transit model shown as a solid gray line. In both sectors, two transits are detected. The bottom panel shows the FEROS RV measurements and the Keplerian model as a gray line.
\ref{fig:TIC332_T_phase} shows the \tess photometry (top left), ground-based photometry (top right), and FEROS RV measurements (bottom) phase folded with the orbital period of 13.3\,days.

\begin{table*}
    \centering
    \caption{Median values of the relevant physical and orbital parameters for TOI-2373\,b, TOI-2416\,b, and TOI-2524\,b alongside their 1$\sigma$ uncertainties.}
    \begin{tabular}{ p{4.0cm} p{5.0cm}  p{5.0cm} l}
    \hline\hline  \noalign{\vskip 0.7mm}
          Parameter & TOI-2373\,b & TOI-2416\,b & TOI-2524\,b \\
    \hline \noalign{\vskip 0.7mm}

  $P$ (d) & $13.33668_{-0.00001}^{+0.00001}$ & $8.275479_{-0.000009}^{+0.000009}$ & $7.18585_{-0.00001}^{+0.00001}$ \\ \noalign{\vskip 0.9mm}
  $e$  & $0.112_{-0.009}^{+0.007}$ & $0.32_{-0.02}^{+0.02}$ & $0_{}^{}$ (fixed) \\ \noalign{\vskip 0.9mm}
  $i$ (degrees) & $89.2_{-0.2}^{+0.6}$ & $90.0_{-0.6}^{+0.6}$ & $89.4_{-0.4}^{+0.4}$ \\ \noalign{\vskip 0.9mm}
  $m$ ($M_{\mathrm{jup}}$) & $9.3_{-0.2}^{+0.2}$ & $3.00_{-0.09}^{+0.10}$ & $0.64_{-0.04}^{+0.04}$ \\ \noalign{\vskip 0.9mm}
  $r$ ($R_{\mathrm{jup}}$) & $0.93_{-0.02}^{+0.02}$ & $0.88_{-0.02}^{+0.02}$ & $1.00_{-0.03}^{+0.02}$ \\ \noalign{\vskip 0.9mm}
  $a$ (au) & $0.112_{-0.001}^{+0.001}$ & $0.0831_{-0.0007}^{+0.0007}$ & $0.0730_{-0.0007}^{+0.0007}$ \\ \noalign{\vskip 0.9mm}
  $\rho$ ($\mathrm{g\,cm}^{-3}$) & $14.4_{-1.0}^{+0.9}$ & $5.4_{-0.3}^{+0.3}$ & $0.79_{-0.08}^{+0.08}$ \\ \noalign{\vskip 0.9mm}
  $T_{\mathrm{eq}}$ (K) & $860_{-10}^{+10}$ & $1080_{-10}^{+10}$ & $1100_{-20}^{+20}$ \\ \noalign{\vskip 0.9mm}

    \hline\hline \noalign{\vskip 0.5mm}
    \end{tabular}
        \label{tab:results}
  \end{table*}

\subsubsection{TOI-2416}
\label{sec:TOI-2416}


As a result of our Bayesian log-evidence-based comparison, the eccentric model is very strongly favored against the circular model ($\Delta\ln{\mathcal{Z}}=27.84$). TOI-2416\,b is a warm Jupiter with a mass of $m_p=3.00_{-0.09}^{+0.10}\,M_{\mathrm{jup}}$, a radius of $r_p=0.88_{-0.02}^{+0.02}\,R_{\mathrm{jup}}$, an orbital eccentricity of $e=0.32_{-0.02}^{+0.02}$ and a bulk density of $\rho_p=5.4_{-0.3}^{+0.3}\,\mathrm{g\,cm}^{-3}$. It orbits its host star at a distance of $a_p=0.08$\,au corresponding to an orbital period of $P=8.28$\,days. Its equilibrium temperature is $T_{\mathrm{eq}}=1080_{-10}^{+10}$ K.

\ref{fig:TIC237_TS} shows a time series of the photometric and RV data along with the residuals underneath. The top panel shows the \tess light curves for Sectors 2, 3, 4, 8, 28, 29, 30, 31 and 38 together with the transit model shown as solid gray line. In all sectors, two to three transits can be observed. The bottom panel shows the FEROS RV measurements and the Keplerian model as a gray line.
\ref{fig:TIC237_T_phase} shows the \tess photometry (top left), ground-based photometry (top right), and FEROS RV measurements (bottom) phase folded with the orbital period of 8.3\,days.

\subsubsection{TOI-2524}
\label{sec:TOI-2524}


Considering a difference of $\Delta\ln{\mathcal{Z}}=-2.23$, the circular model is moderately favored against the eccentric model. Therefore, we chose to adopt the simpler circular model as our final result. TOI-2524\,b is a warm giant with a mass of $m_p=0.64_{-0.04}^{+0.04}\,M_{\mathrm{jup}}$, a radius of $r_p=1.00_{-0.03}^{+0.02}\,R_{\mathrm{jup}}$ and a bulk density of $\rho_p=0.79_{-0.08}^{+0.08}\,\mathrm{g\,cm}^{-3}$. It orbits its host star at a distance of $a_p=0.07$\,au corresponding to an orbital period of $P=7.19$\,days. Its equilibrium temperature is $T_{\mathrm{eq}}=1100_{-20}^{+20}$ K.

\ref{fig:TIC169_TS} shows a time series of the photometric and RV data along with the residuals underneath. The top panel shows the \tess light curves for Sectors 9, 35, 45, and 46 together with the transit model shown as solid gray line. In all sectors, two to four transits can be observed. The bottom panel shows the FEROS RV measurements and the Keplerian model as a gray line.
\ref{fig:TIC169_T_phase} shows the \tess photometry (top left), ground-based photometry (top right), and FEROS RV measurements (bottom) phase folded with the orbital period of 7.2\,days.

\begin{figure*}
    \centering
    \includegraphics[height=6cm]{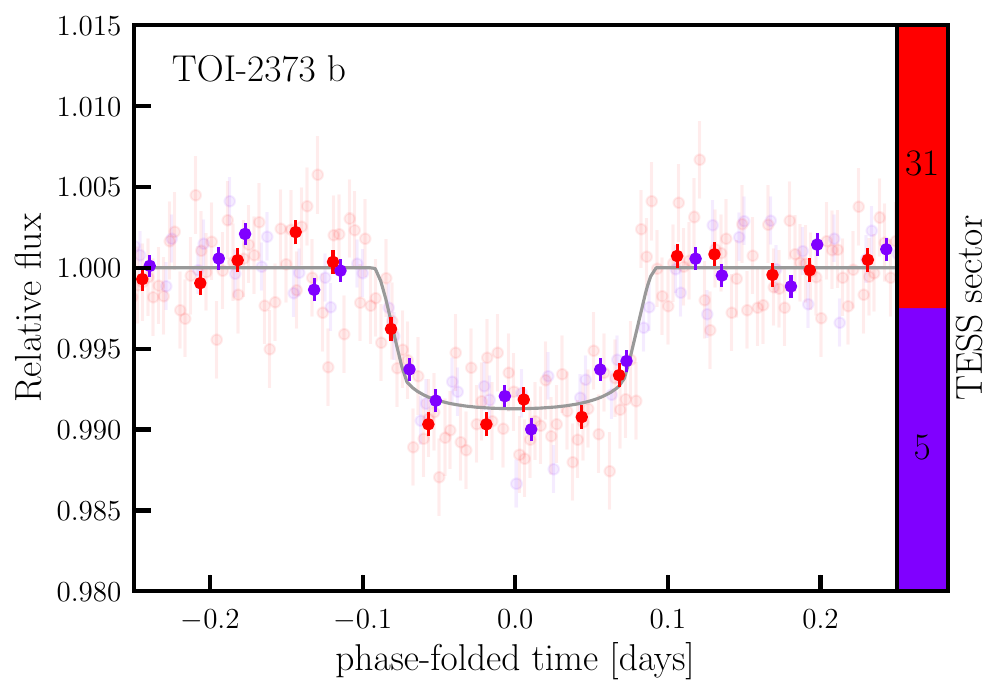}
    \includegraphics[height=6cm]{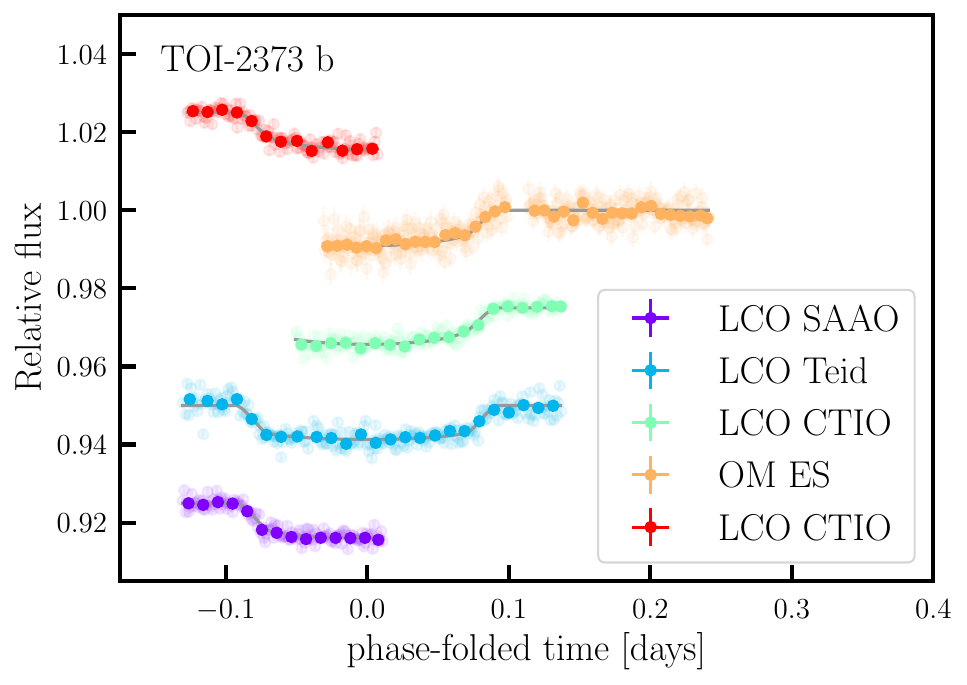}
    \includegraphics[width=.49\textwidth]{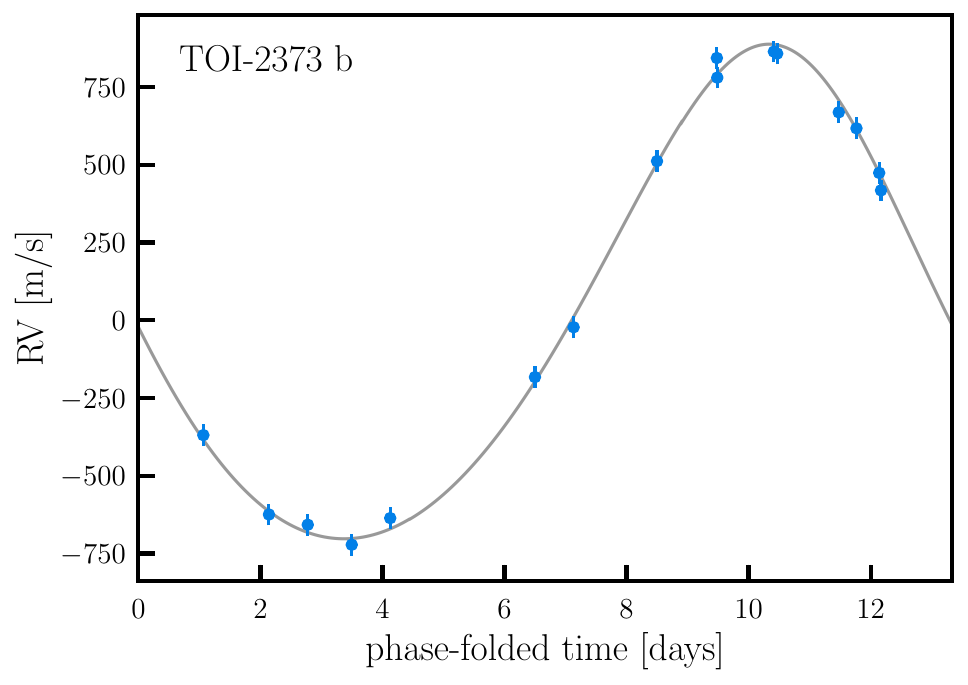}
    \caption{Phase plots for TOI-2373\,b. {\em Top left}: phase-folded light curves for all TESS sectors. Faint points correspond to the unbinned data, the strong points to binned data. {\em Top right}: phase-folded light curves of the photometry from the ground-based facilities. Individual data sets are offset for better visibility. LCO CTIO observed two different transits and thus appears twice.  {\em Bottom}: phase-folded radial velocities.}
    \label{fig:TIC332_T_phase}
\end{figure*}

\begin{figure*}
    \centering
    \includegraphics[height=6cm]{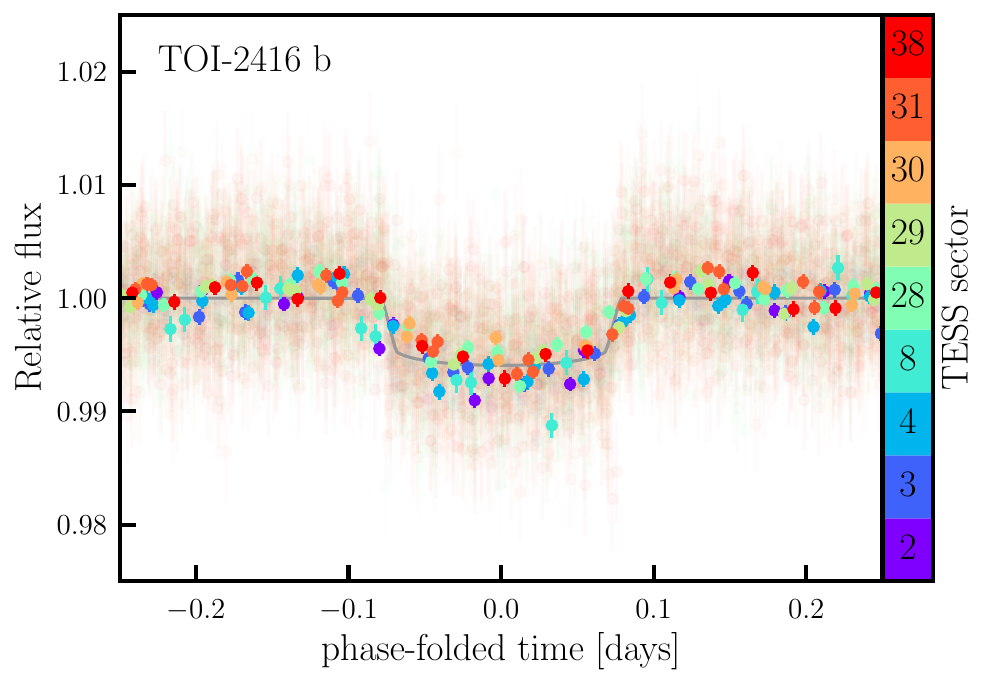}
    \includegraphics[height=6cm]{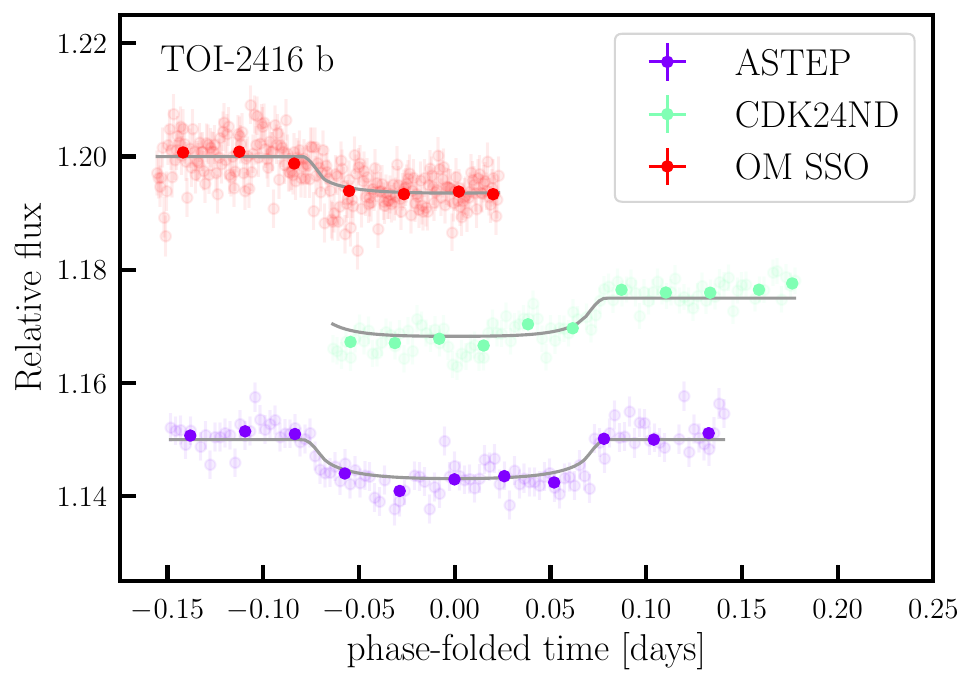}
    \includegraphics[width=.49\textwidth]{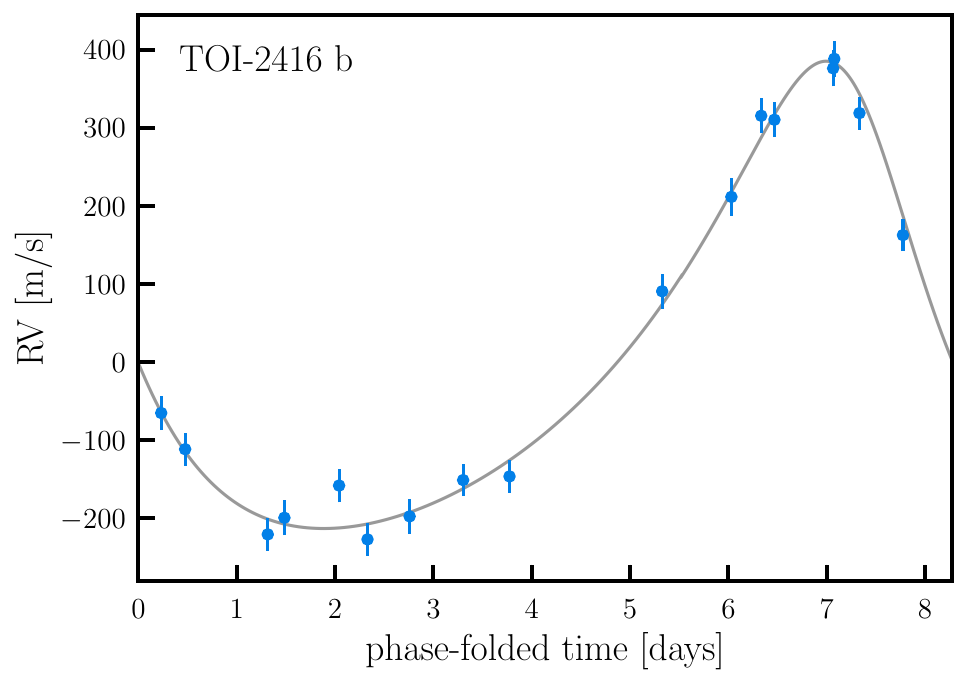}
    \caption{Same as \ref{fig:TIC332_T_phase}, but for  TOI-2416\,b.}
    \label{fig:TIC237_T_phase}
\end{figure*}

\subsection{Heavy element mass}

Following \cite{Thorngren2019}, we determined the bulk metallicity fraction $Z_p$ for TOI-2373\,b, TOI-2416\,b, and TOI-2524\,b as well as the anomalous heating efficiency $\epsilon$ for TOI-2416\,b and TOI-2524\,b. The estimation of the planetary metallicity is based on an Markov Chain Monte Carlo inversion of a planet evolution model \citep[see][]{Thorngren2019}, using the planetary mass, bulk metallicity, age, and log anomalous heating fraction as parameters. The priors are either taken from observations (mass and age) or based on a mass-metallicity relation \citep[for metallicity;][]{Thorngren2016} and equilibrium temperature \citep[for heating;][]{Thorngren2018}. The model uses these parameters to compute a radius which it compares to the observed radius. The anomalous heating efficiency is defined as the logarithm of the heat injected into a planet as a fraction of the incident stellar flux \citep{Thorngren2018}. It is a parameterization of an unknown mechanism transporting energy from irradiation to the interior of the planet and causing the planet to inflate (see Sect. \ref{sec:Discussion}).

We found bulk metallicities of $Z_p=0.11^{+0.06}_{-0.05}$, $Z_p=0.36^{+0.03}_{-0.03}$, and $Z_p=0.24^{+0.03}_{-0.03}$ for TOI-2373\,b, TOI-2416\,b, and TOI-2524\,b, respectively. As well, we estimated heating efficiencies of \textbf{$\epsilon = 0.34^{+0.012}_{-0.018}\%$} and \textbf{$\epsilon = 0.40^{+0.014}_{-0.021}\%$} for TOI-2416\,b and TOI-2524\,b, respectively. TOI-2373\,b has an equilibrium temperature of $860^{+10}_{-10}$\,K and is therefore too cold for any anomalous heating. 

\begin{figure*}
    \centering
     \includegraphics[height=6cm]{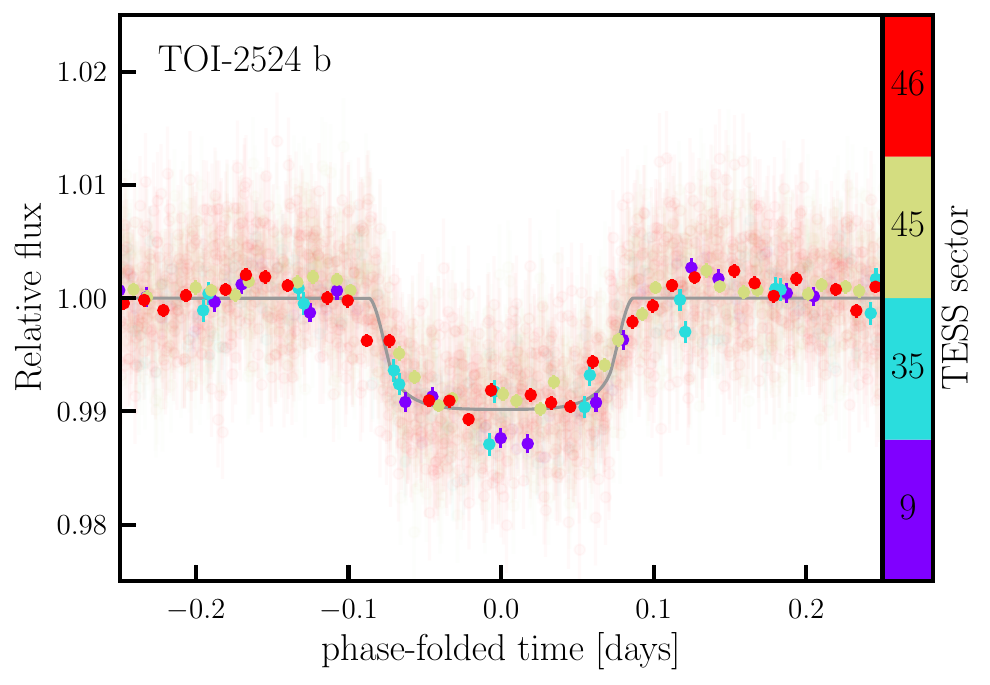}
    \includegraphics[height=6cm]{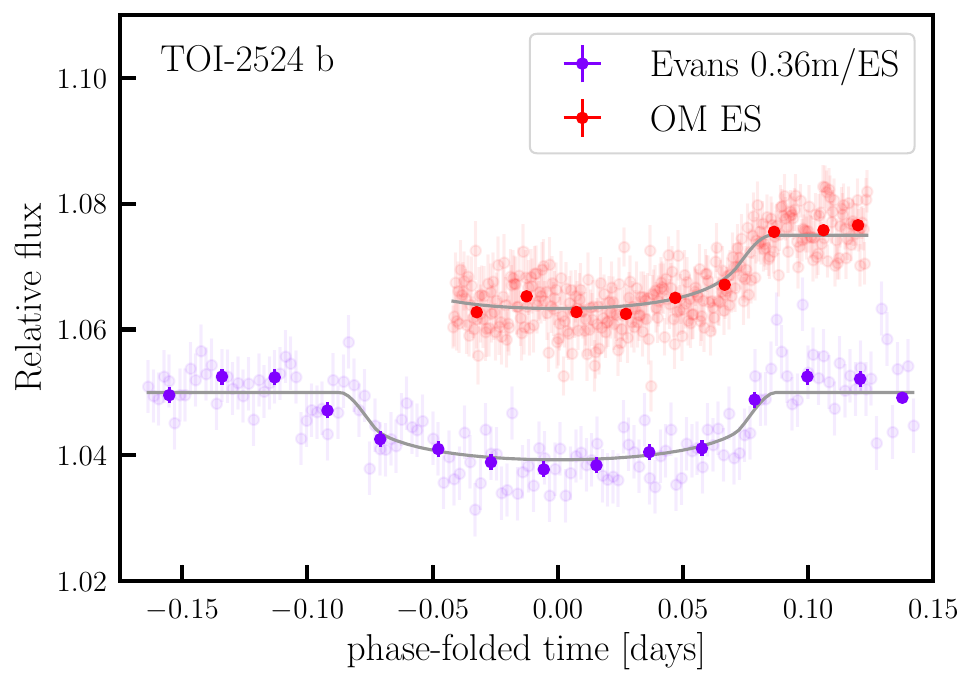}
    \includegraphics[width=.49\textwidth]{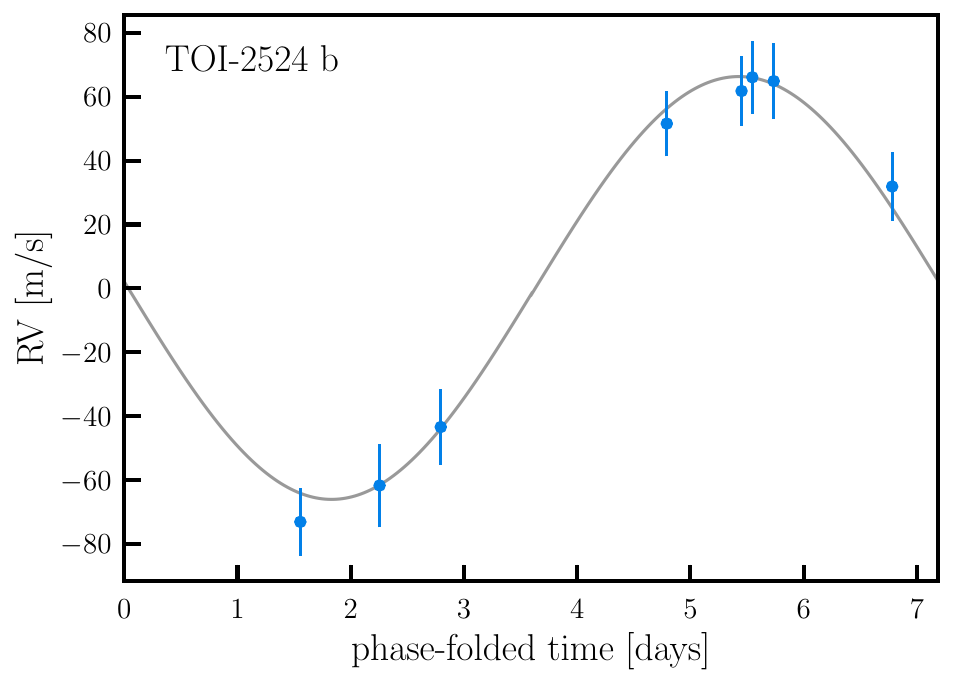}
    \caption{Same as \ref{fig:TIC332_T_phase}, but for  TOI-2524.}
    \label{fig:TIC169_T_phase}
\end{figure*}

\begin{figure}
    \centering
    \includegraphics[width=9cm]{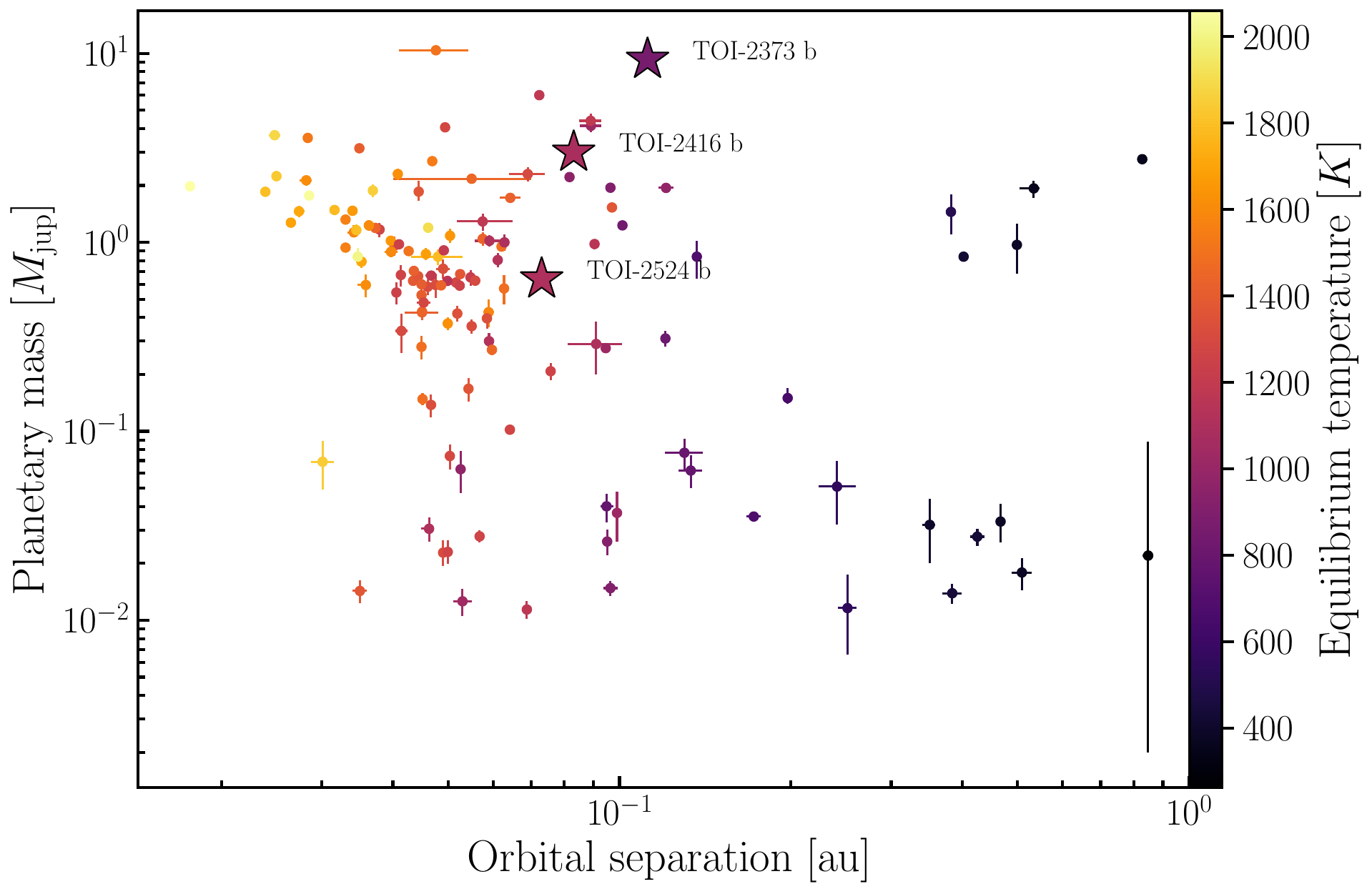}
    \caption{Mass-separation diagram for planets around solar-analog stars (definition according to \cite{Lehmann2022}, see \ref{sec:Introduction}). The new planets TOI-2373\,b, TOI-2416\,b, and TOI-2524\,b are shown as big stars.}
    \label{fig:mass-sep}
\end{figure}

 \section{Discussion}
 \label{sec:Discussion}
 
\begin{figure}
    \centering
    \includegraphics[width=9cm]{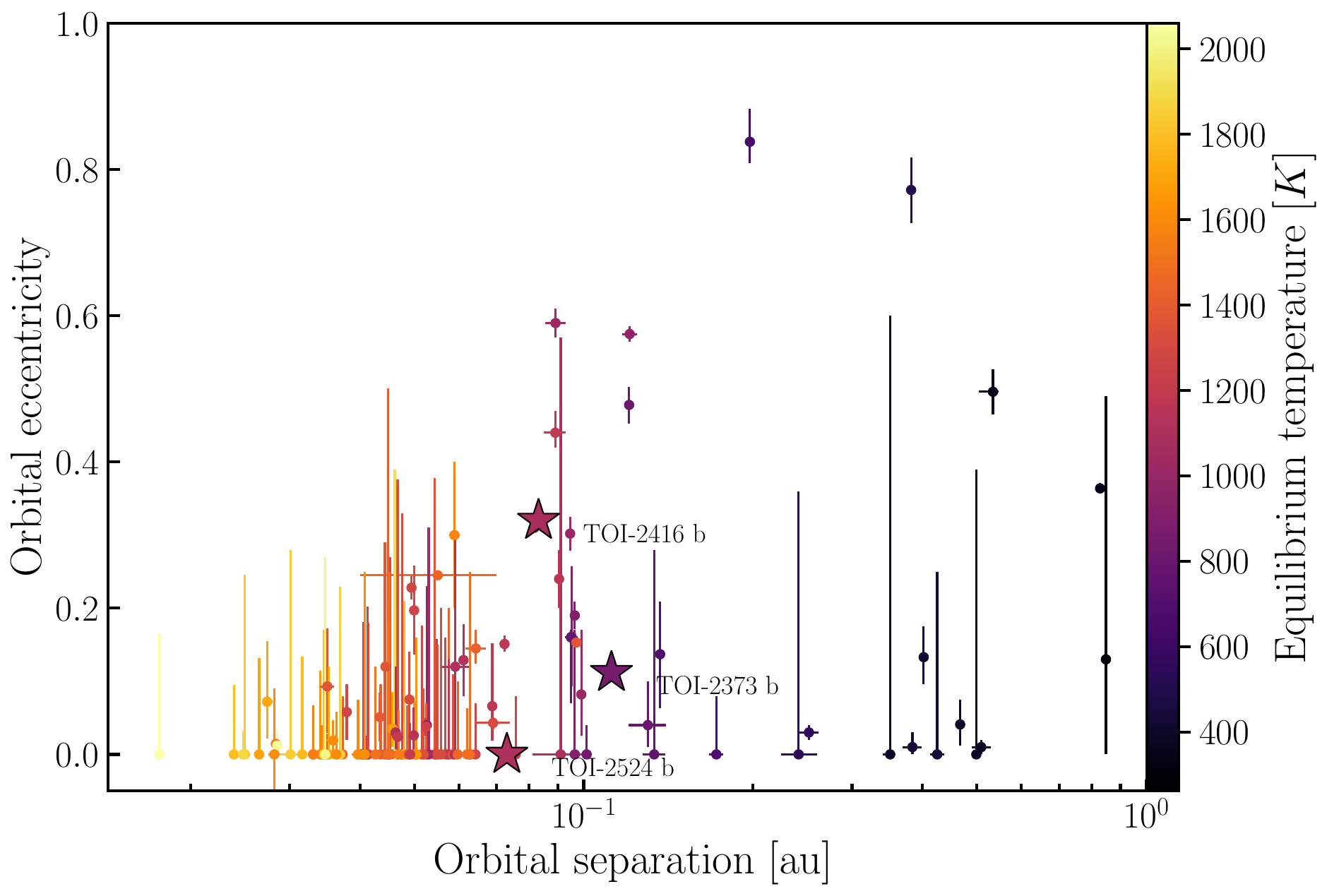}
    \caption{Mass-separation diagram for planets around solar-analog stars (definition according to \cite{Lehmann2022}, see \ref{sec:Introduction}). Shown as big stars are the new planets TOI-2373\,b, TOI-2416\,b, and TOI-2524\,b.}
    \label{fig:ecc-sep}
\end{figure}

\begin{figure*}
    \centering
    \includegraphics[width=\textwidth]{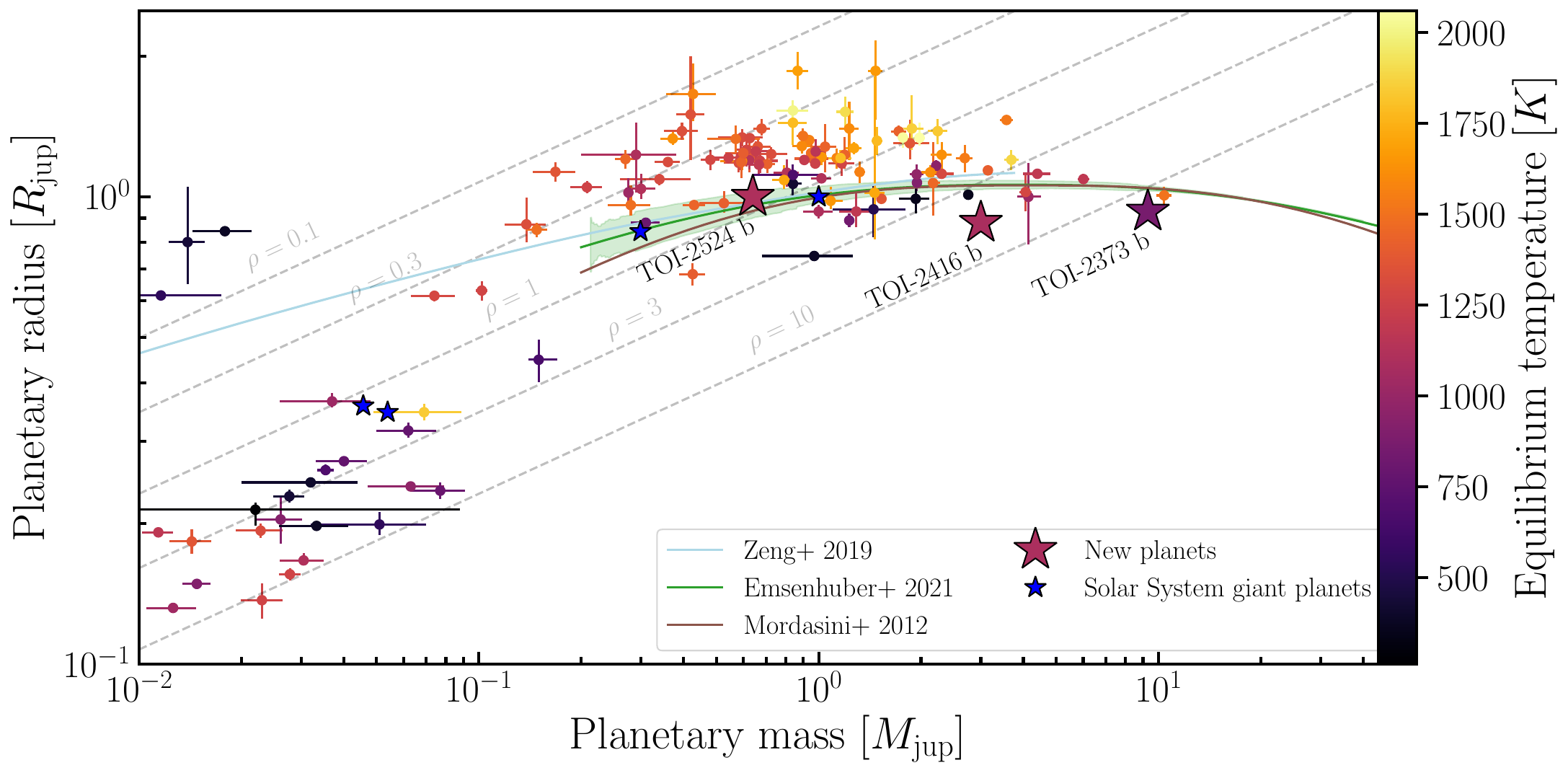}
    \caption{Mass-radius diagram for planets around Solar-analog stars. Shown as blue stars are the solar System giant planets, whereas the new planets TOI-2373\,b, TOI-2416\,b, and TOI-2524\,b are shown as colored stars. Shown as solid lines are exoplanet composition models based on \cite{Mordasini2012, Zeng2019, Emsenhuber2021b} and fixed densities of $\rho$ = 0.1, 0.3, 1.0, 3.0, and 1.0 $\mathrm{g\,cm}^{-3}$ are shown as dashed gray lines. The shaded area around the fit to the data of \cite{Emsenhuber2021b} marks the moving standard deviation of the their synthetic planets.}
    \label{fig:mass-radius}
\end{figure*}

 Figures \ref{fig:mass-sep} - \ref{fig:mass-radius} show the three targets presented in this work in the context of all known transiting exoplanets orbiting solar analog stars, including the solar system planets. The planets have been selected from the TEPCAT catalog \citep{Southworth2011} to match the following selection criteria on the host stars: 4.3 dex $<\log g <$ 4.6 dex, $-0.1<$ [Fe/H] $<0.4$, and 5600 K $<T_{\mathrm{eff}}<$ 5900 K. \ref{fig:mass-sep} shows a mass-separation diagram, with the exoplanets color coded according to their equilibrium temperatures. TOI-2373\,b, TOI-2416\,b, and TOI-2524\,b are shown as big color-coded stars. The three planets, while all having similar semimajor axes, cover a wide range of masses, with TOI-2373\,b being one of the two most massive known planets around a solar analog, featuring a mass of $m_p = 9.3_{-0.2}^{+0.2} M_{\mathrm{jup}}$, the other being CoRoT-27\,b with a mass of $m_p = 10.39\pm0.55 M_{\mathrm{jup}}$\citep{Parviainen2014}.
 
\ref{fig:ecc-sep} shows the same set of planets according to their orbital eccentricities and orbital separations. TOI-2373\,b and TOI-2416\,b have rather eccentric orbits compared to most known hot Jupiters orbiting similar stars. In the regime of warm Jupiters, there are planets with very highly eccentric orbits, however, this regime is only sparsely sampled with known exoplanets. Warm Jupiters with high eccentricities include TOI-2179\,b \citep[$e=0.575$;][]{Schlecker2020}, CoRoT-20\,b \citep[$e=0.59$;][]{Rey2018,Raetz2019}, Kepler-1656\,b \citep[$e=0.838$;][]{Brady2018,Angelo2022}, Kepler-1657\,b \citep[$e=0.496$;][]{Hebrard2019}, and K2-287\,b \citep[$e=0.478$;][]{Jordan2019,Borsato2021}. Understanding the orbital eccentricity demographics is important for constraining formation and migration models, as different evolutionary tracks predict different distributions.

\ref{fig:mass-radius} shows all known exoplanets around Solar analog-stars, including the solar system giant planets, in a mass-radius diagram, allowing for study of their bulk densities. Shown in dashed lines are nominal density curves corresponding to 0.1, 0.3, 1, 3, and 10 $\mathrm{g\,cm}^{-3}$, while exoplanet composition models from \cite{Mordasini2012, Zeng2019, Emsenhuber2021b} are shown as solid lines of varying colors. For the latter model, we fitted their synthetic planet population with orbital periods shorter than 30 days and masses larger than 100 $M_{\mathrm{earth}}$ using the same functional form as \citet{Mordasini2012}
\begin{equation}
    r(m) = \frac{b}{1+\left|\frac{\log_{10}(m/M_0)}{w}\right|^p}
\end{equation}
and found $b=1.061\pm0.002\,R_{\mathrm{jup}}$, $M_0= 3.85\pm0.10\,M_{\mathrm{jup}}$, $w=2.0 \pm 0.2$, and $p=2.5\pm 0.2$. The error of the fit was determined by bootstrapping their sample and taking the standard deviations of the resulting fit parameters. In \ref{fig:mass-radius}, we show as a shaded region the moving standard deviation of the synthetic data with respect to the fit. This visualizes the predicted scatter, which is larger at lower planetary masses.

While TOI-2524\,b closely follows the model predictions, TOI-2373\,b and TOI-2416\,b are denser, containing either more massive cores, or consisting of a larger fraction of heavy elements in their envelopes. Hot Jupiters have frequently been observed to be inflated \citep{Demory2011}, although the exact mechanisms creating these puffy planets are still not entirely understood \citep{Laughlin2011, Sarkis2021, Schneider2022}. Observed inflated hot Jupiters span a temperature range of $1300\,K<T_{\mathrm{eq}}<1500\,K$, and have a mean observed radius of $1.1\pm0.1$ $R_{\mathrm{jup}}$. 
However, the radius inflation effect is predicted to exist for planets with T>1000K \citep{Demory2011,MillerFortney2011}. Therefore, for consistency we include anomalous heating in our model of TOI-2524\,b, which has an equilibrium temperature of 1100$\pm20$\,K. Although this planet has a lower density than the other planets presented in this paper, it does not appear to be significantly inflated. 
Another possible explanation for TOI-2524\,b's lower density compared to TOI-2373\,b and TOI-2416\,b might be a lower abundance of heavy elements, which could be associated to the host star's relatively low metallicity of [Fe/H] = $+0.06\pm0.05$. Assuming a similarly low abundance of heavy elements in the protoplanetary disk as in the star, would result in an equally low abundance of heavy elements in the planet; however, the correlation between stellar metallicity and heavy element content is unclear \citep{MillerFortney2011, Thorngren2016}. The low heavy element content of the planet could be also associated to formation/migration history.

Both TOI-2373 and TOI-2416 have higher metallicities and host denser exoplanets. The comparison to evolved, synthetic planets modeled by \citet{Emsenhuber2021b} shown in \ref{fig:mass-radius} reveals that they are not necessarily reproduced naturally by core accretion formation models. TOI-2373\,b has a bulk metallicity of $Z_p=0.11^{+0.06}_{-0.05}$. A planet as massive as TOI-2373\,b could possibly have formed either by core-accretion or by gravitational instability \citep{Schlaufman2018} and should have the same metallicity as the host star. Postformation accretion of metals could explain its high amount of heavy elements \citep{Ikoma2006,Leconte2009,Cabrera2010,Ginzburg2020,Shibata2022,Morbidelli2023}. TOI-2416 is a warm super-Jupiter and has an even higher bulk metallicity of $Z_p=0.36^{+0.03}_{-0.03}$. This is an unusually high amount of metals and makes it a rare case. While it is still unclear how such a planet can form, similar planets, including CoRoT-10\,b ($m=2.75\,M_{\mathrm{jup}}$ and $Z_p\approx0.23$), HATS-17\,b ($m=1.34\,M_{\mathrm{jup}}$ and $Z_p\approx0.46$), HAT-P-20\,b ($m=7.25\,M_{\mathrm{jup}}$ and $Z_p\approx0.29$) and Kepler-419\,b ($m=2.5\,M_{\mathrm{jup}}$ and $Z_p\approx0.25$) have been presented by \citet[]{Thorngren2016}. Combined with the existing sample of dense exoplanets, this poses a challenge to planet formation and evolution or, alternatively, interior structure models.

 \section{Summary and Conclusion}
 \label{sec:SummaryConclusion}

We report the discovery and characterization of three giant exoplanets orbiting stars similar to the Sun, first identified as candidates using data from the \tess mission and confirmed with RV measurements with \textit{FEROS}. 
TOI-2373\,b is a warm Jupiter with an orbital period of $\sim$ 13.3 days, an estimated mass of $m_p$ = $9.3^{+0.6}_{-0.2}\,M_{\mathrm{jup}}$ and a radius of $r_p$ = $0.93^{+0.2}_{-0.2}\,R{\mathrm{jup}}$.  With a mean density of $\rho = 14.4^{+0.9}_{-1.0}\,\mathrm{g\,cm}^{-3}$, TOI-2373\,b is among the densest planets discovered so far and presents a challenge to current theories of planet formation and evolution.
 TOI-2416\,b is a planet with a short period of $\sim$ 8.3 days, a mass of $m_p$ = 3.0$^{+0.10}_{-0.09}\,M_{\mathrm{jup}}$ and a small radius of $r_p$ = $0.88^{+0.02}_{-0.02}\,R_{\mathrm{jup}}$, also resulting in an anomalously high mean density of $\rho = 5.4^{+0.3}_{-0.4}\,\mathrm{g\,cm}^{-3}$.
  TOI-2524\,b is a low-density warm Jupiter near the hot Jupiter transition region with a period of $\sim$ 7.2 days. For TOI-2524\,b we estimated a mass of $m_p$ = $0.64^{+0.04}_{-0.04}\,M_{\mathrm{jup}}$ and a radius of $r_p$ = $1.00^{+0.02}_{-0.03}\,R_{\mathrm{jup}}$, leading to a mean density of $\rho = 0.79^{+0.08}_{-0.08}\,\mathrm{g\,cm}^{-3}$, similar to that of Saturn.
 
  The newly discovered exoplanets TOI-2373\,b, TOI-2416\,b, and TOI-2524\,b have estimated equilibrium temperatures of $860^{+10}_{-10}$ K, $1010^{+10}_{-10}$ K, and $1100^{+20}_{-20}$ K, respectively, placing them in the sparsely populated transition region between hot and warm Jupiters. This is further supported by \cite{Rice2022}, who empirically defined the boundary between hot and warm Jupiters based on orbital alignment or misalignment, using the relative size of the orbit $a/R_{\star}$, with hot Jupiters satisfying $a/R_{\star}<11$ and warm Jupiters satisfying $a/R_{\star}>11$. With orbital sizes of $a_p/R_{\star}=23.1$, $a_p/R_{\star}=14.4$, and $a_p/R_{\star}=14.3$ for TOI-2373\,b, TOI-2416\,b, and TOI-2524\,b, respectively, these targets reside in the transition region. 

While we do not have measurements for the three targets' RVs during the planets' transits, we can estimate the amplitudes of the Rossiter-McLaughlin effect \citep{Rossiter1924, McLaughlin1924} to be 12.9, 10.4, and 22.8 $\mathrm{m\,s}^{-1}$, respectively, which can be measured by high-resolution spectrographs.
  
  \cite{Rice2022} showed that warm Jupiters tend to be in spin-orbit alignment, whereas hot Jupiters tend not to be aligned. \cite{Attia2023} find the same tendency, with longer-period planets being more aligned than closer-in planets. These findings are also consistent with \cite{Huang2016}, who come to the conclusion that hot Jupiters form separately from other gas giants, while warm Jupiters tend to form via disk migrations, explaining their coplanar orbits. As transiting planets, TOI-2373\,b, TOI-2416\,b, and TOI-2524\,b are well suited to estimate the sky-projected spin-orbit alignment of their host stars through Rossiter-McLaughlin observations, aiding in constraining the formation scenarios of these systems.

\section{Acknowledgments}
\label{sec:Acknowledgments}
This research has made use of the Exoplanet Follow-up Observation Program website, which is operated by the California Institute of Technology, under contract with the National Aeronautics and Space Administration under the Exoplanet Exploration Program. Resources supporting this work were provided by the NASA High-End Computing (HEC) Program through the NASA Advanced Supercomputing (NAS) Division at Ames Research Center for the production of the SPOC data products. We acknowledge the use of public TESS data from pipelines at the TESS Science Office and at the TESS Science Processing Operations Center. This paper includes data collected by the TESS mission, which are publicly available from the Mikulski Archive for Space Telescopes (MAST) operated by the Space Telescope Science Institute (STScI). Funding for the TESS mission is provided by NASA’s Science Mission Directorate. 
This work makes use of observations from the ASTEP telescope. ASTEP benefited from the support of the French and Italian polar agencies IPEV and PNRA in the framework of the Concordia station program, from OCA, INSU, Idex UCAJEDI (ANR- 15-IDEX01), and ESA through the Science Faculty of the European Space Research and Technology Centre (ESTEC). This work made use of \texttt{tpfplotter} by J. Lillo-Box (publicly available in www.github.com/jlillo/tpfplotter), which also made use of the python packages \texttt{astropy}, \texttt{lightkurve}, \texttt{matplotlib} and \texttt{numpy}.
The data used for the analyses presented in this paper can be accessed via \dataset[https://doi.org/10.17909/zsh5-wq95]{https://doi.org/10.17909/zsh5-wq95}, \dataset[https://doi.org/10.17909/edzq-va27]{https://doi.org/10.17909/edzq-va27}, and \dataset[https://doi.org/10.17909/phhz-qv30]{https://doi.org/10.17909/phhz-qv30}.
\textbf{Gaia DR2 data can be accessed via 
\dataset[https://doi.org/10.5270/esa-ycsawu7]{https://doi.org/10.5270/esa-ycsawu7}.}
The results reported herein benefited from collaborations and/or information exchange within NASA’s Nexus for Exoplanet System Science (NExSS) research coordination network sponsored by NASA’s Science Mission Directorate under Agreement No. 80NSSC21K0593 for the program “Alien Earths”.
R.Br. acknowledges support from FONDECYT Project 11200751. A.J., R.Br., F.R., and M.H.\ acknowledge support from ANID -- Millennium  Science  Initiative -- ICN12\_009. A.J.\ acknowledges additional support from FONDECYT project 1210718.
T.T. acknowledges support by the DFG Research Unit FOR 2544 "Blue Planets around Red Stars" project No. KU 3625/2-1.
T.T. further acknowledges support by the BNSF program "VIHREN-2021" project No. КП-06-ДВ/5.
R.Bu. acknowledges the support from DFG under Germany’s Excellence Strategy EXC 2181/1-390900948, Exploratory project EP 8.4 (the Heidelberg STRUCTURES Excellence
Cluster).
T.H. acknowledges support from the European Research Council under the Horizon 2020 Framework Program via the ERC Advanced grant Origins 832428.
The postdoctoral fellowship of K.B. is funded by F.R.S.-FNRS grant T.0109.20 and by the Francqui Foundation.
This research received funding from the European Research Council (ERC) under the European Union's Horizon 2020 research and innovation program (grant agreement n$^\circ$ 803193/BEBOP), and from the Science and Technology Facilities Council (STFC; grant n$^\circ$ ST/S00193X/1).

\facilities{\textit{TESS}, ASTEP, LCOGT, El Sauce, Observatoire Moana, FEROS/MPG2.2m}
\software{\texttt{Exo-Striker}~\citep{Trifonov2019_es}
            \texttt{CERES}~\citep{Brahm2017},
            \texttt{tesseract}~(F. I. Rojas et al. 2023 in preparation.),
          \texttt{ZASPE}~\citep{zaspe},
          \texttt{emcee}~\citep{ForemanMackey2013},
          \texttt{batman}~\citep{Kreidberg2015},
          \texttt{BANZAI}~\citep{McCully_2018SPIE10707E}
          \texttt{AstroImageJ}~\citep{Collins2017},
          }

\newpage

\begin{appendix} 

\label{appendix}
In this appendix, we show additional plots, including TPF plots for all \tess sectors used within this work, and additional time-series plots of the photometric and RV data, as well as tables containing our FEROS RV measurements, and the full results of our analysis.
\ref{tab:FEROS} lists the RV and stellar activity data obtained with the FEROS instrument for all three targets and \ref{tab:results_TOI-2373}, \ref{tab:results_TOI-2416}, and \ref{tab:results_TOI-2524} list the results for all fit parameters and both tested models from the NS for TOI-2373, TOI-2416, and TOI-2524, respectively. \ref{fig:TPF_2373}, \ref{fig:TPF_2416}, and \ref{fig:TPF_2524} show the TPF plots for the \tess sectors used for our analysis but not shown before. \ref{fig:TIC332_TS}, \ref{fig:TIC237_TS}, and \ref{fig:TIC169_TS} show time-series plots of the photometric and RV data for TOI-2373, TOI-2416, and TOI-2524, respectively. These plots are complementary to \ref{fig:TIC332_T_phase}, \ref{fig:TIC237_T_phase}, and \ref{fig:TIC169_T_phase}.

\setcounter{table}{0}
\renewcommand{\thetable}{A\arabic{table}}

\setcounter{figure}{0}
\renewcommand{\thefigure}{A\arabic{figure}}

\begin{table}[]
    \centering
    \caption{FEROS radial velocity and stellar activity measurements along with their 1$\sigma$ uncertainties.}
    \resizebox{\textwidth}{!}{
    \begin{tabular}{lrrrrrrrrrrrrr}
    \hline\hline  \noalign{\vskip 0.7mm}
    BJD  \hspace{0.0 mm}& RV ($\mathrm{m\,s^{-1}}$) & $\sigma_{\mathrm{RV}}$  & BIS ($\mathrm{m\,s^{-1}}$) & $\sigma_{\mathrm{BIS}}$ & H-Alpha & $\sigma_{\mathrm{H-Alpha}}$ & $\log{\mathrm{RHK}}$ & $\sigma_{\log{\mathrm{RHK}}}$ & FWHM & He\,I & $\sigma_{\mathrm{He\,I}}$ & Na\,II & $\sigma_{\mathrm{Na\,II}}$\\
        \hline \noalign{\vskip 0.7mm}
        & \multicolumn{1}{c}{TOI-2373} &\\
        \hline \noalign{\vskip 0.7mm}
2458905.55535 & -898.0 & 9.9 & 1 & 15 & 0.1469 & 0.0047 & -4.8448 & 0.0899 & -3 & 0.4994 & 0.0102 & 0.2507 & 0.0075\\
2458911.54656 & 603.9 & 10.2 & -9 & 15 & 0.1215 & 0.0042 & -4.9407 & 0.1573 & -8 & 0.4950 & 0.0100 & 0.2507 & 0.0075\\
2458913.53698 & 492.4 & 11.1 & -22 & 15 & & & & & & & & &\\
2458917.53471 & -800.8 & 11.1 & -22 & 15 & 0.1239 & 0.0046 & -4.9716 & 0.1593 & -20 & 0.5223 & 0.0112 & 0.2246 & 0.0077\\
2458919.52262 & -812.4 & 13.0 & -28 & 18 & 0.1473 & 0.0054 & -4.6186 & 0.1356 & -31 & 0.4980 & 0.0124 & 0.2221 & 0.0092\\
2458922.52825 & -198.7 & 12.9 & -36 & 18 & 0.1402 & 0.0050 & -4.6925 & 0.1965 & -38 & 0.5245 & 0.0124 & 0.2123 & 0.0089\\
2458927.53614 & 297.6 & 10.5 & -49 & 15 & 0.1249 & 0.0038 & -4.6071 & 0.1995 & -48 & 0.4944 & 0.0093 & 0.2226 & 0.0066\\
2458931.50776 & -833.7 & 10.7 & -29 & 15 & 0.1320 & 0.0042 & -4.7896 & 0.1572 & -29 & 0.4797 & 0.0097 & 0.2439 & 0.0071\\
2459188.62842 & -359.0 & 9.4 & 28 & 14 & 0.1368 & 0.0040 & -4.9199 & 0.0648 & 27 & 0.5253 & 0.0092 & 0.2607 & 0.0068\\
2459190.62923 & 335.4 & 8.8 & -36 & 13 & 0.1290 & 0.0035 & -5.0861 & 0.0809 & -36 & 0.5046 & 0.0085 & 0.2492 & 0.0061\\
2459191.61001 & 667.0 & 10.1 & -06 & 15 & 0.1434 & 0.0047 & -5.0645 & 0.0990 & -6 & 0.5054 & 0.0102 & 0.2600 & 0.0079\\
2459192.59836 & 681.3 & 11.3 & -91 & 16 & 0.1273 & 0.0052 & -5.0354 & 0.1301 & -83 & 0.5048 & 0.0116 & 0.2423 & 0.0088\\
2459207.63706 & 241.4 & 9.9 & -18 & 15 & 0.1283 & 0.0040 & -5.0056 & 0.0963 & -21 & 0.5201 & 0.0095 & 0.2297 & 0.0069\\
2459260.58094 & 440.9 & 12.6 & -1 & 18 & 0.1619 & 0.0060 & - & - & -11 & 0.5083 & 0.0132 & 0.2408 & 0.0098\\
2459272.56488 & 687.6 & 10.7 & -06 & 15 & 0.1516 & 0.0050 & -5.1569 & 0.1731 & -4 & 0.5182 & 0.0110 & 0.2662 & 0.0080\\
2459276.55084 & -545.6 & 11.8 & -30 & 17 & 0.1437 & 0.0050 & -5.4864 & 0.6791 & -31 & 0.4872 & 0.0113 & 0.2166 & 0.0082\\
         \hline \noalign{\vskip 0.7mm}
        & \multicolumn{1}{c}{TOI-2416} &\\
        \hline \noalign{\vskip 0.7mm}
         2458920.53485 & 274.7 & 11.7 & -38 & 16 & 0.1318 & 0.0049 & -5.0250 & 0.2500 & -4 & 0.4995 & 0.0108 & 0.2129 & 0.0079\\
2458924.51734 & -199.3 & 10.1 & 4 & 15 & 0.1111 & 0.0039 & -4.7267 & 0.1015 & 2 & 0.5152 & 0.0101 & 0.1961 & 0.0066\\
2458928.50695 & 170.7 & 14.7 & 26 & 19 & 0.1159 & 0.0064 & -5.3012 & 0.7261 & 26 & 0.5068 & 0.0149 & 0.2122 & 0.0112\\
2458929.55344 & 347.7 & 13.4 & 42 & 18 & 0.1318 & 0.0060 & - & - & 43 & 0.5034 & 0.0135 & 0.2076 & 0.0099\\
2459188.60699 & -261.9 & 9.6 & 2 & 14 & 0.1201 & 0.0035 & -5.1144 & 0.2002 & 2 & 0.5148 & 0.0089 & 0.2043 & 0.0059\\
2459189.62117 & -268.3 & 8.4 & -14 & 13 & 0.1018 & 0.0029 & -4.9003 & 0.0747 & -14 & 0.5030 & 0.0077 & 0.2131 & 0.0051\\
2459190.59374 & -192.3 & 8.2 & -3 & 13 & 0.1251 & 0.0033 & -5.2681 & 0.1344 & -3 & 0.4965 & 0.0080 & 0.1992 & 0.0051\\
2459192.61798 & 49.7 & 11.7 & 0.0 & 17 & 0.1260 & 0.0054 & -4.7975 & 0.0836 & -1 & 0.5009 & 0.0126 & 0.2380 & 0.0090\\
2459194.62382 & 278.1 & 9.6 & 13 & 14 & 0.1168 & 0.1168 & -4.7863 & 0.0793 & 13 & 0.5120 & 0.0097 & 0.2110 & 0.0062\\
2459207.6163 & -187.6 & 9.5 & -22 & 14 & 0.1078 & 0.0035 & -5.3027 & 0.2413 & -23 & 0.5119 & 0.0090 & 0.2065 & 0.0060\\
2459211.61738 & 121.7 & 8.4 & -1 & 13 & 0.1088 & 0.0032 & -5.0988 & 0.1003 & -1 & 0.4938 & 0.0077 & 0.2006 & 0.0053\\
2459213.60256 & -240.6 & 11.7 & 24 & 16 & 0.1132 & 0.0043 & -4.9743 & 0.1494 & 25 & 0.5145 & 0.0111 & 0.2095 & 0.0078\\
2459260.56028 & 335.5 & 13.2 & -31 & 18 & 0.1584 & 0.0060 & -4.6841 & 0.1229 & -34 & 0.4982 & 0.0132 & 0.1797 & 0.0094\\
2459264.52873 & -238.8 & 12.9 & -29 & 18 & 0.1326 & 0.0055 & - & - & -28 & 0.4775 & 0.0125 & 0.2039 & 0.0090\\
2459270.52139 & -152.7 & 9.4 & -5 & 14 & 0.1139 & 0.0033 & -5.0822 & 0.1685 & -7 & 0.5005 & 0.0086 & 0.2153 & 0.0057\\
2459276.51409 & 269.6 & 11.6 & -26 & 16 & 0.1119 & 0.0044 & -4.6902 & 0.1068 & -25 & 0.4835 & 0.0108 & 0.2187 & 0.0079\\
2459278.5535 & -106.3 & 11.5 & -9 & 16 & 0.1217 & 0.0044 & -5.5500 & 0.9654 & -9 & 0.4950 & 0.0110 & 0.2018 & 0.0073\\
         \hline \noalign{\vskip 0.7mm}
        & \multicolumn{1}{c}{TOI-2524} &\\
        \hline \noalign{\vskip 0.7mm}
         2459209.77377 & 52.6 & 11.0 & 17 & 14 & 0.1193 & 0.0038 & -5.0542 & 0.1871 & 18 & 0.5158 & 0.0091 & 0.2650 & 0.0067\\
2459212.77926 & -85.4 & 9.5 & 9 & 13 & 0.1207 & 0.0036 & -4.9892 & 0.0982 & 13 & 0.5210 & 0.0085 & 0.2870 & 0.0064\\
2459216.77064 & 53.8 & 10.5 & 19 & 14 & 0.1277 & 0.0046 & -5.0112 & 0.0925 & 23 & 0.5050 & 0.0104 & 0.2672 & 0.0077\\
2459223.86138 & 49.5 & 9.9 & -2 & 14 & 0.1210 & 0.0036 & -4.9228 & 0.0840 & -15 & 0.5081 & 0.0086 & 0.2657 & 0.0064\\
2459263.78018 & -74.0 & 12.3 & 29 & 15 & 0.1354 & 0.0047 & -4.8204 & 0.0888 & 29 & 0.4997 & 0.0104 & 0.3056 & 0.0083\\
2459278.69207 & -55.7 & 11.0 & -4.3 & 14 & 0.1361 & 0.0039 & -5.2602 & 0.2095 & -42 & 0.5039 & 0.0089 & 0.2918 & 0.0071\\
2459280.68823 & 39.3 & 9.2 & 2 & 13 & 0.1118 & 0.0031 & -5.1824 & 0.1390 & 22 & 0.5040 & 0.0083 & 0.2941 & 0.0060\\
2459282.6783 & 19.6 & 9.7 & -1 & 13 & 0.1086 & 0.0034 & -4.9126 & 0.0715 & -6 & 0.5022 & 0.0086 & 0.3096 & 0.0065\\
         \hline\hline  \noalign{\vskip 0.7mm}
    \end{tabular}}
    \label{tab:FEROS}
\end{table}

\begin{table*}[]
    \centering
    \caption{Results from the NS run for TOI-2373. Listed are likelihood parameters, priors, posteriors and derived parameters for both a circular and eccentric model. $\Delta$ln$\mathcal{Z}$ is defined as ln$\mathcal{Z}_{\mathrm{ecc}}$ - ln$\mathcal{Z}_{\mathrm{circ}}$. \tess sector numbers are denoted as "TX" for transit offset and jitter parameters.}
    \resizebox{\textwidth}{!}{
    \begin{tabular}{lcrlrl}
    \hline\hline  \noalign{\vskip 0.7mm}
          Parameter && \multicolumn{2}{c}{Circular Model} & \multicolumn{2}{c}{Eccentric Model} \\
          \hspace{0.0 mm} &  \hspace{0.0 mm} & Posterior  & Prior  & Posterior  & Prior  \\
    \hline \noalign{\vskip 0.7mm}
ln$\mathcal{Z}$ &  & \multicolumn{2}{c}{24,878.63} & \multicolumn{2}{c}{24,895.34} \\ \noalign{\vskip 0.9mm}
$\Delta$ln$\mathcal{Z}$ &  & \multicolumn{4}{c}{16.71} \\ \noalign{\vskip 0.9mm}
\hline \noalign{\vskip 0.7mm}

RV$_{\mathrm{offset}}$ & $\mathrm{m\,s}^{-1}$ & $4632_{-36}^{+36}$ & $\mathcal{U}(4500.0, 4800.0)$ & $4596_{-8}^{+8}$ & $\mathcal{U}(4500.0, 4700.0)$  \\ \noalign{\vskip 0.9mm}
RV$_{\mathrm{jitter}}$ & $\mathrm{m\,s}^{-1}$ & $149_{-24}^{+27}$ & $\mathcal{U}(50.0, 200.0)$ & $36_{-9}^{+15}$ & $\mathcal{U}(-20.0, 100.0)$  \\ \noalign{\vskip 0.9mm}
$K$ & $\mathrm{m\,s}^{-1}$ & $778_{-47}^{+47}$ & $\mathcal{U}(600.0, 900.0)$ & $784_{-10}^{+10}$ & $\mathcal{U}(650.0, 850.0)$  \\ \noalign{\vskip 0.9mm}
$P$ & days & $13.33669_{-0.00002}^{+0.00002}$ & $\mathcal{U}(13.25, 13.4)$ & $13.33668_{-0.00001}^{+0.00001}$ & $\mathcal{U}(13.25, 13.4)$  \\ \noalign{\vskip 0.9mm}
$e$ &  & $0$ & fixed & $0.112_{-0.009}^{+0.007}$ & $\mathcal{U}(0.0, 0.22)$  \\ \noalign{\vskip 0.9mm}
$\omega$ & degrees & $-$ & undefined & $15_{-6}^{+7}$ & $\mathcal{U}(-90.0, 360.0)$  \\ \noalign{\vskip 0.9mm}
$i$ & degrees & $88.6_{-0.2}^{+0.2}$ & $\mathcal{U}(82.0, 92.0)$ & $89.2_{-0.2}^{+0.6}$ & $\mathcal{U}(85.0, 92.0)$  \\ \noalign{\vskip 0.9mm}
$t_0$ & days & $2458448.325_{-0.002}^{+0.002}$ & $\mathcal{U}(2458448.0, 2458449.0)$ & $2458448.326_{-0.001}^{+0.001}$ & $\mathcal{U}(2458448.0, 2458449.0)$  \\ \noalign{\vskip 0.9mm}
$a/R_\star$ &  & $21.9_{-0.8}^{+0.8}$ & $\mathcal{U}(0.0, 30.0)$ & $23.1_{-0.5}^{+0.6}$ & $\mathcal{U}(0.0, 30.0)$  \\ \noalign{\vskip 0.9mm}
$r/R_\star$ & & $0.0897_{-0.0010}^{+0.0009}$ & $\mathcal{U}(0.0, 0.15)$ & $0.0867_{-0.0009}^{+0.0009}$ & $\mathcal{U}(0.0, 0.15)$  \\ \noalign{\vskip 0.9mm}
transit$_{\mathrm{offset,T5}}$ & & $0.00014_{-0.00004}^{+0.00004}$ & $\mathcal{U}(-0.001, 0.001)$ & $0.00012_{-0.00003}^{+0.00003}$ & $\mathcal{U}(-0.001, 0.001)$  \\ \noalign{\vskip 0.9mm}
transit$_{\mathrm{offset,T31}}$ & & $0.00031_{-0.00004}^{+0.00004}$ & $\mathcal{U}(-0.001, 0.001)$ & $0.00031_{-0.00003}^{+0.00002}$ & $\mathcal{U}(-0.001, 0.001)$  \\ \noalign{\vskip 0.9mm}
transit$_{\mathrm{offset,LCO-SAAO}}$ & & $0.0000_{-0.0001}^{+0.0001}$ & $\mathcal{U}(-0.001, 0.001)$ & $0.00006_{-0.00008}^{+0.00009}$ & $\mathcal{U}(-0.001, 0.001)$  \\ \noalign{\vskip 0.9mm}
transit$_{\mathrm{offset,LCO-Teid}}$ & & $0.00121_{-0.00008}^{+0.00008}$ & $\mathcal{U}(-0.001, 0.001)$ & $0.00131_{-0.00005}^{+0.00006}$ & $\mathcal{U}(-0.001, 0.001)$  \\ \noalign{\vskip 0.9mm}
transit$_{\mathrm{offset,LCO-CTIO1}}$ & & $0.00007_{-0.00010}^{+0.00010}$ & $\mathcal{U}(-0.001, 0.001)$ & $-0.00005_{-0.00007}^{+0.00007}$ & $\mathcal{U}(-0.001, 0.001)$  \\ \noalign{\vskip 0.9mm}
transit$_{\mathrm{offset,LCO-CTIO2}}$ & & $0.00026_{-0.00010}^{+0.00010}$ & $\mathcal{U}(-0.001, 0.001)$ & $0.00027_{-0.00008}^{+0.00008}$ & $\mathcal{U}(-0.001, 0.001)$  \\ \noalign{\vskip 0.9mm}
transit$_{\mathrm{offset,OM-ES}}$ & & $-0.0002_{-0.0001}^{+0.0001}$ & $\mathcal{U}(-0.001, 0.001)$ & $-0.00014_{-0.00012}^{+0.00009}$ & $\mathcal{U}(-0.001, 0.001)$  \\ \noalign{\vskip 0.9mm}
transit$_{\mathrm{jitter,T5}}$ & & $0.000994_{-0.000009}^{+0.000005}$ & $\mathcal{U}(-0.01, 0.01)$ & $0.000988_{-0.000008}^{+0.000007}$ & $\mathcal{U}(-0.01, 0.01)$  \\ \noalign{\vskip 0.9mm}
transit$_{\mathrm{jitter,T31}}$ & & $0.000993_{-0.000010}^{+0.000005}$ & $\mathcal{U}(-0.01, 0.01)$ & $0.000987_{-0.000015}^{+0.000009}$ & $\mathcal{U}(-0.01, 0.01)$  \\ \noalign{\vskip 0.9mm}
transit$_{\mathrm{jitter,LCO-SAAO}}$ & & $0.000993_{-0.000010}^{+0.000005}$ & $\mathcal{U}(-0.01, 0.01)$ & $0.00097_{-0.00002}^{+0.00002}$ & $\mathcal{U}(-0.01, 0.01)$  \\ \noalign{\vskip 0.9mm}
transit$_{\mathrm{jitter,LCO-Teid}}$ & & $0.0009989_{-0.0000015}^{+0.0000008}$ & $\mathcal{U}(-0.01, 0.01)$ & $0.000996_{-0.000003}^{+0.000002}$ & $\mathcal{U}(-0.01, 0.01)$  \\ \noalign{\vskip 0.9mm}
transit$_{\mathrm{jitter,LCO-CTIO1}}$ & & $0.000995_{-0.000007}^{+0.000003}$ & $\mathcal{U}(-0.01, 0.01)$ & $0.000990_{-0.000008}^{+0.000006}$ & $\mathcal{U}(-0.01, 0.01)$  \\ \noalign{\vskip 0.9mm}
transit$_{\mathrm{jitter,LCO-CTIO2}}$ & & $0.000996_{-0.000006}^{+0.000003}$ & $\mathcal{U}(-0.01, 0.01)$ & $0.000990_{-0.000010}^{+0.000006}$ & $\mathcal{U}(-0.01, 0.01)$  \\ \noalign{\vskip 0.9mm}
transit$_{\mathrm{jitter,OM-ES}}$ & & $0.0009_{-0.0003}^{+0.0001}$ & $\mathcal{U}(-0.01, 0.01)$ & $0.00007_{-0.00005}^{+0.00054}$ & $\mathcal{U}(-0.01, 0.01)$  \\ \noalign{\vskip 0.9mm}
$u_{1, \mathrm{TESS}}$ & & $0.2_{-0.1}^{+0.2}$ & $\mathcal{U}(0.0, 1.0)$ & $0.2_{-0.1}^{+0.1}$ & $\mathcal{U}(0.0, 1.0)$  \\ \noalign{\vskip 0.9mm}
$u_{2, \mathrm{TESS}}$ & & $0.3_{-0.2}^{+0.2}$ & $\mathcal{U}(0.0, 1.0)$ & $0.5_{-0.2}^{+0.2}$ & $\mathcal{U}(0.0, 1.0)$  \\ \noalign{\vskip 0.9mm}
$u_{1, \mathrm{LCO-SAAO}}$ & & $0.2_{-0.1}^{+0.1}$ & $\mathcal{U}(0.0, 1.0)$ & $0.31_{-0.09}^{+0.07}$ & $\mathcal{U}(0.0, 1.0)$  \\ \noalign{\vskip 0.9mm
}$u_{2, \mathrm{LCO-SAAO}}$ & & $0.3_{-0.2}^{+0.2}$ & $\mathcal{U}(0.0, 1.0)$ & $0.3_{-0.1}^{+0.2}$ & $\mathcal{U}(0.0, 1.0)$  \\ \noalign{\vskip 0.9mm}
$u_{1, \mathrm{LCO-Teid}}$ & & $0.2_{-0.1}^{+0.1}$ & $\mathcal{U}(0.0, 1.0)$ & $0.28_{-0.07}^{+0.06}$ & $\mathcal{U}(0.0, 1.0)$  \\ \noalign{\vskip 0.9mm}
$u_{2, \mathrm{LCO-Teid}}$ & & $0.3_{-0.2}^{+0.2}$ & $\mathcal{U}(0.0, 1.0)$ & $0.3_{-0.1}^{+0.1}$ & $\mathcal{U}(0.0, 1.0)$  \\ \noalign{\vskip 0.9mm}
$u_{1, \mathrm{LCO-CTIO}}$ & & $0.5_{-0.2}^{+0.1}$ & $\mathcal{U}(0.0, 1.0)$ & $0.55_{-0.08}^{+0.07}$ & $\mathcal{U}(0.0, 1.0)$  \\ \noalign{\vskip 0.9mm}
$u_{2, \mathrm{LCO-CTIO}}$ & & $0.3_{-0.2}^{+0.2}$ & $\mathcal{U}(0.0, 1.0)$ & $0.3_{-0.1}^{+0.1}$ & $\mathcal{U}(0.0, 1.0)$  \\ \noalign{\vskip 0.9mm}
$u_{1, \mathrm{OM-ES}}$ & & $0.5_{-0.2}^{+0.2}$ & $\mathcal{U}(0.0, 1.0)$ & $0.2_{-0.1}^{+0.2}$ & $\mathcal{U}(0.0, 1.0)$  \\ \noalign{\vskip 0.9mm}
$u_{2, \mathrm{OM-ES}}$ & & $0.2_{-0.2}^{+0.3}$ & $\mathcal{U}(0.0, 1.0)$ & $0.4_{-0.2}^{+0.2}$ & $\mathcal{U}(0.0, 1.0)$  \\ \noalign{\vskip 0.9mm}
\hline \noalign{\vskip 0.7mm}
$m$ & $M_{\mathrm{jup}}$ & \multicolumn{2}{c}{$9.3_{-0.6}^{+0.6}$} & \multicolumn{2}{c}{$9.3_{-0.2}^{+0.2}$} \\ \noalign{\vskip 0.9mm}

$r$ & $R_{\mathrm{jup}}$ & \multicolumn{2}{c}{$0.96_{-0.02}^{+0.02}$} & \multicolumn{2}{c}{$0.93_{-0.02}^{+0.02}$} \\ \noalign{\vskip 0.9mm}

$a$ & au & \multicolumn{2}{c}{$0.112_{-0.001}^{+0.001}$} & \multicolumn{2}{c}{$0.112_{-0.001}^{+0.001}$} \\ \noalign{\vskip 0.9mm}

$\rho$ & $\mathrm{g\,cm}^{-3}$ & \multicolumn{2}{c}{$13_{-1}^{+1}$} & \multicolumn{2}{c}{$14.4_{-1.0}^{+0.9}$} \\ \noalign{\vskip 0.9mm}

$T_{\mathrm{eq}}$ & K & \multicolumn{2}{c}{$860_{-10}^{+10}$} & \multicolumn{2}{c}{$860_{-10}^{+10}$} \\ \noalign{\vskip 0.9mm}

    \hline\hline \noalign{\vskip 0.5mm}
    \end{tabular}}
        \label{tab:results_TOI-2373}
    \end{table*}

    \begin{table*}[]
        \centering
        \caption{Results from the NS run for TOI-2416. Listed are likelihood parameters, priors, posteriors and derived parameters for both a circular and eccentric model. $\Delta$ln$\mathcal{Z}$ is defined as ln$\mathcal{Z}_{\mathrm{ecc}}$ - ln$\mathcal{Z}_{\mathrm{circ}}$. \tess sector numbers are denoted as "TX" for transit offset and jitter parameters.}
        \resizebox{0.95\textwidth}{!}{
        \begin{tabular}{lcrlrl}
        \hline\hline  \noalign{\vskip 0.7mm}
              Parameter && \multicolumn{2}{c}{Circular Model} & \multicolumn{2}{c}{Eccentric Model} \\
              \hspace{0.0 mm} &  \hspace{0.0 mm} & Posterior  & Prior  & Posterior  & Prior  \\
        \hline \noalign{\vskip 0.7mm}
    ln$\mathcal{Z}$ &  & \multicolumn{2}{c}{326,957.01} & \multicolumn{2}{c}{326,984.85} \\ \noalign{\vskip 0.9mm}
    $\Delta$ln$\mathcal{Z}$ &  & \multicolumn{4}{c}{27.84} \\ \noalign{\vskip 0.9mm}
    \hline \noalign{\vskip 0.7mm}
    
    RV$_{\mathrm{offset}}$ & $\mathrm{m\,s}^{-1}$ & $22190_{-17}^{+18}$ & $\mathcal{U}(22100.0, 22250.0)$ & $22145_{-6}^{+6}$ & $\mathcal{U}(22100.0, 22250.0)$  \\ \noalign{\vskip 0.9mm}
    RV$_{\mathrm{jitter}}$ & $\mathrm{m\,s}^{-1}$ & $99_{-9}^{+9}$ & $\mathcal{U}(0.0, 120.0)$ & $21_{-5}^{+7}$ & $\mathcal{U}(-20.0, 100.0)$  \\ \noalign{\vskip 0.9mm}
    $K$ & $\mathrm{m\,s}^{-1}$ & $285_{-27}^{+27}$ & $\mathcal{U}(150.0, 400.0)$ & $296_{-8}^{+8}$ & $\mathcal{U}(150.0, 400.0)$  \\ \noalign{\vskip 0.9mm}
    $P$ & days & $8.275478_{-0.000010}^{+0.000010}$ & $\mathcal{U}(8.24, 8.3)$ & $8.275479_{-0.000009}^{+0.000009}$ & $\mathcal{U}(8.24, 8.3)$  \\ \noalign{\vskip 0.9mm}
    $e$ &  & $0$ & fixed & $0.32_{-0.02}^{+0.02}$ & $\mathcal{U}(0.0, 0.5)$  \\ \noalign{\vskip 0.9mm}
    $\omega$ & degrees & $-$ & undefined & $30_{-5}^{+5}$ & $\mathcal{U}(-90.0, 360.0)$  \\ \noalign{\vskip 0.9mm}
    $i$ & degrees & $89.8_{-0.3}^{+0.2}$ & $\mathcal{U}(80.0, 90.0)$ & $90.0_{-0.6}^{+0.6}$ & $\mathcal{U}(80.0, 92.0)$  \\ \noalign{\vskip 0.9mm}
    $t_0$ & days & $2458359.4885_{-0.0008}^{+0.0010}$ & $\mathcal{U}(2458359.0, 2458360.0)$ & $2458359.4884_{-0.0009}^{+0.0009}$ & $\mathcal{U}(2458359.0, 2458360.0)$  \\ \noalign{\vskip 0.9mm}
    $a/R_\star$ &  & $17.5_{-0.2}^{+0.1}$ & $\mathcal{U}(0.0, 45.0)$ & $14.4_{-0.5}^{+0.4}$ & $\mathcal{U}(0.0, 45.0)$  \\ \noalign{\vskip 0.9mm}
    $r/R_\star$ &  & $0.0734_{-0.0007}^{+0.0007}$ & $\mathcal{U}(0.0, 0.25)$ & $0.0732_{-0.0008}^{+0.0008}$ & $\mathcal{U}(0.0, 0.25)$  \\ \noalign{\vskip 0.9mm}
    transit$_{\mathrm{offset,T2}}$ &  & $0.00025_{-0.00003}^{+0.00003}$ & $\mathcal{U}(-0.001, 0.001)$ & $0.00023_{-0.00003}^{+0.00003}$ & $\mathcal{U}(-0.001, 0.001)$  \\ \noalign{\vskip 0.9mm}
    transit$_{\mathrm{offset,T3}}$ &  & $0.00027_{-0.00004}^{+0.00004}$ & $\mathcal{U}(-0.001, 0.001)$ & $0.00026_{-0.00004}^{+0.00005}$ & $\mathcal{U}(-0.001, 0.001)$  \\ \noalign{\vskip 0.9mm}
    transit$_{\mathrm{offset,T4}}$ &  & $0.00025_{-0.00004}^{+0.00004}$ & $\mathcal{U}(-0.001, 0.001)$ & $0.00026_{-0.00005}^{+0.00004}$ & $\mathcal{U}(-0.001, 0.001)$  \\ \noalign{\vskip 0.9mm}
    transit$_{\mathrm{offset,T8}}$ &  & $0.00030_{-0.00007}^{+0.00008}$ & $\mathcal{U}(-0.001, 0.001)$ & $0.00030_{-0.00007}^{+0.00007}$ & $\mathcal{U}(-0.001, 0.001)$  \\ \noalign{\vskip 0.9mm}
    transit$_{\mathrm{offset,T28}}$ &  & $0.00029_{-0.00003}^{+0.00003}$ & $\mathcal{U}(-0.001, 0.001)$ & $0.00030_{-0.00003}^{+0.00003}$ & $\mathcal{U}(-0.001, 0.001)$  \\ \noalign{\vskip 0.9mm}
    transit$_{\mathrm{offset,T29}}$ &  & $0.00028_{-0.00003}^{+0.00003}$ & $\mathcal{U}(-0.001, 0.001)$ & $0.00027_{-0.00003}^{+0.00003}$ & $\mathcal{U}(-0.001, 0.001)$  \\ \noalign{\vskip 0.9mm}
    transit$_{\mathrm{offset,T30}}$ &  & $0.00030_{-0.00003}^{+0.00003}$ & $\mathcal{U}(-0.001, 0.001)$ & $0.00030_{-0.00003}^{+0.00003}$ & $\mathcal{U}(-0.001, 0.001)$  \\ \noalign{\vskip 0.9mm}
    transit$_{\mathrm{offset,T31}}$ &  & $0.00032_{-0.00003}^{+0.00003}$ & $\mathcal{U}(-0.001, 0.001)$ & $0.00033_{-0.00003}^{+0.00003}$ & $\mathcal{U}(-0.001, 0.001)$  \\ \noalign{\vskip 0.9mm}
    transit$_{\mathrm{offset,T38}}$ &  & $0.00030_{-0.00003}^{+0.00003}$ & $\mathcal{U}(-0.001, 0.001)$ & $0.00031_{-0.00003}^{+0.00003}$ & $\mathcal{U}(-0.001, 0.001)$  \\ \noalign{\vskip 0.9mm}
    transit$_{\mathrm{offset,ASTEP}}$ &  & $0.0011_{-0.0002}^{+0.0002}$ & $\mathcal{U}(-0.001, 0.001)$ & $0.0011_{-0.0003}^{+0.0002}$ & $\mathcal{U}(-0.001, 0.001)$  \\ \noalign{\vskip 0.9mm}
    transit$_{\mathrm{offset,CDK24ND}}$ &  & $0.0020_{-0.0003}^{+0.0002}$ & $\mathcal{U}(-0.001, 0.001)$ & $0.0020_{-0.0002}^{+0.0002}$ & $\mathcal{U}(-0.001, 0.001)$  \\ \noalign{\vskip 0.9mm}
    transit$_{\mathrm{offset,OM-SSO}}$ &  & $-0.0034_{-0.0002}^{+0.0002}$ & $\mathcal{U}(-0.001, 0.001)$ & $-0.0034_{-0.0002}^{+0.0002}$ & $\mathcal{U}(-0.001, 0.001)$  \\ \noalign{\vskip 0.9mm}
    transit$_{\mathrm{jitter,T2}}$ &  & $-0.00088_{-0.00004}^{+0.00004}$ & $\mathcal{U}(-0.01, 0.01)$ & $-0.00086_{-0.00006}^{+0.00174}$ & $\mathcal{U}(-0.01, 0.01)$  \\ \noalign{\vskip 0.9mm}
    transit$_{\mathrm{jitter,T3}}$ &  & $-0.00097_{-0.00006}^{+0.00194}$ & $\mathcal{U}(-0.01, 0.01)$ & $-0.00098_{-0.00004}^{+0.00004}$ & $\mathcal{U}(-0.01, 0.01)$  \\ \noalign{\vskip 0.9mm}
    transit$_{\mathrm{jitter,T4}}$ &  & $0.00109_{-0.00217}^{+0.00006}$ & $\mathcal{U}(-0.01, 0.01)$ & $0.00112_{-0.00004}^{+0.00004}$ & $\mathcal{U}(-0.01, 0.01)$  \\ \noalign{\vskip 0.9mm}
    transit$_{\mathrm{jitter,T8}}$ &  & $-0.0002_{-0.0005}^{+0.0009}$ & $\mathcal{U}(-0.01, 0.01)$ & $-0.0003_{-0.0005}^{+0.0009}$ & $\mathcal{U}(-0.01, 0.01)$  \\ \noalign{\vskip 0.9mm}
    transit$_{\mathrm{jitter,T28}}$ &  & $0.00000_{-0.00007}^{+0.00007}$ & $\mathcal{U}(-0.01, 0.01)$ & $0.00001_{-0.00009}^{+0.00010}$ & $\mathcal{U}(-0.01, 0.01)$  \\ \noalign{\vskip 0.9mm}
    transit$_{\mathrm{jitter,T29}}$ &  & $-0.00001_{-0.00008}^{+0.00009}$ & $\mathcal{U}(-0.01, 0.01)$ & $-0.00000_{-0.00010}^{+0.00010}$ & $\mathcal{U}(-0.01, 0.01)$  \\ \noalign{\vskip 0.9mm
    }transit$_{\mathrm{jitter,T30}}$ &  & $-0.00000_{-0.00009}^{+0.00009}$ & $\mathcal{U}(-0.01, 0.01)$ & $0.00000_{-0.00010}^{+0.00009}$ & $\mathcal{U}(-0.01, 0.01)$  \\ \noalign{\vskip 0.9mm}
    transit$_{\mathrm{jitter,T31}}$ &  & $-0.00002_{-0.00009}^{+0.00010}$ & $\mathcal{U}(-0.01, 0.01)$ & $-0.0000_{-0.0001}^{+0.0001}$ & $\mathcal{U}(-0.01, 0.01)$  \\ \noalign{\vskip 0.9mm}
    transit$_{\mathrm{jitter,T38}}$ &  & $0.00005_{-0.00008}^{+0.00009}$ & $\mathcal{U}(-0.01, 0.01)$ & $0.00001_{-0.00009}^{+0.00009}$ & $\mathcal{U}(-0.01, 0.01)$  \\ \noalign{\vskip 0.9mm
    }transit$_{\mathrm{jitter,ASTEP}}$ &  & $-0.0018_{-0.0003}^{+0.0003}$ & $\mathcal{U}(-0.01, 0.01)$ & $0.0018_{-0.0002}^{+0.0002}$ & $\mathcal{U}(-0.01, 0.01)$  \\ \noalign{\vskip 0.9mm}
    transit$_{\mathrm{jitter,CDK24ND}}$ &  & $0.0022_{-0.0002}^{+0.0002}$ & $\mathcal{U}(-0.01, 0.01)$ & $0.0022_{-0.0002}^{+0.0002}$ & $\mathcal{U}(-0.01, 0.01)$  \\ \noalign{\vskip 0.9mm}
    transit$_{\mathrm{jitter,OM-SSO}}$ &  & $0.0030_{-0.0059}^{+0.0002}$ & $\mathcal{U}(-0.01, 0.01)$ & $-0.0030_{-0.0002}^{+0.0002}$ & $\mathcal{U}(-0.01, 0.01)$  \\ \noalign{\vskip 0.9mm}
    $u_{1, \mathrm{TESS}}$ & & $0.2_{-0.1}^{+0.1}$ & $\mathcal{U}(0.0, 1.0)$ & $0.27_{-0.12}^{+0.10}$ & $\mathcal{U}(0.0, 1.0)$  \\ \noalign{\vskip 0.9mm}
    $u_{2, \mathrm{TESS}}$ & & $0.2_{-0.1}^{+0.2}$ & $\mathcal{U}(0.0, 1.0)$ & $0.1_{-0.1}^{+0.2}$ & $\mathcal{U}(0.0, 1.0)$  \\ \noalign{\vskip 0.9mm}
    $u_{1, \mathrm{ASTEP}}$ & & $0.4_{-0.2}^{+0.1}$ & $\mathcal{U}(0.0, 1.0)$ & $0.4_{-0.2}^{+0.2}$ & $\mathcal{U}(0.0, 1.0)$  \\ \noalign{\vskip 0.9mm}
    $u_{2, \mathrm{ASTEP}}$ & & $0.6_{-0.2}^{+0.2}$ & $\mathcal{U}(0.0, 1.0)$ & $0.6_{-0.4}^{+0.3}$ & $\mathcal{U}(0.0, 1.0)$  \\ \noalign{\vskip 0.9mm}
    $u_{1, \mathrm{CDK24ND}}$ & & $0.6_{-0.2}^{+0.2}$ & $\mathcal{U}(0.0, 1.0)$ & $0.6_{-0.2}^{+0.2}$ & $\mathcal{U}(0.0, 1.0)$  \\ \noalign{\vskip 0.9mm}
    $u_{2, \mathrm{CDK24ND}}$ & & $0.6_{-0.2}^{+0.2}$ & $\mathcal{U}(0.0, 1.0)$ & $0.6_{-0.3}^{+0.3}$ & $\mathcal{U}(0.0, 1.0)$  \\ \noalign{\vskip 0.9mm}
    $u_{1, \mathrm{OM-SSO}}$ & & $0.2_{-0.1}^{+0.2}$ & $\mathcal{U}(0.0, 1.0)$ & $0.14_{-0.10}^{+0.14}$ & $\mathcal{U}(0.0, 1.0)$  \\ \noalign{\vskip 0.9mm}
    $u_{2, \mathrm{OM-SSO}}$ & & $0.3_{-0.2}^{+0.2}$ & $\mathcal{U}(0.0, 1.0)$ & $0.4_{-0.2}^{+0.3}$ & $\mathcal{U}(0.0, 1.0)$  \\ \noalign{\vskip 0.9mm}
    \hline \noalign{\vskip 0.7mm}
    $m$ & $M_{\mathrm{jup}}$ & \multicolumn{2}{c}{$3.1_{-0.3}^{+0.3}$} & \multicolumn{2}{c}{$3.00_{-0.09}^{+0.10}$} \\ \noalign{\vskip 0.9mm}
    
    $r$ & $R_{\mathrm{jup}}$ & \multicolumn{2}{c}{$0.88_{-0.02}^{+0.02}$} & \multicolumn{2}{c}{$0.88_{-0.02}^{+0.02}$} \\ \noalign{\vskip 0.9mm}
    
    $a$ & au & \multicolumn{2}{c}{$0.0831_{-0.0007}^{+0.0007}$} & \multicolumn{2}{c}{$0.0831_{-0.0007}^{+0.0007}$} \\ \noalign{\vskip 0.9mm}
    
    $\rho$ & $\mathrm{g\,cm}^{-3}$ & \multicolumn{2}{c}{$5.5_{-0.6}^{+0.6}$} & \multicolumn{2}{c}{$5.4_{-0.4}^{+0.3}$} \\ \noalign{\vskip 0.9mm}
    
    $T_{\mathrm{eq}}$ & K & \multicolumn{2}{c}{$1080_{-10}^{+10}$} & \multicolumn{2}{c}{$1080_{-10}^{+10}$} \\ \noalign{\vskip 0.9mm}
    
        \hline\hline \noalign{\vskip 0.5mm}
        \end{tabular}}
            \label{tab:results_TOI-2416}
        \end{table*}

        \begin{table*}[]
            \centering
            \caption{Results from the NS run for TOI-2524. Listed are likelihood parameters, priors, posteriors and derived parameters for both a circular and eccentric model. $\Delta$ln$\mathcal{Z}$ is defined as ln$\mathcal{Z}_{\mathrm{ecc}}$ - ln$\mathcal{Z}_{\mathrm{circ}}$. \tess sector numbers are denoted as "TX" for transit offset and jitter parameters.}
            \resizebox{\textwidth}{!}{
            \begin{tabular}{lcrlrl}
            \hline\hline  \noalign{\vskip 0.7mm}
                  Parameter && \multicolumn{2}{c}{Circular Model} & \multicolumn{2}{c}{Eccentric Model} \\
                  \hspace{0.0 mm} &  \hspace{0.0 mm} & Posterior  & Prior  & Posterior  & Prior  \\
            \hline \noalign{\vskip 0.7mm}
        ln$\mathcal{Z}$ &  & \multicolumn{2}{c}{148,587.37} & \multicolumn{2}{c}{148,585.14} \\ \noalign{\vskip 0.9mm}
        $\Delta$ln$\mathcal{Z}$ &  & \multicolumn{4}{c}{-2.23} \\ \noalign{\vskip 0.9mm}
        \hline \noalign{\vskip 0.7mm}
        
        RV$_{\mathrm{offset}}$ & $\mathrm{m\,s}^{-1}$ & $9434_{-3}^{+3}$ & $\mathcal{U}(9400.0, 9450.0)$ & $9436_{-4}^{+6}$ & $\mathcal{U}(9400.0, 9450.0)$  \\ \noalign{\vskip 0.9mm}
        RV$_{\mathrm{jitter}}$ & $\mathrm{m\,s}^{-1}$ & $-0_{-5}^{+5}$ & $\mathcal{U}(-10.0, 30.0)$ & $8_{-4}^{+7}$ & $\mathcal{U}(-10.0, 30.0)$  \\ \noalign{\vskip 0.9mm}
        $K$ & $\mathrm{m\,s}^{-1}$ & $67_{-4}^{+4}$ & $\mathcal{U}(45.0, 90.0)$ & $67_{-6}^{+4}$ & $\mathcal{U}(45.0, 90.0)$  \\ \noalign{\vskip 0.9mm}
        $P$ & days & $7.18585_{-0.00001}^{+0.00001}$ & $\mathcal{U}(7.185, 7.187)$ & $7.18585_{-0.00003}^{+0.00002}$ & $\mathcal{U}(7.185, 7.187)$  \\ \noalign{\vskip 0.9mm}
        $e$ &  & $0$ & fixed & $0.06_{-0.03}^{+0.03}$ & $\mathcal{U}(0.0, 0.5)$  \\ \noalign{\vskip 0.9mm}
        $\omega$ & degrees & $-$ & undefined & $101_{-92}^{+130}$ & $\mathcal{U}(-90.0, 360.0)$  \\ \noalign{\vskip 0.9mm}
        $i$ & degrees & $89.4_{-0.4}^{+0.4}$ & $\mathcal{U}(85.0, 92.0)$ & $90.2_{-0.5}^{+0.4}$ & $\mathcal{U}(85.0, 92.0)$  \\ \noalign{\vskip 0.9mm}
        $t_0$ & days & $2458550.155_{-0.001}^{+0.002}$ & $\mathcal{U}(2458550.0, 2458550.3)$ & $2458550.156_{-0.002}^{+0.003}$ & $\mathcal{U}(2458550.0, 2458550.3)$  \\ \noalign{\vskip 0.9mm}
        $a/R_\star$ &  & $14.3_{-0.3}^{+0.2}$ & $\mathcal{U}(5.0, 30.0)$ & $14.5_{-0.4}^{+0.9}$ & $\mathcal{U}(5.0, 30.0)$  \\ \noalign{\vskip 0.9mm}
        $r/R_\star$ &  & $0.092_{-0.001}^{+0.001}$ & $\mathcal{U}(0.0, 0.15)$ & $0.0928_{-0.0007}^{+0.0011}$ & $\mathcal{U}(0.0, 0.15)$  \\ \noalign{\vskip 0.9mm}
        transit$_{\mathrm{offset,T9}}$ &  & $0.00022_{-0.00006}^{+0.00006}$ & $\mathcal{U}(-0.001, 0.001)$ & $0.00021_{-0.00005}^{+0.00006}$ & $\mathcal{U}(-0.001, 0.001)$  \\ \noalign{\vskip 0.9mm
        }transit$_{\mathrm{offset,T35}}$ &  & $0.00044_{-0.00005}^{+0.00004}$ & $\mathcal{U}(-0.001, 0.001)$ & $0.00042_{-0.00005}^{+0.00004}$ & $\mathcal{U}(-0.001, 0.001)$  \\ \noalign{\vskip 0.9mm}
        transit$_{\mathrm{offset,T45}}$ &  & $0.00030_{-0.00003}^{+0.00003}$ & $\mathcal{U}(-0.001, 0.001)$ & $0.00028_{-0.00002}^{+0.00002}$ & $\mathcal{U}(-0.001, 0.001)$  \\ \noalign{\vskip 0.9mm}
        transit$_{\mathrm{offset,T46}}$ &  & $0.00030_{-0.00002}^{+0.00003}$ & $\mathcal{U}(-0.001, 0.001)$ & $0.00029_{-0.00002}^{+0.00003}$ & $\mathcal{U}(-0.001, 0.001)$  \\ \noalign{\vskip 0.9mm}
        transit$_{\mathrm{offset,El-Sauce}}$ &  & $0.0008_{-0.0004}^{+0.0003}$ & $\mathcal{U}(-0.001, 0.001)$ & $0.0011_{-0.0002}^{+0.0003}$ & $\mathcal{U}(-0.001, 0.001)$  \\ \noalign{\vskip 0.9mm}
        transit$_{\mathrm{offset,OM-ES}}$ &  & $0.0014_{-0.0002}^{+0.0002}$ & $\mathcal{U}(-0.001, 0.001)$ & $0.0014_{-0.0002}^{+0.0002}$ & $\mathcal{U}(-0.001, 0.001)$  \\ \noalign{\vskip 0.9mm}
        transit$_{\mathrm{jitter,T9}}$ &  & $-0.00131_{-0.00006}^{+0.00006}$ & $\mathcal{U}(-0.02, 0.02)$ & $-0.00130_{-0.00006}^{+0.00007}$ & $\mathcal{U}(-0.02, 0.02)$  \\ \noalign{\vskip 0.9mm}
        transit$_{\mathrm{jitter,T35}}$ &  & $0.00254_{-0.00512}^{+0.00008}$ & $\mathcal{U}(-0.02, 0.02)$ & $-0.0025_{-0.0001}^{+0.0051}$ & $\mathcal{U}(-0.02, 0.02)$  \\ \noalign{\vskip 0.9mm}
        transit$_{\mathrm{jitter,T45}}$ &  & $0.00002_{-0.00009}^{+0.00010}$ & $\mathcal{U}(-0.02, 0.02)$ & $0.0000_{-0.0001}^{+0.0001}$ & $\mathcal{U}(-0.02, 0.02)$  \\ \noalign{\vskip 0.9mm}
        transit$_{\mathrm{jitter,T46}}$ &  & $0.0000_{-0.0002}^{+0.0001}$ & $\mathcal{U}(-0.02, 0.02)$ & $0.00002_{-0.00011}^{+0.00008}$ & $\mathcal{U}(-0.02, 0.02)$  \\ \noalign{\vskip 0.9mm}
        transit$_{\mathrm{jitter,El-Sauce}}$ &  & $0.0037_{-0.0077}^{+0.0004}$ & $\mathcal{U}(-0.02, 0.02)$ & $0.0038_{-0.0002}^{+0.0002}$ & $\mathcal{U}(-0.02, 0.02)$  \\ \noalign{\vskip 0.9mm}
        transit$_{\mathrm{jitter,OM-ES}}$ &  & $0.0019_{-0.0039}^{+0.0003}$ & $\mathcal{U}(-0.02, 0.02)$ & $0.0020_{-0.0004}^{+0.0003}$ & $\mathcal{U}(-0.02, 0.02)$  \\ \noalign{\vskip 0.9mm}
        $u_{1, \mathrm{TESS}}$ & & $0.13_{-0.08}^{+0.08}$ & $\mathcal{U}(0.0, 1.0)$ & $0.15_{-0.07}^{+0.09}$ & $\mathcal{U}(0.0, 1.0)$  \\ \noalign{\vskip 0.9mm}
        $u_{2, \mathrm{TESS}}$ & & $0.5_{-0.2}^{+0.2}$ & $\mathcal{U}(0.0, 1.0)$ & $0.31_{-0.08}^{+0.08}$ & $\mathcal{U}(0.0, 1.0)$  \\ \noalign{\vskip 0.9mm}
        $u_{1, \mathrm{El-Sauce}}$ & & $0.5_{-0.1}^{+0.1}$ & $\mathcal{U}(0.0, 1.0)$ & $0.44_{-0.10}^{+0.10}$ & $\mathcal{U}(0.0, 1.0)$  \\ \noalign{\vskip 0.9mm}
        $u_{2, \mathrm{El-Sauce}}$ & & $0.6_{-0.3}^{+0.3}$ & $\mathcal{U}(0.0, 1.0)$ & $0.5_{-0.2}^{+0.2}$ & $\mathcal{U}(0.0, 1.0)$  \\ \noalign{\vskip 0.9mm}
        $u_{1, \mathrm{OM-ES}}$ & & $0.5_{-0.1}^{+0.1}$ & $\mathcal{U}(0.0, 1.0)$ & $0.6_{-0.2}^{+0.2}$ & $\mathcal{U}(0.0, 1.0)$  \\ \noalign{\vskip 0.9mm}
        $u_{2, \mathrm{OM-ES}}$ & & $0.7_{-0.2}^{+0.2}$ & $\mathcal{U}(0.0, 1.0)$ & $0.5_{-0.2}^{+0.2}$ & $\mathcal{U}(0.0, 1.0)$  \\ \noalign{\vskip 0.9mm}
        \hline \noalign{\vskip 0.7mm}
        $m$ & $M_{\mathrm{jup}}$ & \multicolumn{2}{c}{$0.64_{-0.04}^{+0.04}$} & \multicolumn{2}{c}{$0.64_{-0.04}^{+0.06}$} \\ \noalign{\vskip 0.9mm}
        
        $r$ & $R_{\mathrm{jup}}$ & \multicolumn{2}{c}{$1.00_{-0.03}^{+0.03}$} & \multicolumn{2}{c}{$1.01_{-0.02}^{+0.02}$} \\ \noalign{\vskip 0.9mm}
        
        $a$ & au & \multicolumn{2}{c}{$0.0730_{-0.0007}^{+0.0007}$} & \multicolumn{2}{c}{$0.0730_{-0.0007}^{+0.0007}$} \\ \noalign{\vskip 0.9mm}
        
        $\rho$ & $\mathrm{g\,cm}^{-3}$ & \multicolumn{2}{c}{$0.79_{-0.09}^{+0.08}$} & \multicolumn{2}{c}{$0.76_{-0.08}^{+0.08}$} \\ \noalign{\vskip 0.9mm}
        
        $T_{\mathrm{eq}}$ & K & \multicolumn{2}{c}{$1100_{-20}^{+20}$} & \multicolumn{2}{c}{$1100_{-20}^{+20}$} \\ \noalign{\vskip 0.9mm}
        
            \hline\hline \noalign{\vskip 0.5mm}
            \end{tabular}}
                \label{tab:results_TOI-2524}
            \end{table*}

\begin{figure}
    \centering
    \includegraphics[width=0.24\textwidth]{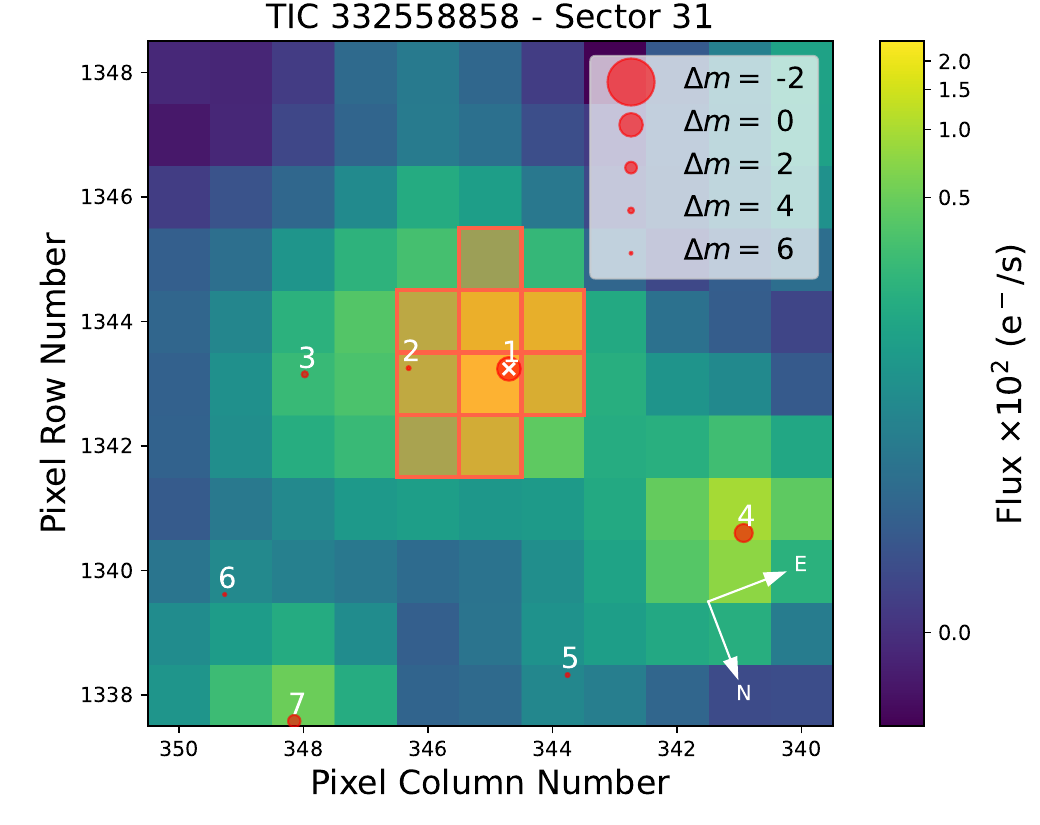}
    \caption{TPF plot for Sector 31 for TOI-2373}
    \label{fig:TPF_2373}
\end{figure}

\begin{figure*}
    \centering
    \includegraphics[width=0.19\textwidth]{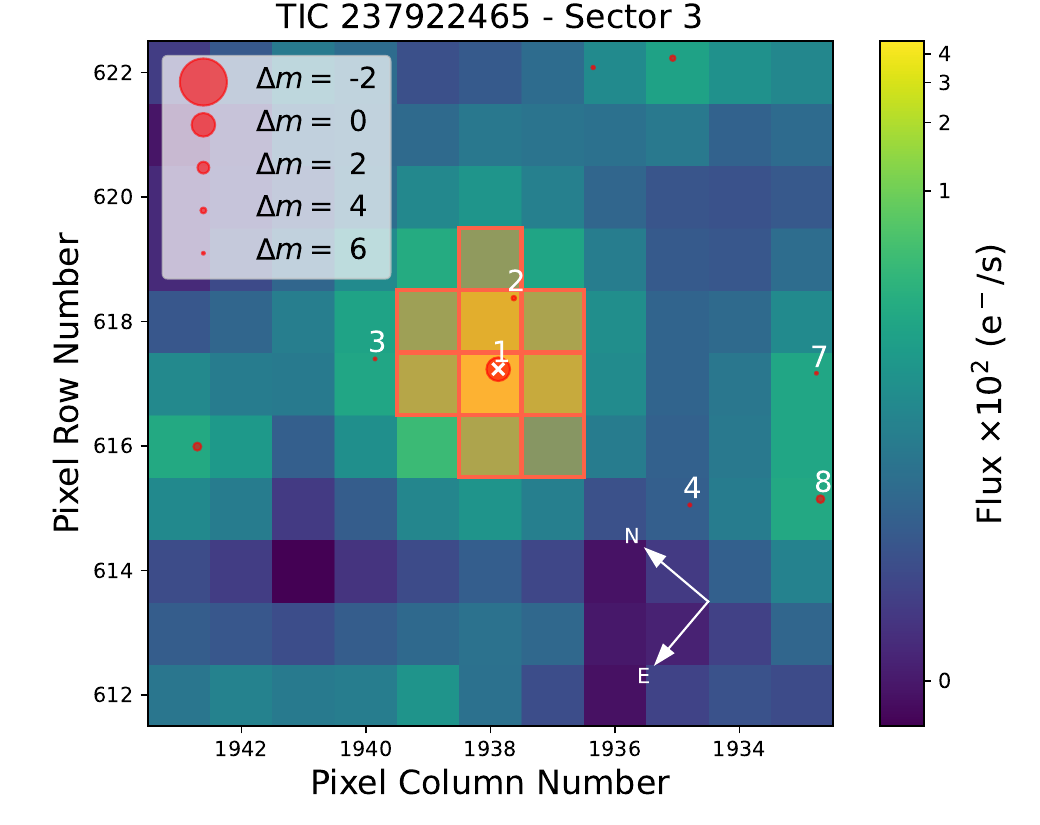}
    \includegraphics[width=0.19\textwidth]{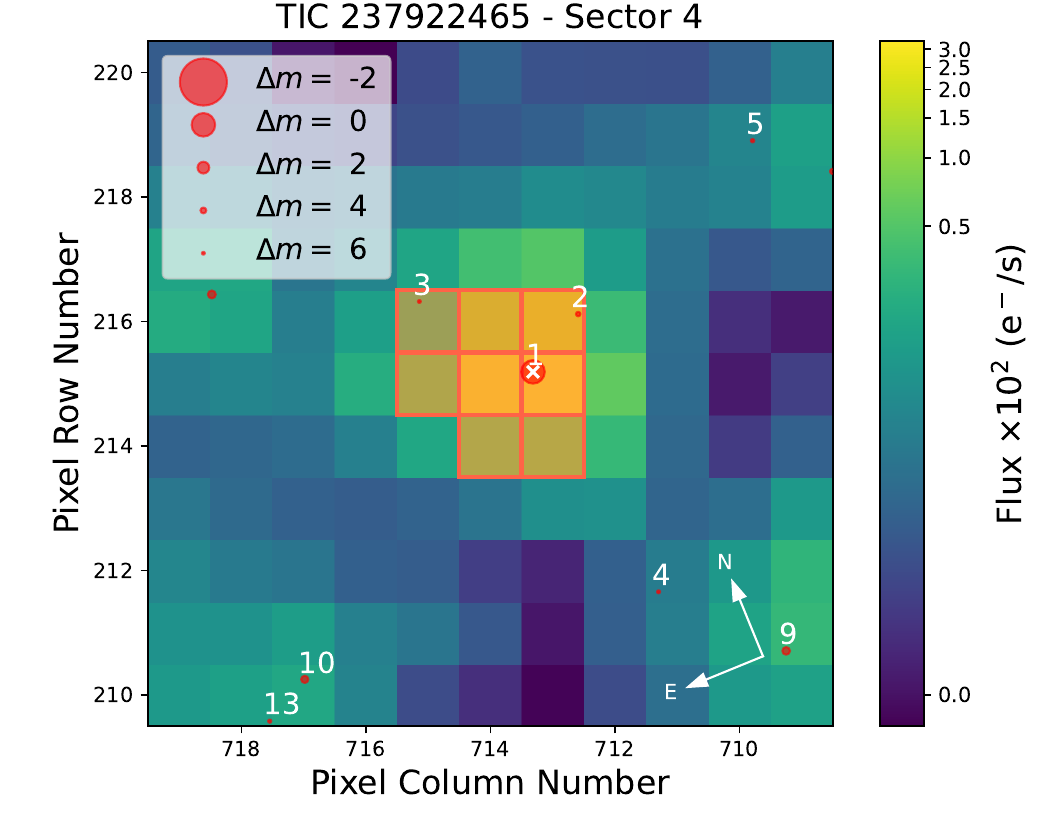}
    \includegraphics[width=0.19\textwidth]{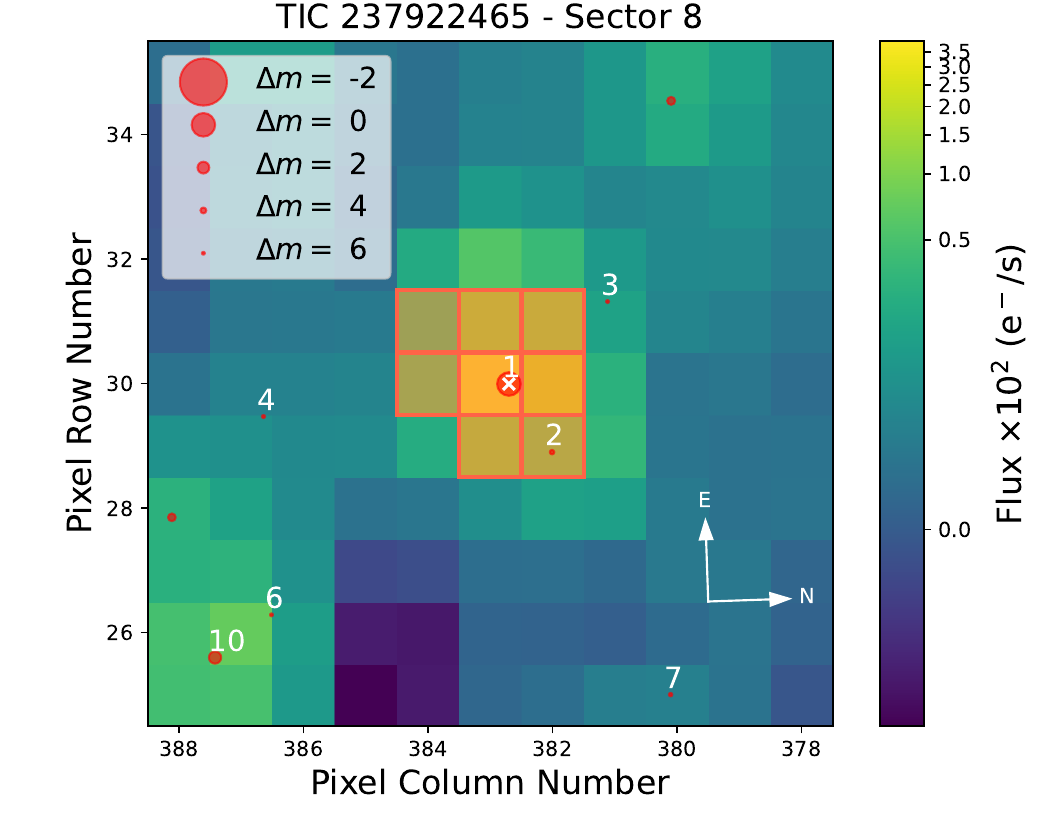}
    \includegraphics[width=0.19\textwidth]{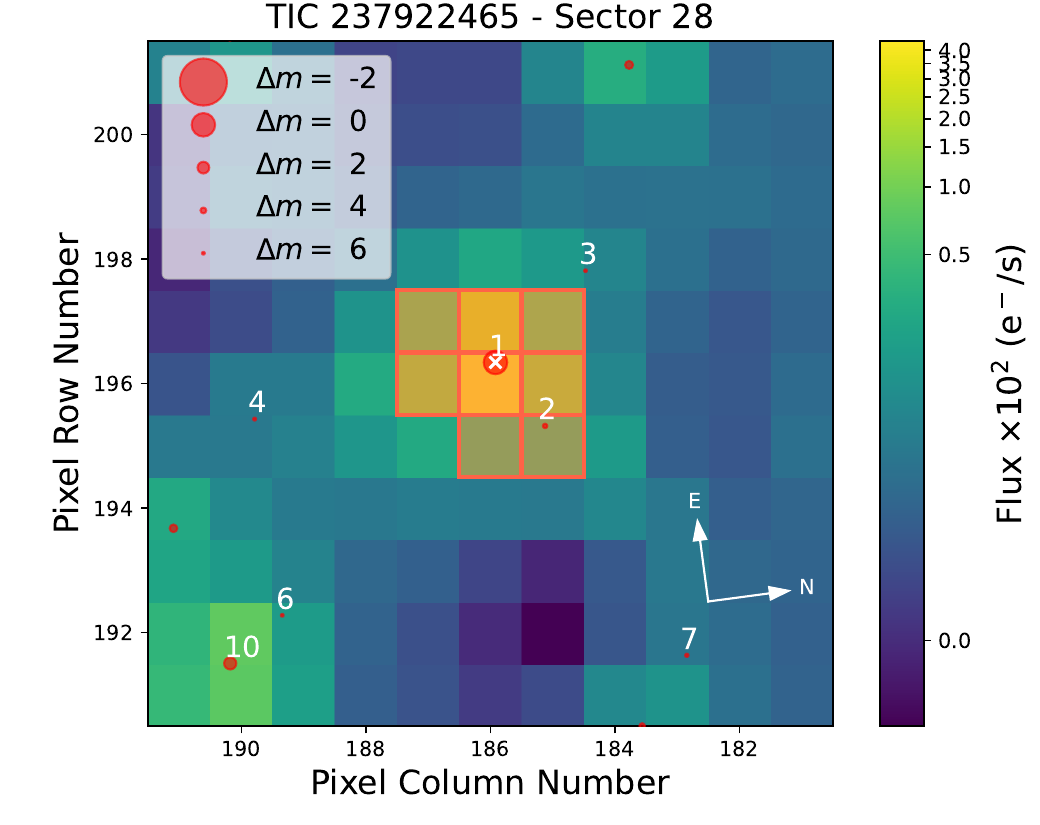}\\
    \includegraphics[width=0.19\textwidth]{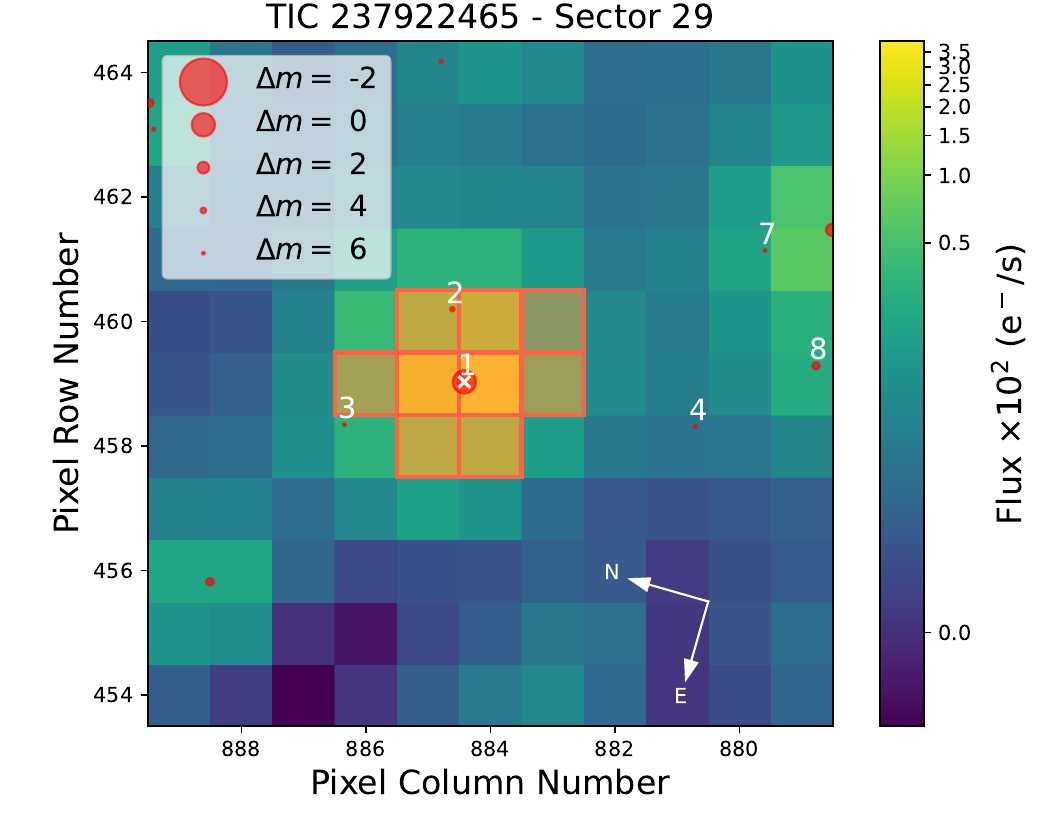}
    \includegraphics[width=0.19\textwidth]{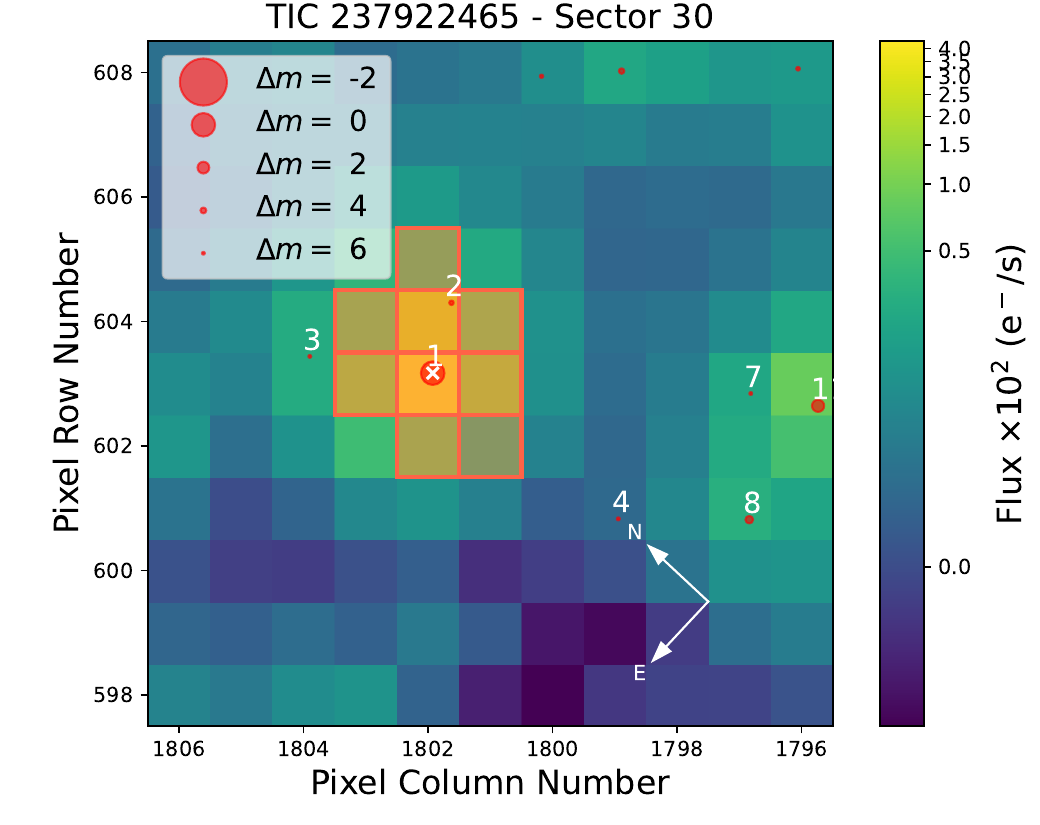}
    \includegraphics[width=0.19\textwidth]{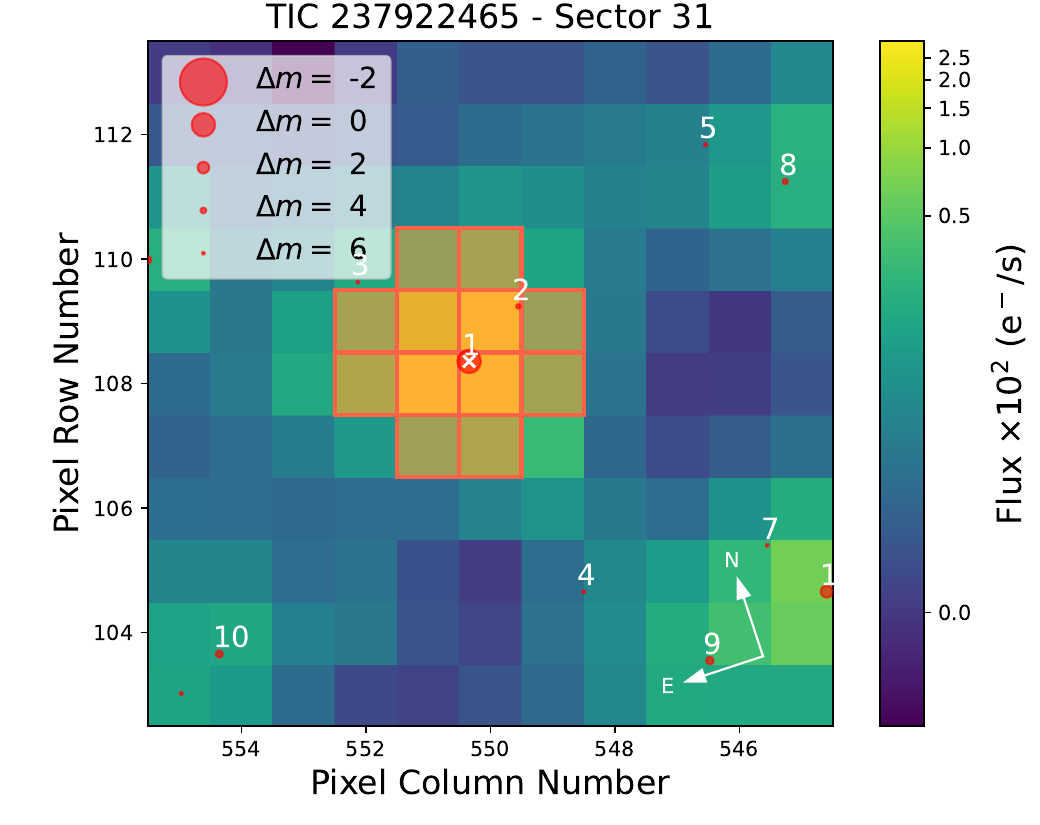}
    \includegraphics[width=0.19\textwidth]{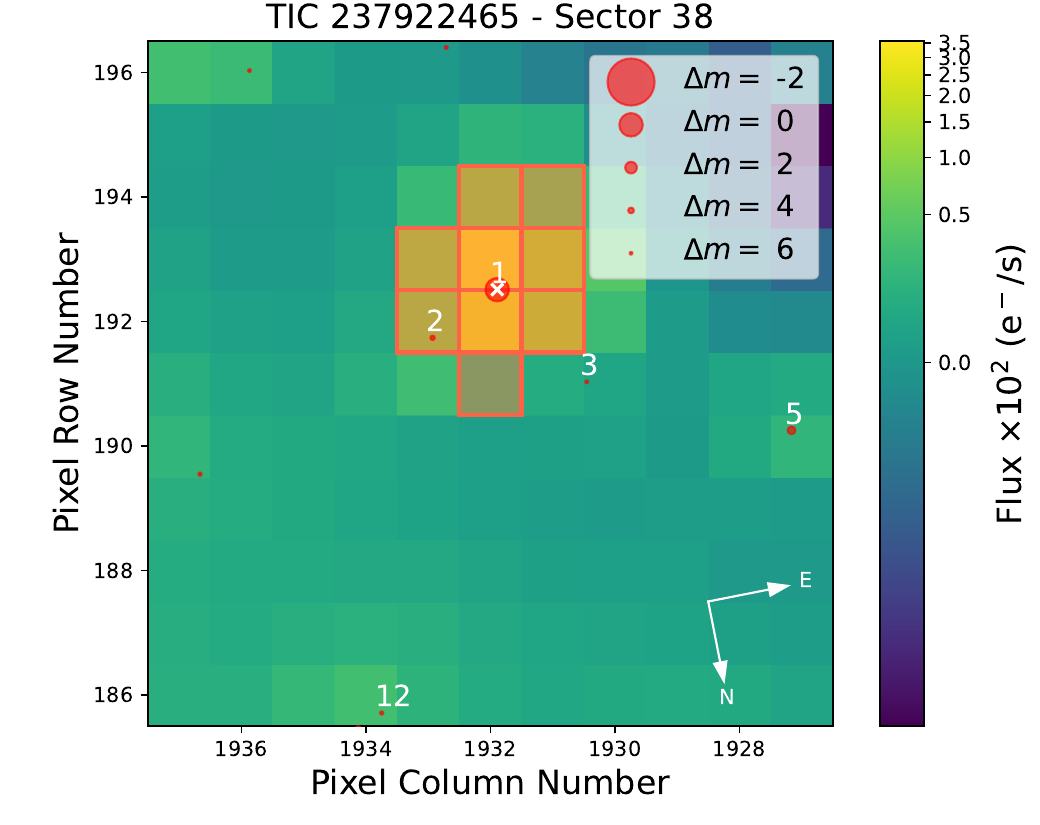}
    \caption{TPF plots for Sectors 3, 4, 8, 28-31, and 38 for TOI-2416}
    \label{fig:TPF_2416}
\end{figure*}

\begin{figure}
    \centering
    \includegraphics[width=0.23\textwidth]{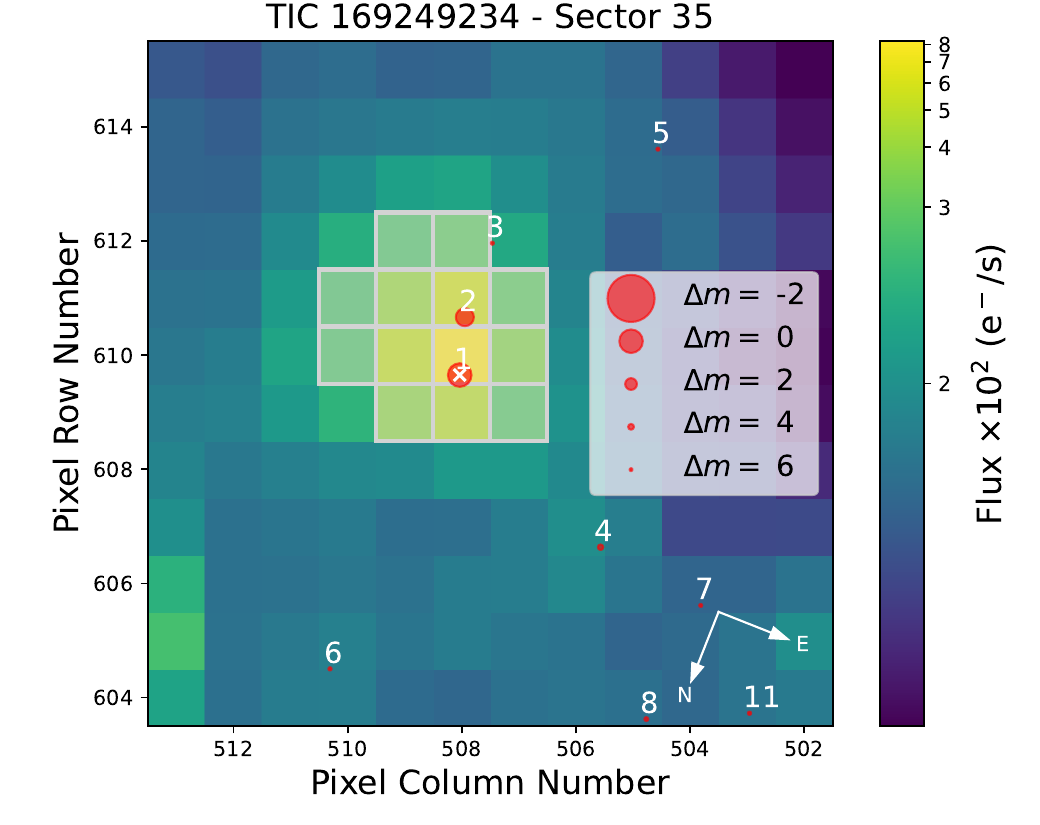}
    \includegraphics[width=0.23\textwidth]{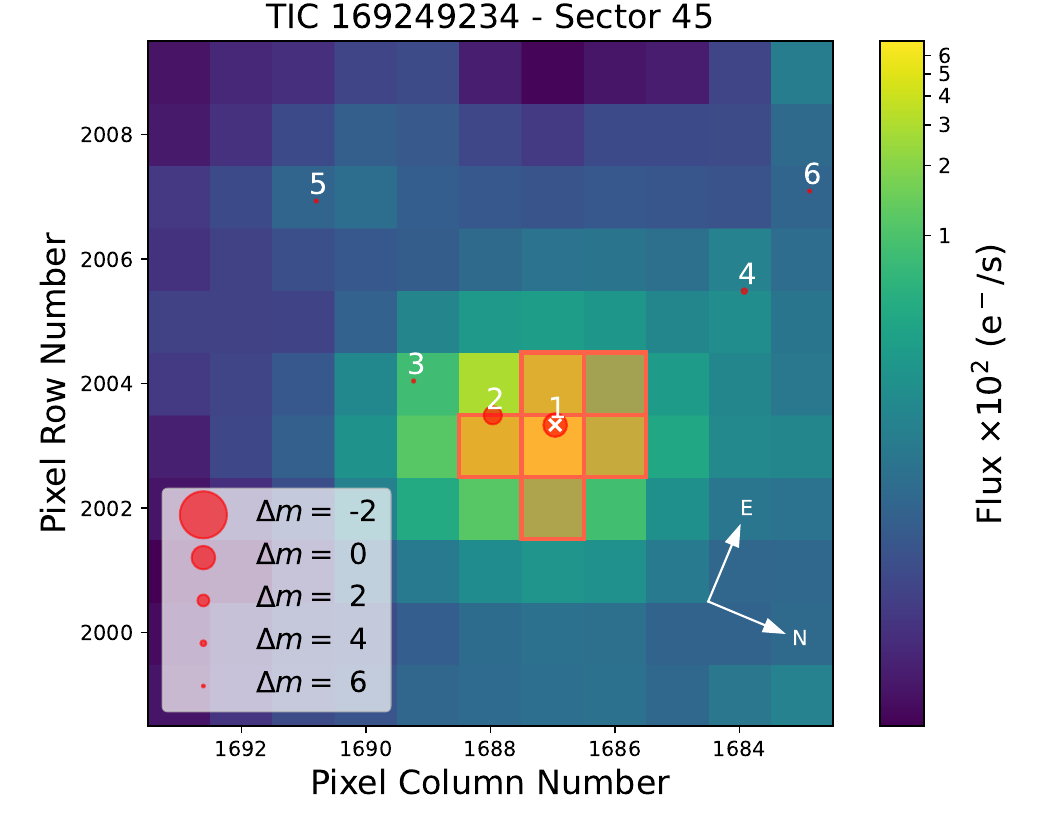}
    \includegraphics[width=0.23\textwidth]{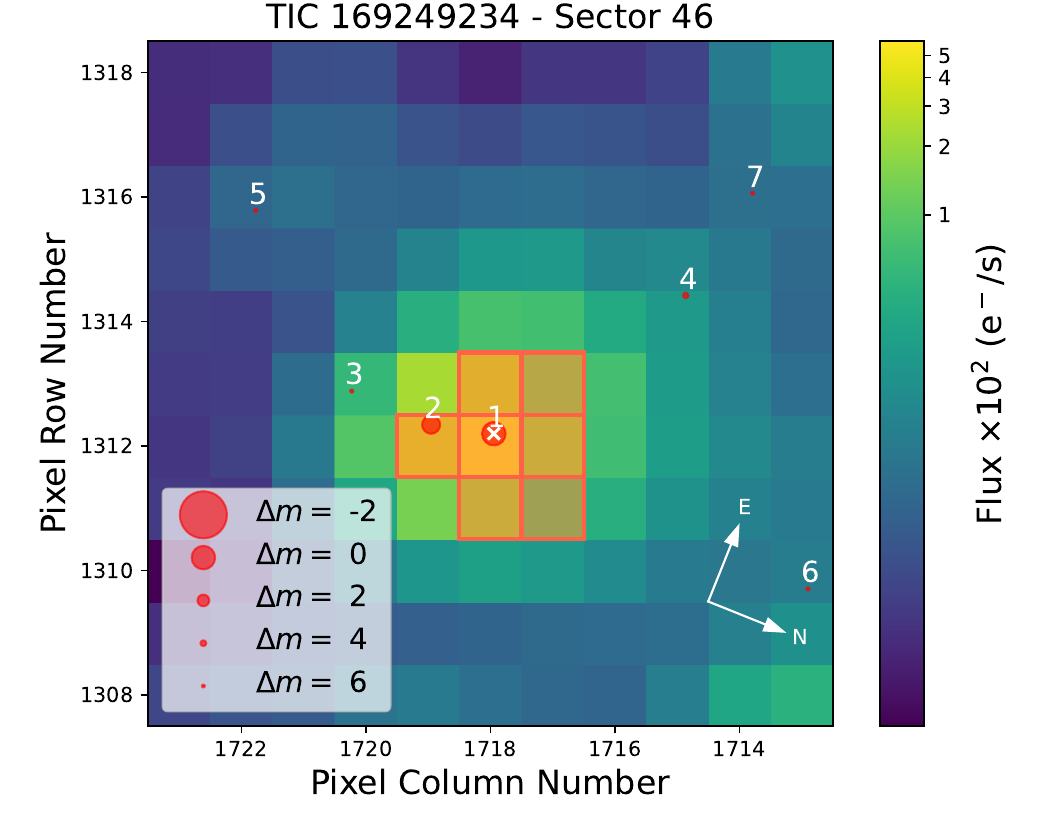}
    \caption{TPF plots for Sectors 35, 45, and 46 for TOI-2524}
    \label{fig:TPF_2524}
\end{figure}

\begin{figure*}
    \centering
    \includegraphics[width=\textwidth]{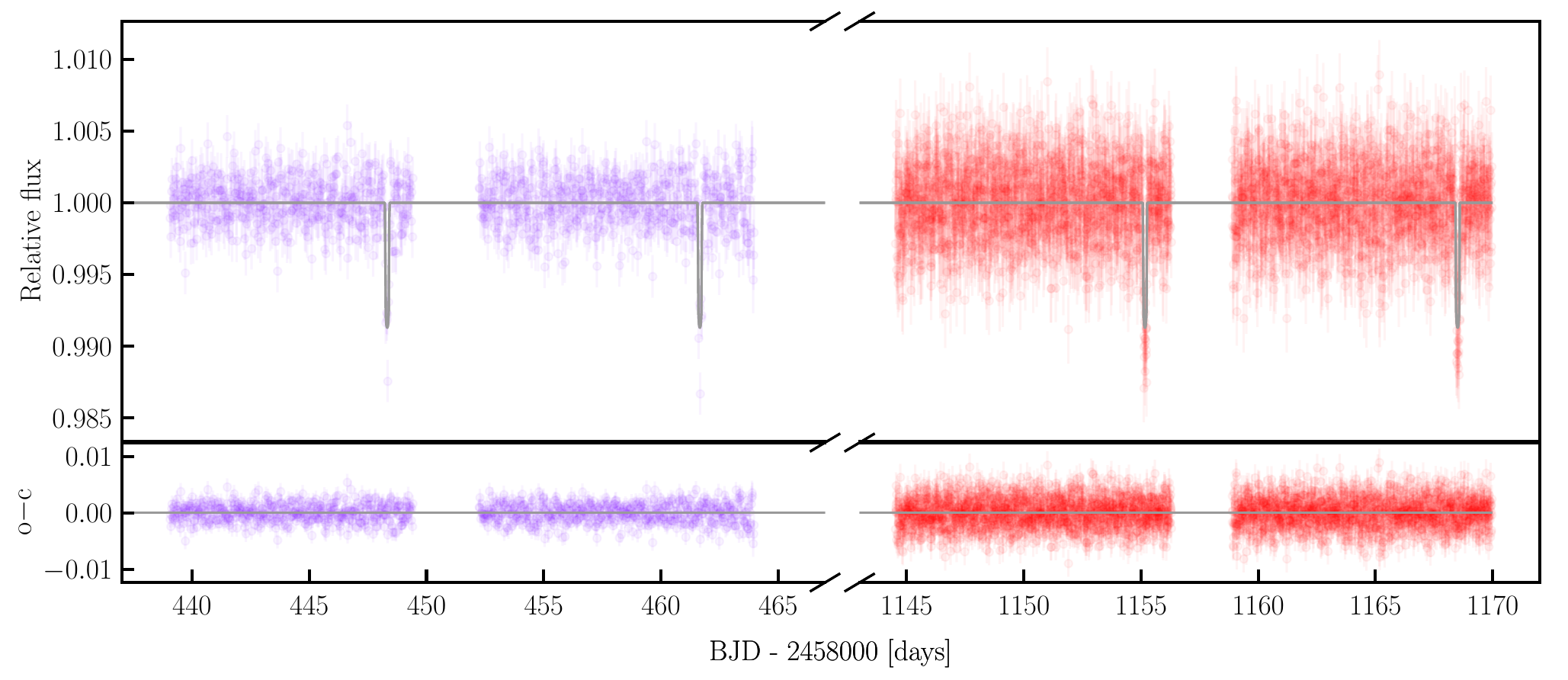}\\
    \includegraphics[width=\textwidth]{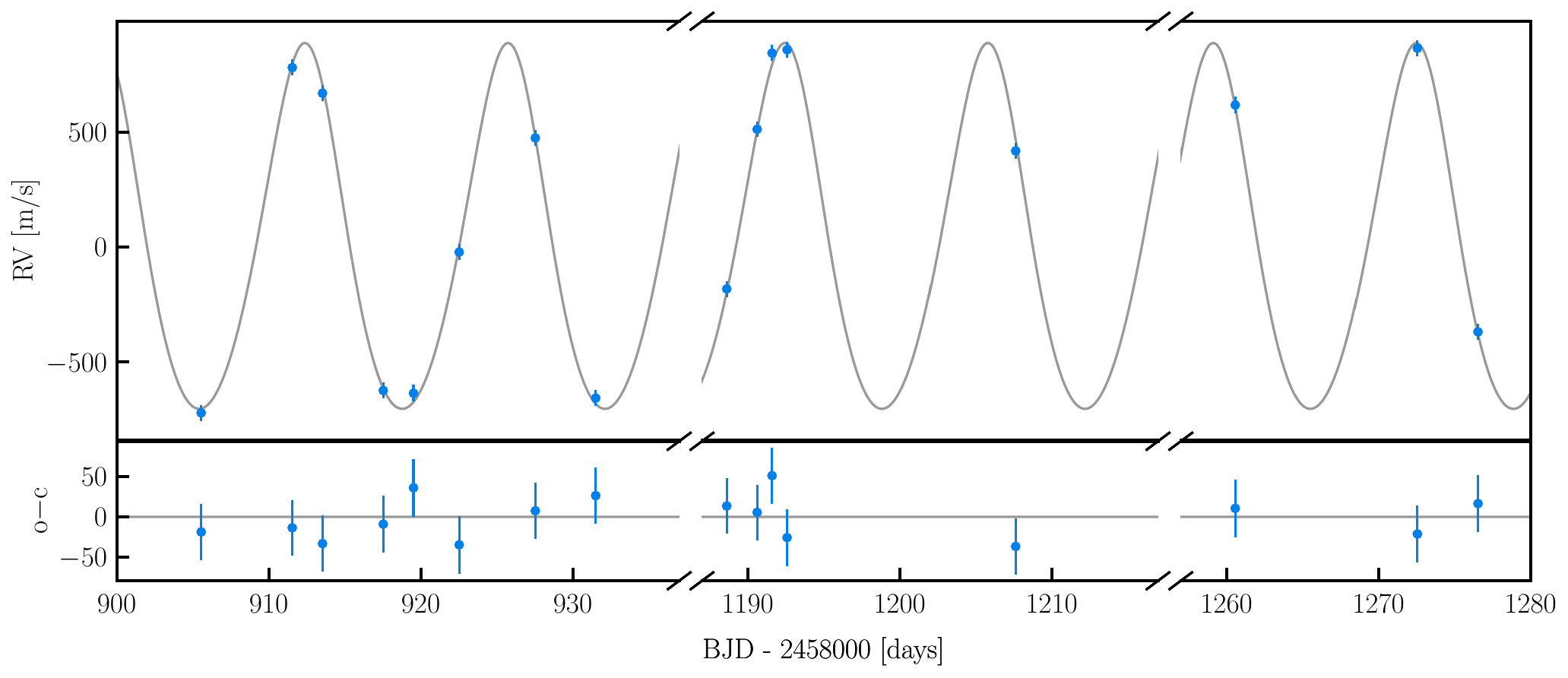}
    \caption{Time series of TOI-2373. {\em Top}: photometric time series. {\em Bottom}: RV time series.}
    \label{fig:TIC332_TS}
\end{figure*}

\begin{figure*}
    \centering
    \includegraphics[width=\textwidth]{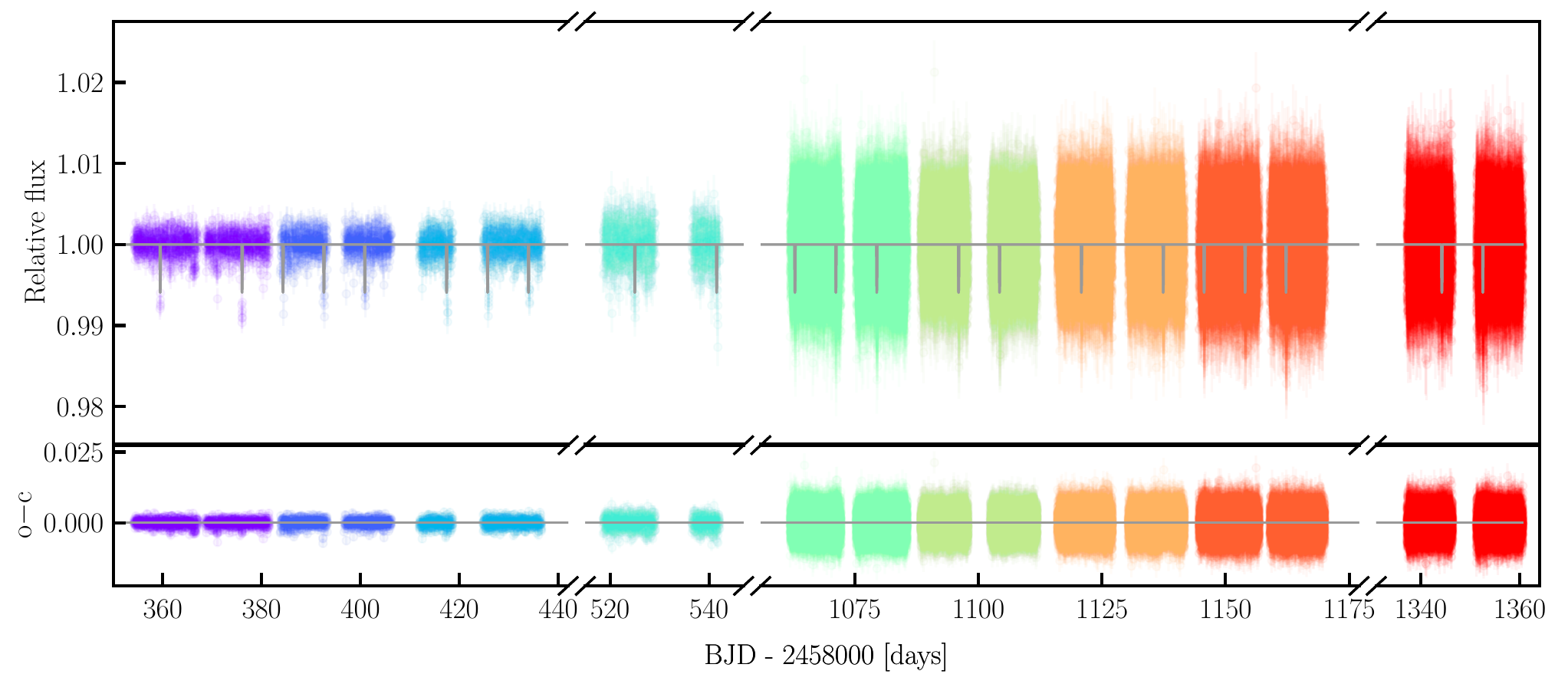}\\
    \includegraphics[width=\textwidth]{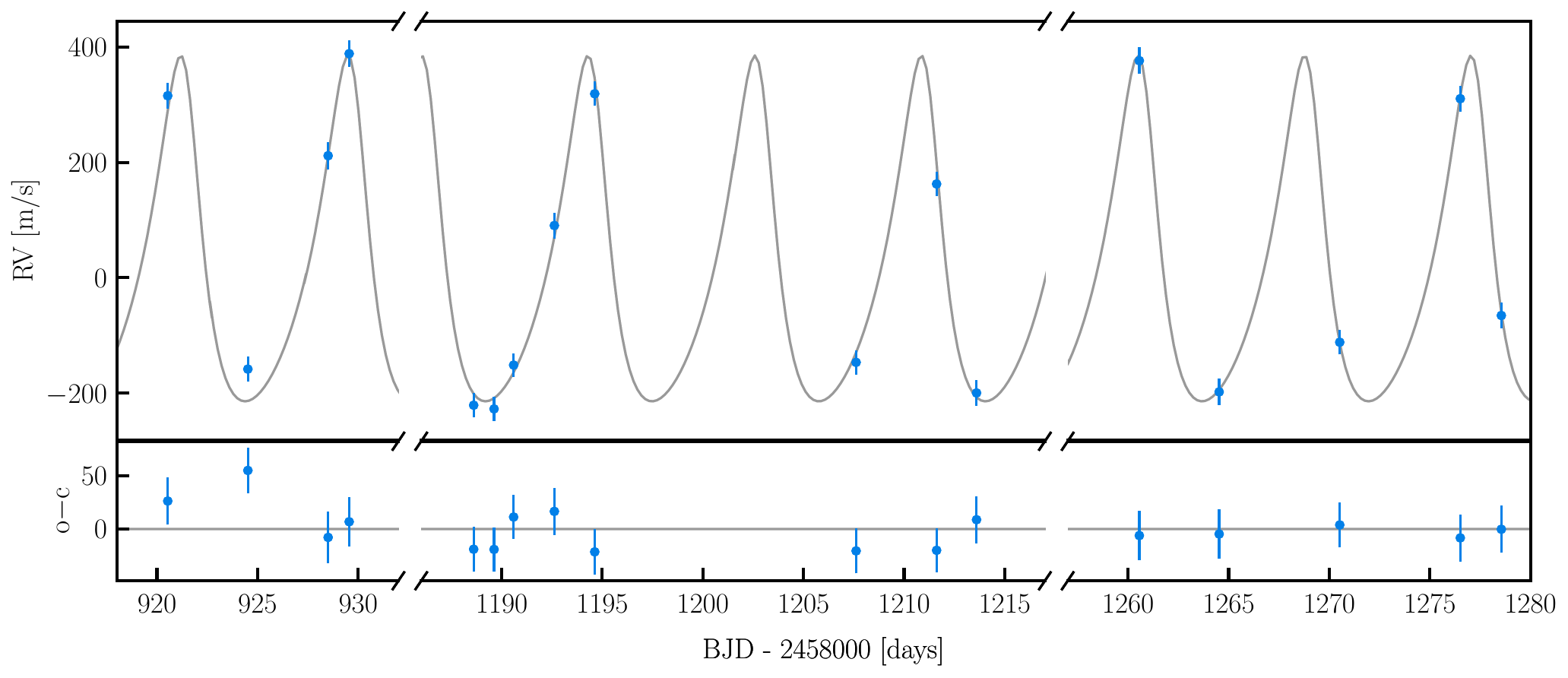}
    \caption{Time series of TOI-2416. {\em Top}: photometric time series. {\em Bottom}: RV time series.}
    \label{fig:TIC237_TS}
\end{figure*}

\begin{figure*}
    \centering
    \includegraphics[width=\textwidth]{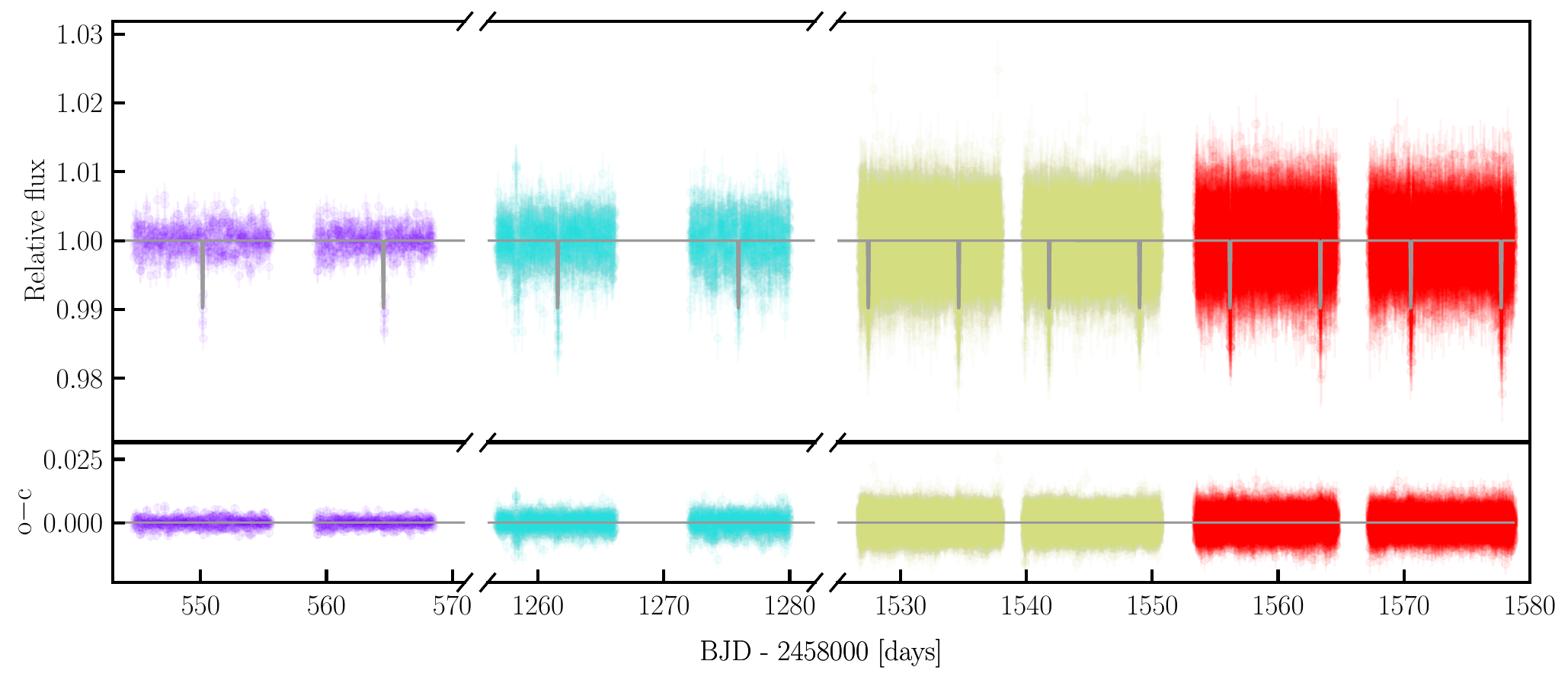}\\
    \includegraphics[width=\textwidth]{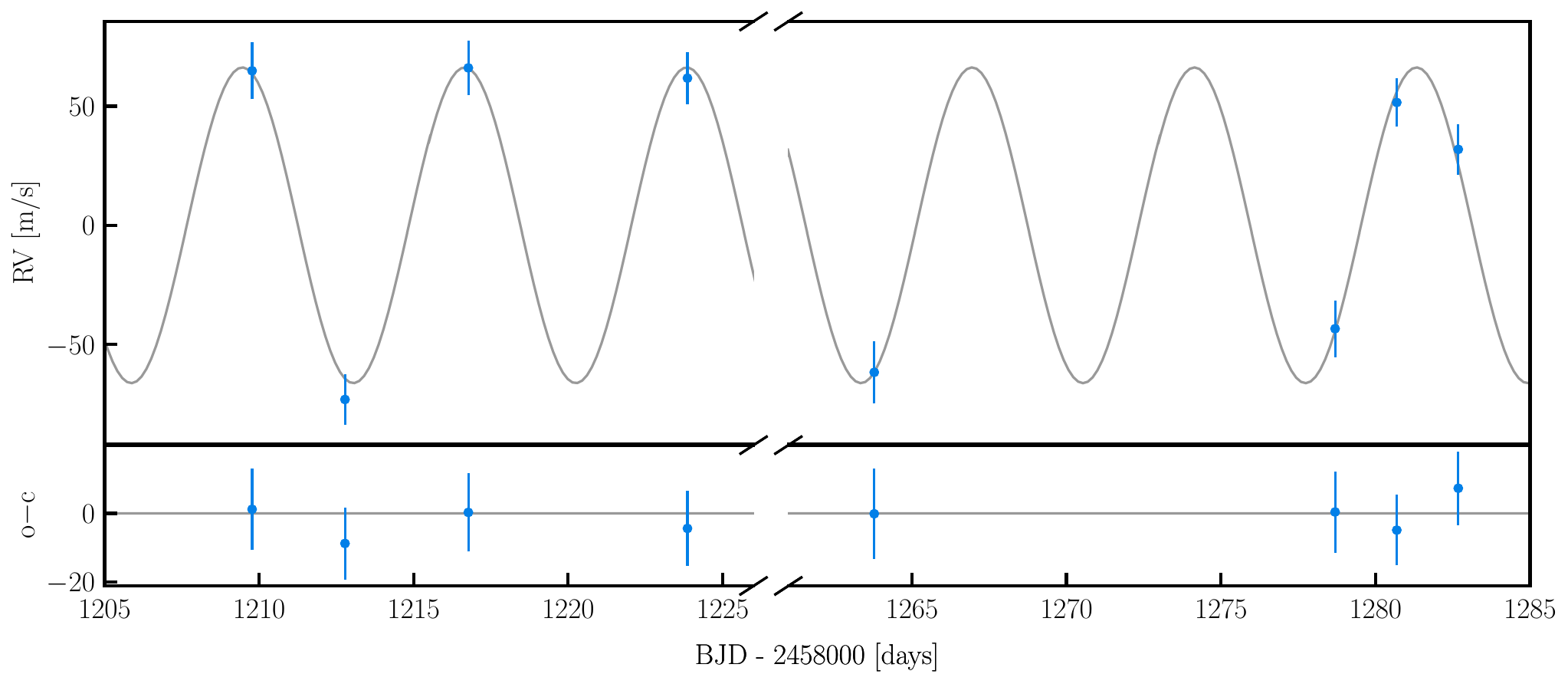}
        \caption{Time series of TOI-2524. {\em Top}: photometric time series. {\em Bottom}: RV time series.}
    \label{fig:TIC169_TS}
\end{figure*}




\end{appendix}

\bibliography{WINE}{}
\bibliographystyle{aasjournal}

\end{document}